\documentclass[12pt,preprint]{aastex}

\shorttitle{DISKS: OBSERVATIONS}
\shortauthors{Farihi}

\begin{document}

\title{CIRCUMSTELLAR DISKS AT WHITE DWARFS: OBSERVATIONS}
		
\author{J. Farihi\altaffilmark{1}}
	
\altaffiltext{1}{Department of Physics \& Astronomy,
			University of Leicester,
			Leicester LE1 7RH, 
			UK; jf123@star.le.ac.uk}
			
\section{INTRODUCTION}

A circumstellar disk or ring is particulate matter that surrounds a star and is primarily confined
to the plane of stellar rotation.  Thus, disks distinguish themselves from spherical clouds or 
envelopes of gas (and dust) that typically surround protostellar objects and evolved giant stars.
Circumstellar disks appear at every stage of stellar evolution, though the origin of the orbiting
material is not always clearly understood.  Pre-main sequence stars accrete material from a
disk that is the flattened remnant of the cloud out of which they formed.  Disks found at young 
stars in subsequent evolutionary stages are the likely site of planet formation, migration, and
sometimes destruction.  Mature main-sequence stars exhibit dusty disks owing to recent or
ongoing energetic collisions among orbiting asteroid or comet analogs, commonly referred 
to as the ``Vega Phenomenon'' (named after the first star observed to have orbiting, non-stellar 
material).  Several first-ascent and asymptotic giant stars are also known to have circumstellar 
disks, through their origins are still debated with hypotheses ranging from debris in a cold 
cometary cloud to consumed stellar companions.  For a thorough (pre-{\em Spitzer}) review, 
see \citet{zuc01}.

White dwarfs are a relatively recent addition to the list of stellar objects with circumstellar disks,
as their intrinsic faintness and the infrared-bright sky have conspired to keep them hidden.  The 
last six years have seen a profusion of white dwarf disk discoveries, due primarily to the 2003 
launch and unprecedented infrared performance of the {\em Spitzer Space Telescope}, while 
ground-based projects with large sky coverage such as the Sloan Digital Sky Survey and the
Two Micron All Sky Survey have also played important roles.  Given the small radii of white 
dwarfs, and because circumstellar material derives its luminosity from the central star, disk 
observations at white dwarfs are challenging.

\section{HISTORY AND BACKGROUND}

Developed in the fifties and sixties, the first photoelectric detectors for infrared\footnote{Here, 
the terms near-, mid-, and far-infrared refer to the wavelength ranges $1-5\,\mu$m, $5-30\,\mu
$m, and $30-200\,\mu$m, though alternative divisions and definitions exist \citep{mcl97}.}  
astronomy were revolutionary \citep{mcl97}, but restricted to luminous sources, especially 
at wavelengths beyond 3\,$\mu$m where the sky itself is bright and variable.  Even so, the 
circumstellar disk at Vega, one of the brightest stars in the sky, was not discovered until 1983 
with the launch of the {\em Infrared Astronomical Satellite (IRAS)}.  Despite the unprecedented 
sensitivity and all-sky coverage of {\em IRAS}, the detection of dust at white dwarfs had to await 
the development of more sensitive infrared detectors and arrays.

\subsection{Early Searches}

Infrared excess emission associated with a star can arise from heated circumstellar dust or
from a self-luminous companion.  Owing to their compact nature, white dwarfs can be easily
outshone by low mass stellar and brown dwarf companions.  This fact led \citet{pro81} to 
search a large sample of nearby white dwarfs for near-infrared $JHK$ photometric excess 
to study the luminosity function of the lowest mass stars and brown dwarfs \citep{pro83,pro82}. 
This was the first search for infrared excess at white dwarfs, and Probst deserves the credit 
for an insight that is now taken for granted and which fostered an abundance of subsequent 
infrared work on white dwarfs.

The first mid-infrared search for excess emission from white dwarfs was similarly motivated 
by the potential identification of brown dwarf companions.  \citet{shi86} carried out a cross 
correlation of {\em IRAS} catalog point sources with known, nearby white dwarfs, but the 
search did not produce any detections.  This is perhaps unsurprising given that {\em IRAS} 
was only sensitive to point sources brighter than 500\,mJy at its shortest wavelength bandpass
of 12$\mu$m, while the brightest white dwarf in the sky (Sirius B) should only be around 5\,mJy 
at that wavelength.

\subsection{The Discovery of Infrared Excess at G29-38}

\begin{figure}
\figurenum{1}
\plotone{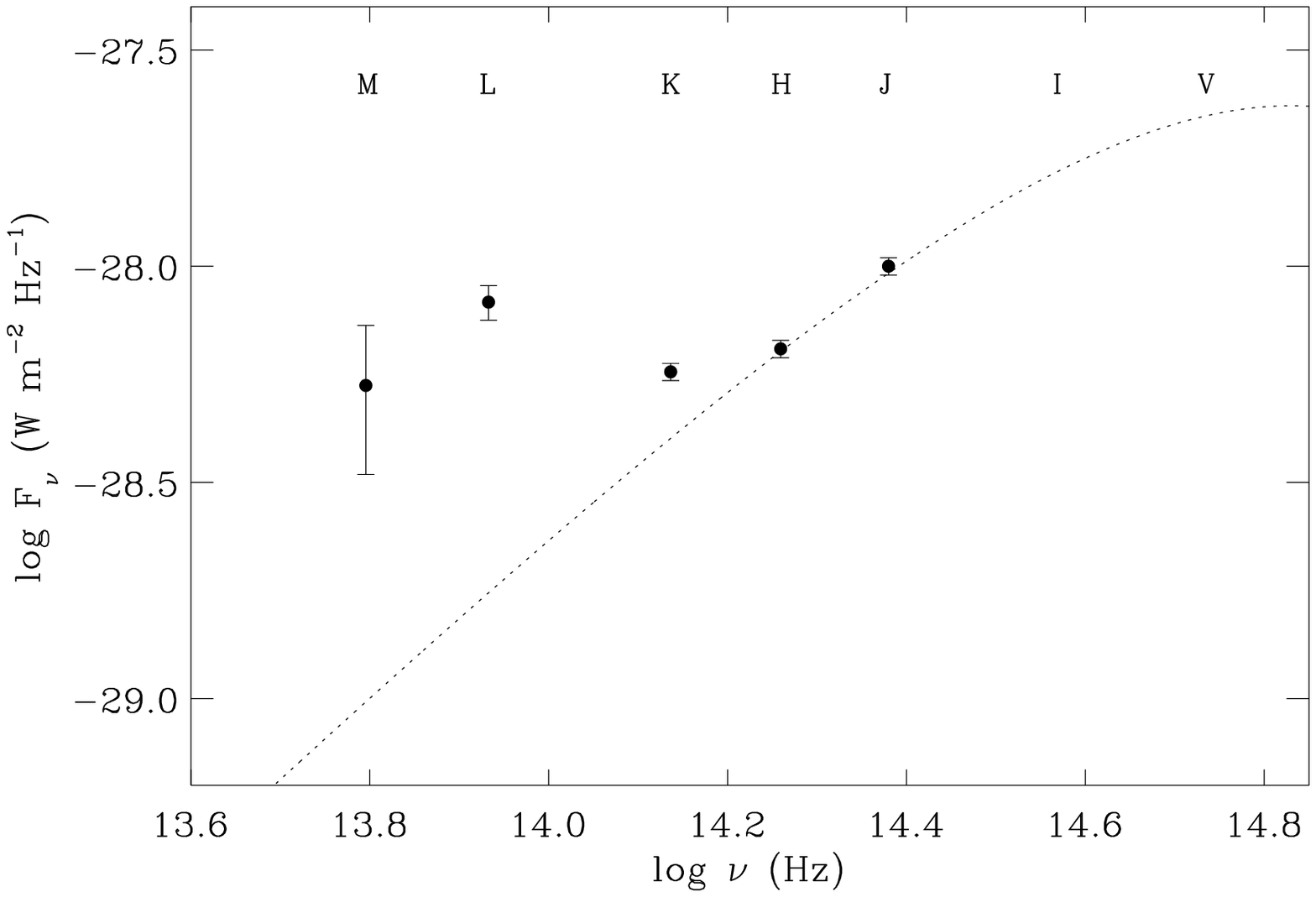}
\caption{The discovery of infrared excess emission from the white dwarf G29-38 made at 
the IRTF.  The plot is a reproduction of the figure presented in \citet{zuc87} and shows their 
measurements with errors in five infrared bandpasses labelled in the figure $J$ (1.25\,$\mu$m), 
$H$ (1.65\,$\mu$m), $K$ (2.2\,$\mu$m), $L$ (3.5\,$\mu$m), and $M$ (4.8\,$\mu$m).  The dotted 
line is a blackbody plotted through the two shortest wavelength fluxes, consistent with the stellar 
photosphere.
\label{fig1}}
\end{figure}

Inspired by the work of Probst and armed with a rapidly evolving set of infrared detectors
atop Mauna Kea at the NASA Infrared Telescope Facility (IRTF), \citet{zuc87} detected the 
first white dwarf with infrared excess that was not associated with a stellar companion; G29-38.  
Photometric observations at three bandpasses longward of 2\,$\mu$m revealed flux in excess 
of that expected for the stellar photosphere of this relatively cool, $T_{\rm eff}=11\,500$\,K white 
dwarf (Figure \ref{fig1}).  Because the prime motivation behind the observations was to search 
for substellar companions, the infrared excess was attributed to a spatially unresolved brown 
dwarf.  Interestingly, circumstellar dust as warm as 1000\,K was not favored due to the likelihood
of rapid dissipation due to ongoing accretion and radiation drag.  \citet{zuc87} prophetically 
noted that if a disk of material were orbiting close enough to reach such high temperatures, 
then spectral signatures of accretion should be seen (at the time, its atmospheric metals had 
not yet been detected).

Over the next few years, the infrared emission at G29-38 was studied intensely by many 
groups, and its unique properties sparked interest across many subfields of astrophysics
research: brown dwarf and planet hunters, astroseismologists, infrared astronomers, and
white dwarf pundits.  Observational evidence gradually began to disfavor a brown dwarf as
the source of infrared emission.  First, some of the first near-infrared imaging arrays revealed 
G29-38 to be a point source in several bandpasses.  Second, near-infrared spectroscopy
measured a continuum flux source \citep{tok88}, whereas a very cool atmosphere was
expected to exhibit absorption features.  Third, the detection of optical stellar pulsations 
echoed in the near-infrared were difficult to reconcile with a brown dwarf secondary 
\citep{pat91,gra90}.  Fourth, significant 10\,$\mu$m emission was detected at G29-38, 
a few times greater than expected for a cool object with the radius of Jupiter, essentially
ruling out the brown dwarf companion hypothesis \citep{tok90,tel90}.

\begin{figure}
\figurenum{2}
\plotone{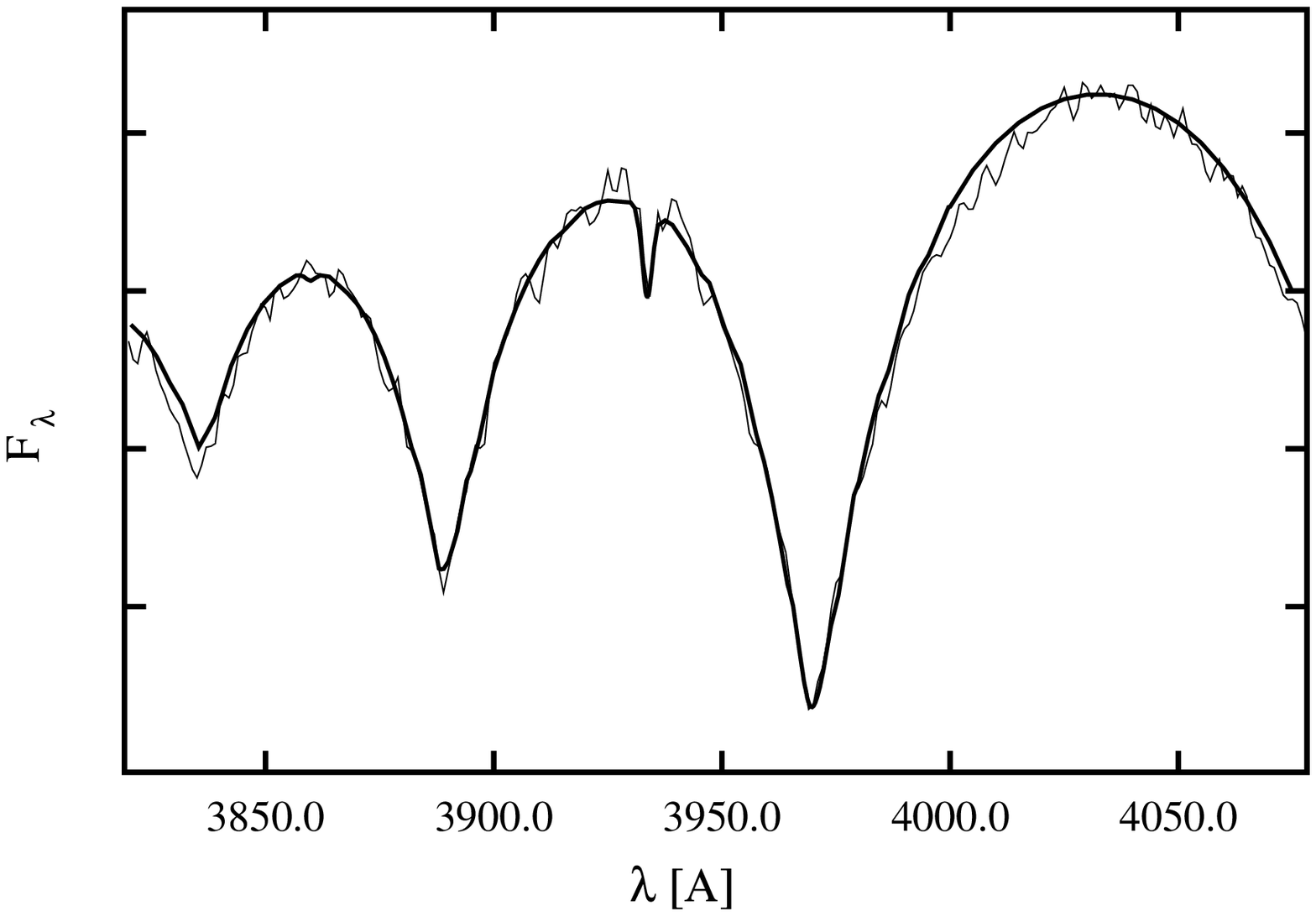}
\caption{The discovery of photospheric calcium in the optical spectrum of G29-38 \citep{koe97},
at the same atomic transition that is strongest in the Sun (the Fraunhofer K line).  The thin line is
the actual data while the thick line is the best model fit.  It is noteworthy that the absorption line 
is quite weak despite the large calcium abundance ($\log\,$(Ca/H)$=-6.8$; \citealt{zuc03}), and 
is due to the relatively high opacity of hydrogen-rich white dwarf atmospheres.  All else being 
equal, a similar calcium abundance produces a line equivalent width roughly 100 times greater 
in a helium-dominated atmosphere (see Figure 3).
\label{fig2}}
\end{figure}

Some lingering doubt remained that G29-38 was indeed surrounded by very warm dust,
but variations seen in radial velocity \citep{bar92} and pulse arrival times \citep{kle94} were 
never successfully attributed to an orbiting companion.  A decade after the discovery of its
infrared excess, the optical and ultraviolet spectroscopic detection of multiple metal species
in the atmosphere of G29-38 \citep{koe97} made it clear the star is currently accreting from
its circumstellar environs (Figure \ref{fig2}).

\subsection{The Polluted Nature of Metal-Rich White Dwarfs}

\begin{figure}
\figurenum{3}
\plotone{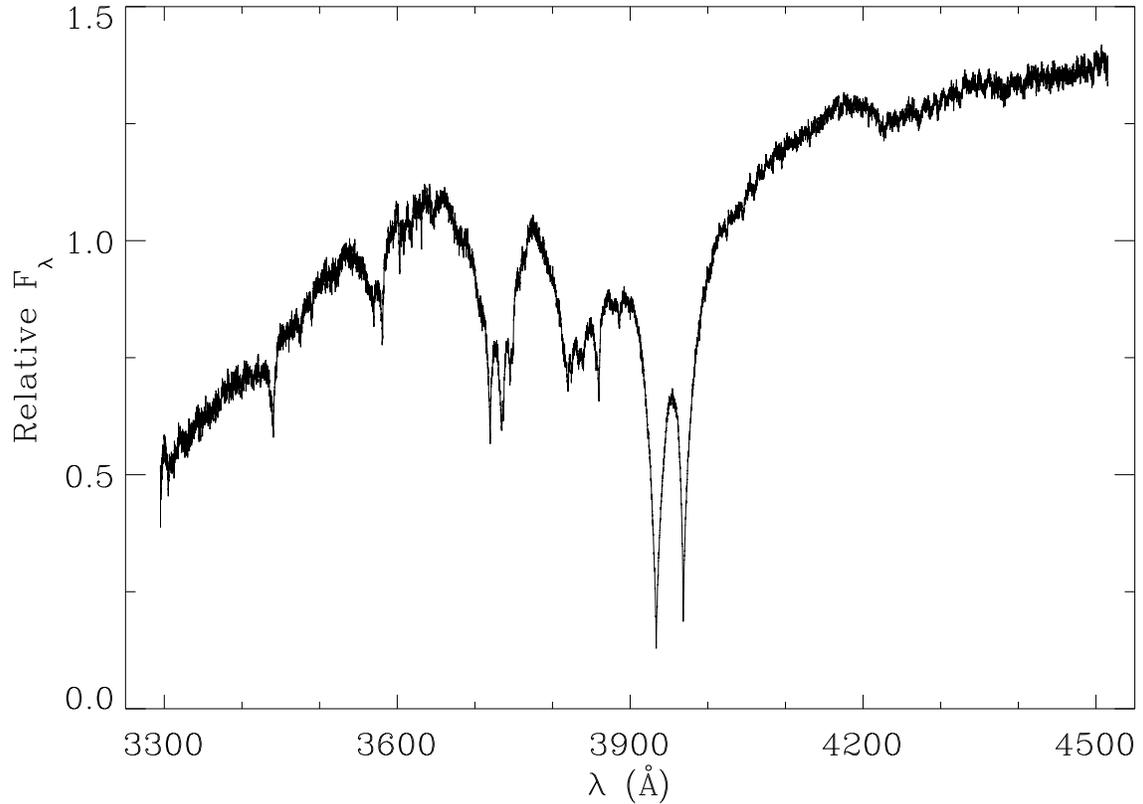}
\caption{The prototype metal-lined white dwarf vMa\,2 (van Maanen's star; \citealt{van17}).
In contrast to hydrogen-rich atmospheres, metal absorption features in helium-rich stars can 
be quite prominent, and often dominate their optical spectra as in this case.  Plotted is an
unpublished spectrum taken with the UVES spectrograph on the Very Large Telescope for the 
SPY program (PI: R.\ Napiwotzki).  All salient features are absorption due to iron, magnesium,
or calcium.
\label{fig3}}
\end{figure}

It would not be possible to tell the story of G29-38 and subsequent disk detections at white 
dwarfs without introducing the phenomenon of atmospheric metal contamination. The origin 
and abundances of photospheric metals in isolated white dwarfs has been an astrophysical 
curiosity dating back to the era when the first few white dwarfs were finally understood to be 
subluminous via the combination of spectra and parallax \citep{van19}.  In a half page journal 
entry, \citet{van17} noted that his accidentally discovered faint star with large proper motion 
had a spectral type of ``about F0'', almost certainly based on its strong calcium H and K 
absorption features (Figure \ref{fig3}).  Only four decades later did it become clear that vMa\,2 
was metal-poor with respect to the Sun \citep{wei60}.  Over the next decade and a half, it 
became gradually understood that white dwarfs (as a class) had metal abundances a few to 
several orders of magnitude below solar \citep{weh75,weg72}.

Any primordial heavy elements in white dwarfs can only be sustained in their photospheres 
for the brief period while the degenerate is still rather hot and contracting, and then only to a 
certain degree \citep{cha95}.  For $T_{\rm eff}<25,000$ K, gravitational settling is enhanced by 
the onset of convection and heavy elements sink rapidly in the high surface gravity atmospheres 
of white dwarfs \citep{alc80,fon79,vau79} leaving behind only hydrogen or helium.  Downward 
diffusion timescales for heavy elements in cool white dwarfs are always orders of magnitude 
shorter than their evolutionary (cooling) timescales \citep{paq86}, and thus external sources
are responsible for the presence of any metals within their photospheres.

The term ``metal-rich'' is used (somewhat ironically) to refer to cool white dwarfs that have 
trace abundances of atmospheric heavy elements.  These are either hydrogen- or helium-rich
atmosphere white dwarfs whose optical spectra exhibit the calcium K absorption line; the same 
atomic transition that is strongest in the Sun.  While iron and magnesium absorption features are
detected in the optical spectra for a substantial fraction of these stars, and additional elements 
are seen in a few cases, all currently known metal-rich white dwarfs display the calcium K line.

[White dwarfs in binary systems may accrete heavy elements from a companion star via Roche 
lobe overflow or wind capture (e.g.\ cataclysmic variables), but the discussion here is restricted 
to white dwarfs that lack close stellar companions.  The topic of gaseous accretion disks at white 
dwarfs resulting from binary mass transfer is not discussed here.]

\subsection{Interstellar or Circumstellar Matter}

There are two possible sources for the atmospheric metals seen in cool, single white dwarfs: 
accretion from the interstellar medium or from its immediate circumstellar environment.  The
latter case refers to material physically associated with the white dwarf and its formation (i.e.
in simplest terms, a remnant planetary system) as opposed to a local accumulation of matter 
with distinct origins.  In both cases, the accretion of heavy elements necessary to enrich the
white dwarf atmosphere may be accompanied by the formation of a circumstellar disk.  Thus,
for cool white dwarfs the phenomena of photospheric metals and circumstellar disks are likely
to have a profound physical connection.

Historically, accretion from the interstellar medium was the most widely accepted hypothesis 
for the metals detected in cool white dwarfs.  This was perhaps all or for the most part due to
the fact that until 1983, all metal absorption features detected in cool white dwarfs were the
result of relatively transparent, helium-dominated atmospheres \citep{sio90b}.  Such stars 
(referred to here as type DBZ) have relatively deep convection zones and commensurately 
long timescales for the downward diffusion of heavy elements, up to 10$^6$\,yr \citep{paq86}.
This allows for the possibility that their extant photospheric metals could be remnants of a
interstellar cloud encounter several diffusion timescales prior \citep{dup93a,dup93b,dup92}.  
However, the general lack of significant hydrogen in DBZ stars has been a continually 
recognized and glaring drawback for the interstellar accretion hypothesis \citep{aan93,
wes79,koe76}. 

The confirmation of the first cool, hydrogen atmosphere white dwarf with metal absorption 
(G74-7, type DAZ; \citealt{lac83}) presented a new challenge to the idea of interstellar cloud 
accretion.  It took some time for robust stellar models to emerge, but it was basically understood
that DA white dwarfs have relatively thin convection zones and correspondingly short metal 
diffusion timescales.  These span a wide range from a matter of days in warmer stars like 
G29-38 all the way up to a few 10$^3$\,yr for relatively cool stars such as G74-7 \citep{paq86}.  
Compared to all previously known metal-enriched white dwarfs, it was clear that the first DAZ 
star had experienced a recent accretion event.  This led to the idea that comet impacts could 
be responsible for the photospheric metals in polluted white dwarfs \citep{alc86}.

The strength of the cometary impact model was that it capitalized on the the hydrogen-poor 
nature of the accreted material in the numerous DBZ stars \citep{sio90a}, yet it was difficult to 
reconcile with the lack of detected DAZ stars, as only G74-7 was known at the time \citep{alc86}.
This apparent dearth of DAZ stars was eventually understood as an observational bias due to 
the relatively high opacity of hydrogen atmospheres compared to those composed primarily
of helium (\citealt{dup93b}; see Figures \ref{fig2} and \ref{fig3}).  The eventual detection of 
atmospheric metals in numerous cool, hydrogen-rich white dwarfs required the combination 
of large telescopes and high-resolution spectroscopy \citep{zuc98}.  While these detections 
presented a challenge to the interstellar accretion hypothesis, they failed to breathe new 
life into the cometary impact model.  For example, the second DAZ white dwarf to be found,
G238-44 \citep{hol97}, has a metal diffusion timescale of only a few days, and an unlikely,
continuous rain of comets is needed to account for its metal abundance \citep{hol97}.

\subsection{G29-38 and the Asteroid Accretion Model}

Thus did the study of metal-enriched white dwarfs come to somewhat of a historic crossroads 
circa 2003, a few months prior to the launch of the {\em Spitzer Space Telescope}.  With the 
sole exception of G29-38 there was a distinct lack of reliable (infrared) data on the circumstellar 
environments of white dwarfs, yet a growing profusion of stars contaminated by metals, and 
problems with existing hypotheses \citep{zuc03}.  Because of their long metal dwell times, 
the contamination measured in DBZ stars with trace hydrogen could be made consistent with 
interstellar accretion models.  But the marked lack of marked lack of dense interstellar clouds 
within 100\,pc of the Sun (i.e.\ the Local Bubble; \citealt{wel99,wel94}) made this picture difficult 
to reconcile with the existence of DAZ stars.  In particular, the necessity for ongoing, high-rate 
accretion of heavy elements at some DAZ stars (e.g.\ G238-44) rendered both the cometary 
impact and the interstellar cloud models unattractive and unlikely.  Lacking a detailed model, 
\citet{sio90a} had speculated that asteroidal or planetary debris could be the ultimate source
of the photospheric metals in hydrogen-poor DBZ white dwarfs.

In a short but seminal paper, \citet{jur03} modeled the observed properties of G29-38 by 
invoking a tidally-destroyed minor planet (i.e.\ asteroid) that generates an opaque, flat 
ring of dust analogous to the rings of Saturn.  Rather than impacting the star, an asteroid 
perturbed into a highly eccentric orbit makes a close approach to the white dwarf, passes 
within its Roche limit and is torn apart by tidal gravity.  Ensuing collisions reduce the fragments 
to rubble and dust, and the resulting disk of material rapidly relaxes into a flat configuration 
owing to a range of very short ($P\sim1$\,hr) orbital periods.  The closely orbiting dust is 
heated by the star producing an infrared excess, and slowly rains down onto the stellar 
surface, polluting its otherwise-pristine atmosphere with heavy elements.  The bulk of a 
flat disk is shielded from the full light of the central star, allowing dust grains to persist for 
timescales longer than permitted by radiation drag forces.  This model compared well to 
all the available infrared data on G29-38, including {\em Infrared Space Observatory (ISO)} 
7 and 15\,$\mu$m photometry \citep{cha99}.

\section{PRE-{\em SPITZER} AND GROUND-BASED OBSERVATIONS}

\subsection{Photometric Searches for Near-Infrared Excess}

The infrared excess and photospheric metals in G29-38 gave astronomers an empirical 
model to test at other white dwarfs prior to the asteroid accretion model.  Because the dust 
emission at G29-38 is prominent beginning at 2\,$\mu$m (i.e.\ the $K$ band), an obvious 
starting point would be to search for white dwarfs with similar near-infrared excess detectable 
with ground-based photometry.  As mentioned previously, the excess at G29-38 was found in
just such a survey, but aimed at identifying unevolved, low mass companions such as brown
dwarfs.  Although the authors did not state the result, \citet{zuc92} found no candidate analogs
to G29-38 among $JHK$ photometry of roughly 200 white dwarfs searched for near-infrared 
excess.  Although Zuckerman \& Becklin continued to take similar data for white dwarfs over 
the next several years, very little was published on the topic until the {\em Spitzer} era.

Three photometric studies again aimed at identifying low mass stellar and brown dwarf 
companions to white dwarfs were published beginning in 2000.  \citet{gre00} surveyed 
around 60 extreme ultraviolet-selected white dwarfs at $JK$ and identified a few stars 
with near-infrared excess consistent with stellar companions.  These stars are too hot for 
any photospheric metals to be considered pollutants, and they did not search for mild 
$K$-band excesses that might be expected from circumstellar dust.  \citet{far05} published 
the cumulative results of a decade and a half of near-infrared observations of white dwarfs 
begun by Zuckerman \& Becklin.  Among 371 white dwarfs, the study included over two dozen 
white dwarfs with a $K$-band excess from cool secondary stars, but no candidates for dusty 
white dwarfs.  However, only roughly one third of the sample stars had independent $JK$ 
photometry while data for the remainder of the sample was taken from the Two Micron All 
Sky Survey (2MASS; \citealt{skr06}).

\citet{wac03} and later \citet{hoa07} published the results of a cross-correlation between the 
2249 entries in the white dwarf catalog of \citet{mcc99} and the 2MASS point source catalog 
second and final data releases.  This enormous undertaking found a few dozen previously 
unidentified white dwarfs with infrared excess owing to cool stellar companions.  But although 
G29-38 satisfied their criteria for infrared excess, the search found no other candidates with 
similar colors; a startling result when taken at face value.  However, the sensitivity limits of 
2MASS, particularly in the $K_s$ band, often prevent reliable flux estimates for white dwarfs.  
A typical catalog entry in \citet{mcc99} has $V\sim15$\,mag and near zero optical and infrared 
colors, so that its predicted near-infrared magnitudes should be similar.  Unfortunately 2MASS 
$K_s$-band data becomes increasingly unreliable for sources fainter than $K_s=14$\,mag, 
severely limiting a robust dust search at a large number of nearby white dwarfs \citep{far09a}.

\subsection{Metal-Polluted White Dwarf Discoveries}

Over a period of time spilling over into the first couple of years after the launch of {\em Spitzer},
there were two major surveys of white dwarfs using high-resolution optical spectrographs on
the world's largest telescopes.  \citet{zuc03} published a survey of nearly 120 cool DA white
dwarfs with the HIRES spectrograph on the Keck I telescope.  The study specifically aimed to 
detect photospheric metals in $T_{\rm eff}\la10,000$\,K, hydrogen-rich white dwarfs and was 
highly successful.  Overall, 24 new DAZ stars were identified via the calcium K-line, including
5 that were known to have detached, low mass main-sequence companions \citep{zuc03}.  The 
study of \citet{zuc03} underscored the fact that photospheric metals in DA white dwarf produce 
weak optical absorption features that require high-powered instruments in order to be detected; 
nearly all of their detections had equivalent widths smaller than 0.5\,\AA.

\begin{figure}
\figurenum{4}
\includegraphics[scale=0.65,angle=-90]{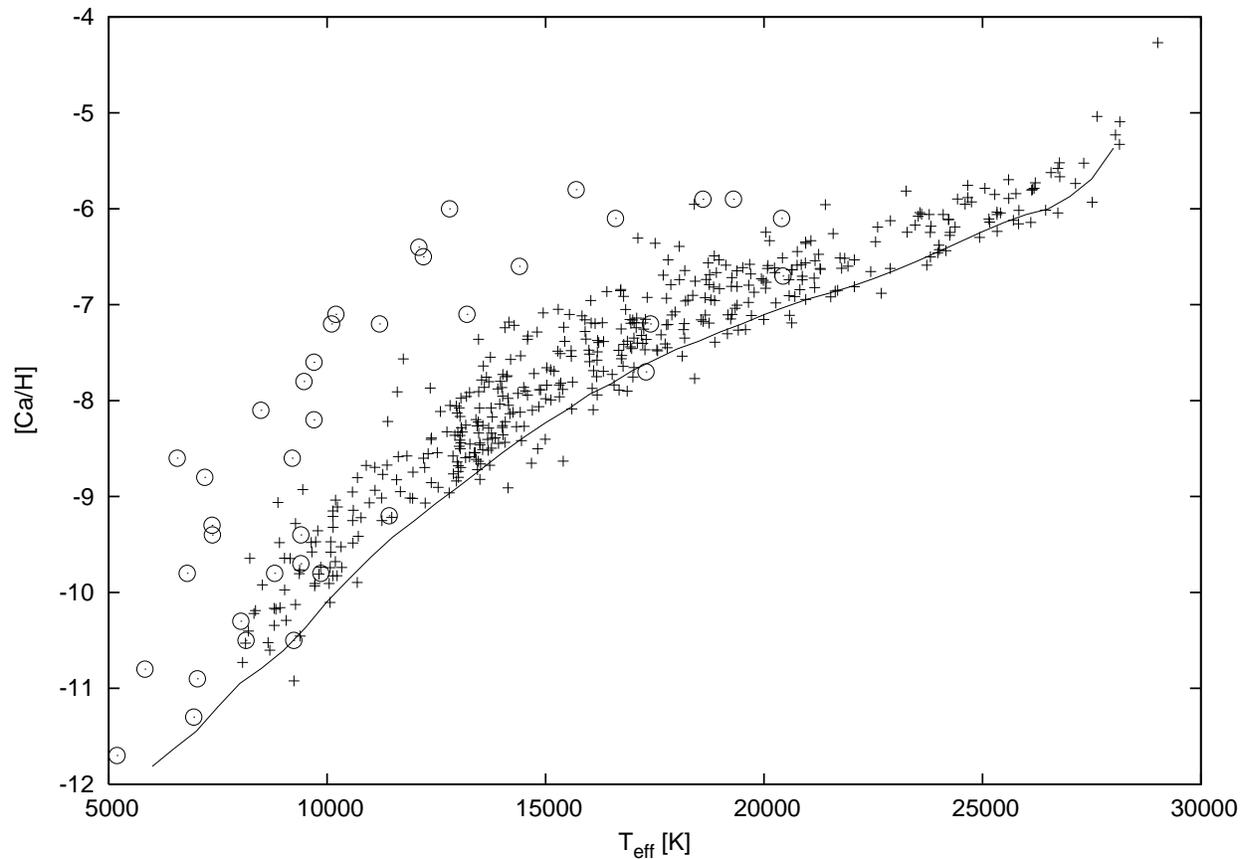}
\caption{Calcium-to-hydrogen abundances (circles) and upper limits (plus symbols) for over 550 
DA white dwarfs from the surveys of \citet{koe05b} and \citet{zuc03}.  The plot demonstrates the 
observational bias precluding the detection of modest metal abundances in warmer white dwarfs.  
The solid line represents an equivalent width detection limit of 15\,m\AA \ \citep{koe06}.
\label{fig4}}
\end{figure}

Less than two years later, \citet{koe05b} published a sub-sample of stars from the Supernova
Progenitor Survey (SPY; \citealt{nap03}) which observed over one thousand white dwarfs with
the UVES spectrograph on the Very Large Telescope (VLT) unit 2.  This extensive survey aimed 
to identify radial velocity variable, double degenerate binaries and thus required high spectral
resolution.  But as a by-product, the search uncovered 18 new DAZ and nine new DBZ stars 
among warmer, $T_{\rm eff}\ga10,000$\,K white dwarfs, all displaying weak calcium K lines 
with equivalent widths less than 0.3\,\AA \ \citep{koe05a,koe05b}.  Importantly, nearly 500 
DA stars of various temperatures were searched for calcium K line absorption and upper 
limit abundances determined for null detections (Figure \ref{fig4}).  These results provide a 
visually straightforward demonstration of the observational bias against the optical detection 
of metals in warmer white dwarfs.

\subsection{The Spectacular Case of GD\,362}

\begin{figure}
\figurenum{5}
\plotone{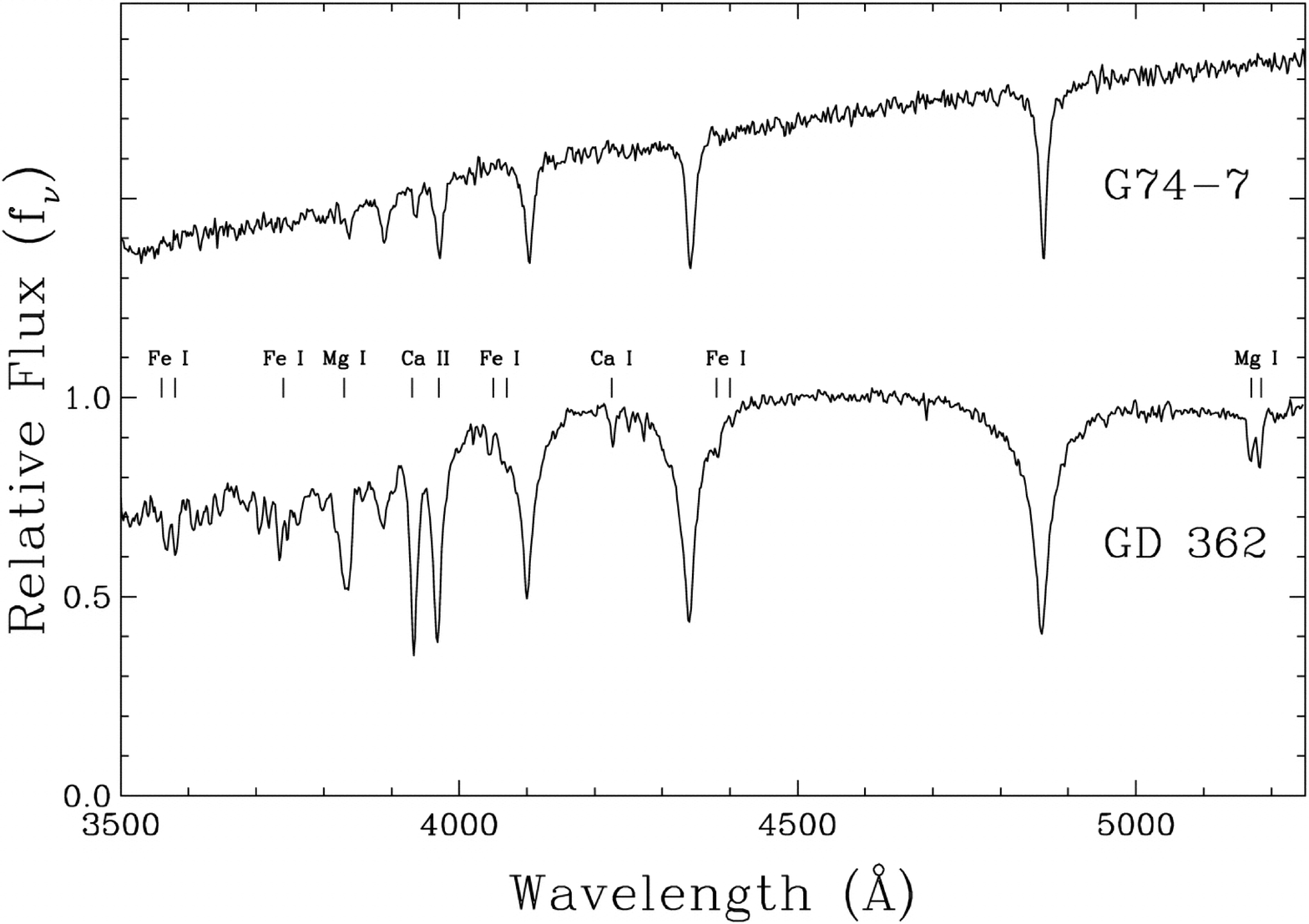}
\caption{The highly metal-rich optical spectrum of GD\,362 exhibits strong lines of calcium,
magnesium, and iron in addition to hydrogen Balmer lines, thus appearing as a DAZ-type star
\citep{gia04}.  As it turns out, this spectacularly polluted star has an atmosphere dominated by 
helium rather than hydrogen \citep{zuc07}.  Notably, the calcium H line is sufficiently strong that 
it overwhelms H$\epsilon$.  Shown for comparison is G74-7, the prototype DAZ white dwarf.
\label{fig5}}
\end{figure}

The second white dwarf discovered to have circumstellar dust came 18 years(!) after 
G29-38 and resulted from two groups simultaneously recognizing the significance of its very 
metal-rich spectrum.  \citet{gia04} reported the optical spectrum of GD\,362 had the strongest 
calcium lines seen in any DA star to date (Figure \ref{fig5}), strong lines of magnesium and 
iron, and nearly {\em solar} abundances of these elements\footnote{I was on the phone to my 
collaborators within minutes of reading the GD\,362 discovery paper, making sure the group 
planned to observe the star when it rose again a few months later.}.  The authors concluded 
the cool star was too nearby to have attained its spectacular metal content from an interstellar 
cloud, but otherwise offered no explanation for their amazing find.

The significance of the spectacular pollution in this star was evident to at least two groups 
of astronomers, each involved in the early stages of {\em Spitzer} programs on white dwarfs.
Using different observational methods, both teams simultaneously published evidence for a 
circumstellar disk at GD\,362 in the same issue of the Astrophysical Journal.  Chronologically,
\citet{kil05} observed the white dwarf first, using low-resolution near-infrared spectroscopy at 
$0.8-2.5$\,$\mu$m.  Their spectrum revealed continuum excess beginning near 2.0\,$\,mu$m,
matched well by adding 700\,K blackbody radiation to the expected stellar flux.  The authors 
inferred that the infrared emission is due to heated circumstellar dust but neither modeled nor
constrained its physical characteristics or origin.

\begin{figure}
\figurenum{6}
\epsscale{0.7}
\plotone{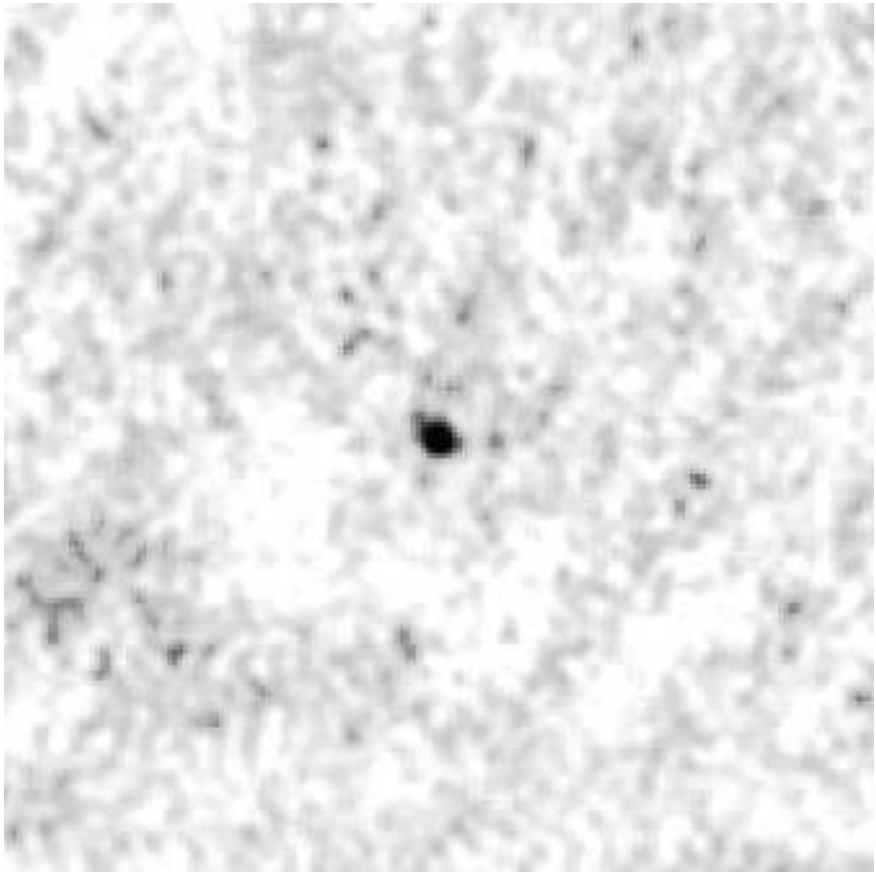}
\epsscale{1.0}
\caption{The $N'$-band (11.3\,$\mu$m) imaging discovery of unresolved infrared excess at 
GD\,362, acquired with MICHELLE \citep{gla97} on Gemini North \citep{bec05}.  This image 
has been processed and is somewhat historic as it represents a remarkable ground-based 
detection of 1.4\,mJy ($4.5\sigma$), representing less than 1 hour on source.  Contrast this 
with the multi-hour exposures necessary to detect the 11\,mJy flux from G29-38 at a similar 
wavelength with earlier instruments 
\citep{tel90,tok90,gra90}
\label{fig6}}
\end{figure}

\citet{bec05} performed ground-based, mid-infrared photometric observations of GD\,362 and
detected the source in the $N$' band (11.3\,$\mu$m) at 1.4\,mJy (Figure \ref{fig6}.  Because 
the expected flux from the stellar photosphere at this wavelength is only 0.01\,mJy, the detection 
by itself is strong evidence for warm dust.  \citet{bec05} also obtained near-infrared $JHKL'$ 
photometry, constraining the spectral energy distribution of the infrared excess, and revealing
that the emitting surface area was too large for a substellar companion.  They employed the
flat dust ring model of \citet{jur03} to the infrared emission at GD\,362 and showed that the
innermost dust is located within 10 stellar radii and has a temperature around 1200\,K, where 
typical dust grains rapidly sublimate.  Thus, the location of the dust was found to be consistent 
with a tidally disrupted minor planet, and the resulting circumstellar disk the probable source
of the accreted metals.

The $JHK$ photometry for GD\,362 was probably insufficient to confidently establish the
presence of circumstellar dust.  While the $K$-band photometry reveals a 10\% excess over
the expected photospheric flux, the confidence was less than $3\sigma$, and hence the need
for the $L'$- and $N'$-band measurements.  Whereas the spectroscopy constrains the shape
of any excess, whether a continuum (e.g.\ dust) or having absorption features (e.g.\ substellar 
companions).

\subsection{Spectroscopic Searches for Near-Infrared Excess}

\begin{figure}
\figurenum{7}
\plotone{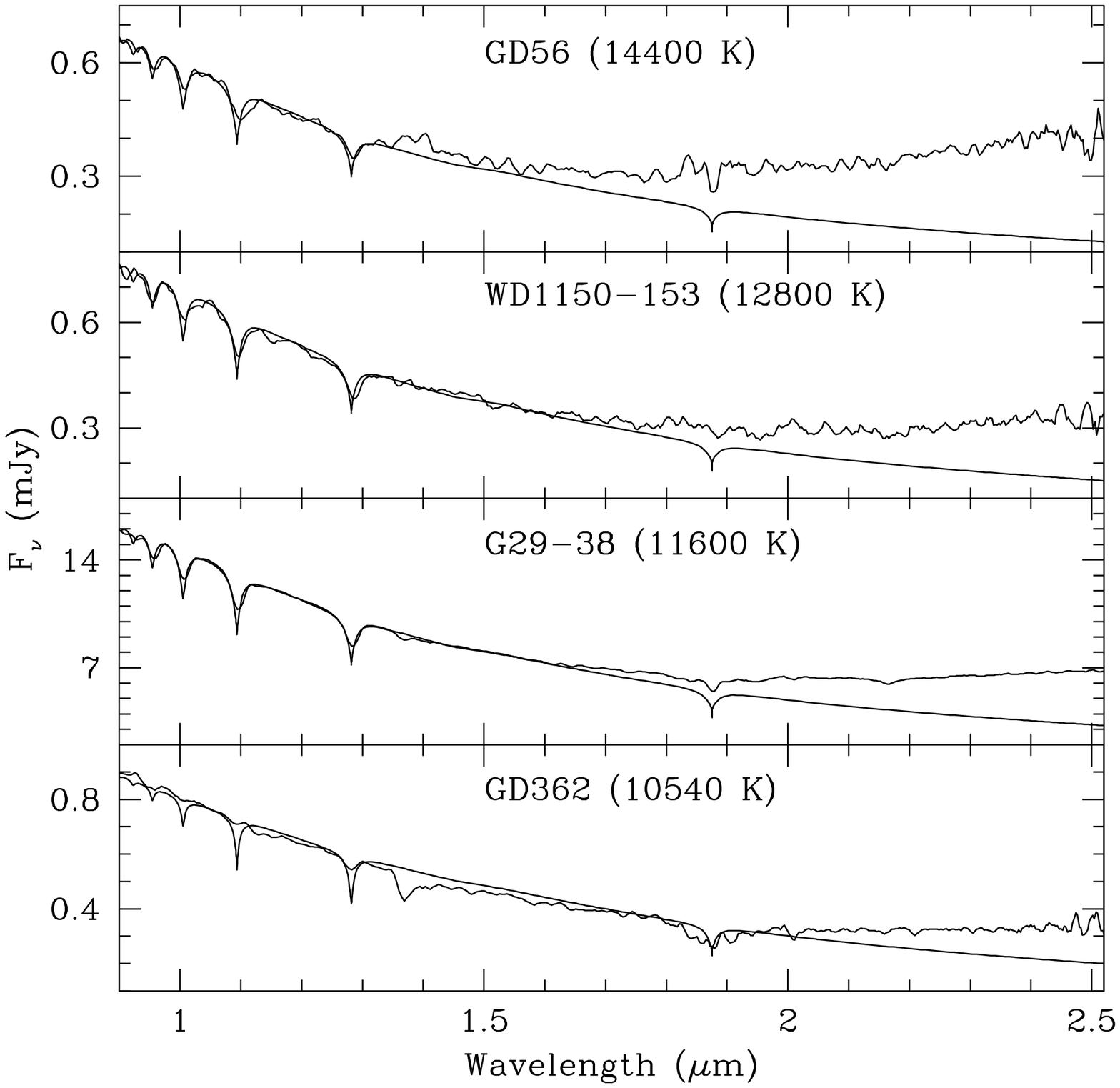}
\caption{Low-resolution, near-infrared spectra of four metal-rich white dwarfs with circumstellar 
dust \citep{kil07,kil06,kil05} taken with the SpeX spectrograph \citep{ray03} and plotted against 
model stellar atmospheres.  Though the measured spectra vary somewhat in slope and strength, 
the inner dust temperature at all of these stars is likely to be essentially the same, as the excess 
emission is largely a function of the emitting solid angle as seen from Earth.  The $H$-band excess 
at GD\,56 is perhaps unique among white dwarfs with dust and may be due to a warp in the inner 
disk region.
\label{fig7}}
\end{figure}

The success of the near-infrared spectroscopic detection of excess emission at GD\,362 was 
soon repeated.  \citet{kil06} detected a very strong continuum excess at the metal-rich white 
dwarf GD\,56 (Figure \ref{fig7}), sufficient to produce a photometric excess in the $H$-band 
\citep{far09a} and stronger overall than the near-infrared excesses at both G29-38 and GD$\,
362$.  This discovery accompanied the first published survey for near-infrared excess that 
specifically targeted metal-rich white dwarfs.  Among 18 DAZ stars, the search identified two 
additional targets with marginal $K$-band excesses; GD\,133 and PG\,1015+161.  In both
cases the potential excess was considered too uncertain due to relatively low signal-to-noise
(S/N) or nearby sources of potential confusion \citep{kil06}, and no conclusions were made 
for these stars.  Using the same technique roughly one year later, \citet{kil07} identified a 
continuum $K$-band excess at one more metal-enriched white dwarf, EC\,11507$-$1519 
(also Figure \ref{fig7}).

Generally, near-infrared data alone cannot distinguish between various dust emission 
models, as the excess at these wavelengths only represents `the tip of the iceberg' (as 
shown later).  The near-infrared excesses measured at GD\,56 and EC\,11507$-$1519 
were attributed to $T\approx900$\,K dust rather than substellar companions, the distribution 
of the orbiting material was not modeled \citep{kil07,kil06}.  Hence, the science that emerged 
from these few years of discoveries made in the near-infrared were generally qualitative and 
provided some statistical limits on the frequency of DAZ dust disk emission similar to that of 
G29-38.  An identical search of 15 DBZ stars \citep{kil08} failed to produce any new dust disks, 
but again provided some limiting statistics on the frequency of metal-rich white dwarfs with warm 
circumstellar dust.  With hindsight, there were as many as six white dwarfs with circumstellar dust 
that went unidentified in these near-infrared spectroscopic observations.

\subsection{Spectroscopy at Longer Wavelengths}

\begin{figure}
\figurenum{8}
\plotone{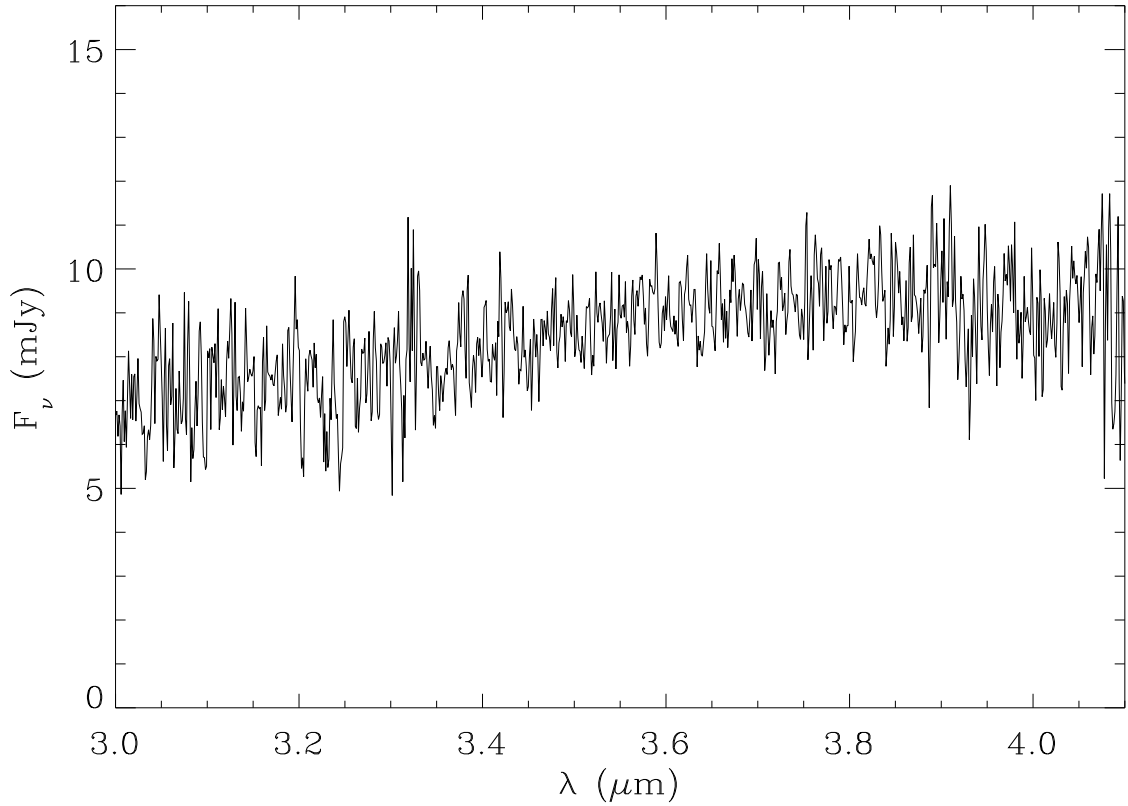}
\caption{$L'$-band grism, $R\approx500$ spectrum of G29-38 obtained using NIRI
\citep{hod03} on the Gemini North 8\,m telescope in approximately 2\,hr of exposure time. The 
S/N of the featureless spectrum is a strong function of wavelength due to varying atmospheric 
transparency, but is typically around 10 \citep{far08b}.
\label{fig8}}
\end{figure}

There are few sufficiently bright, metal-rich white dwarfs that can be observed from the ground 
in a variety of infrared modes.  \citet{tok90} obtained a noisy yet pioneering, IRTF low-resolution 
spectrum of G29-38 between roughly 3 and 4\,$\mu$m with an early generation near-infrared
spectrometer.  The authors concluded no spectral features were present, specifically not those 
typically seen in comets.  By modern standards, the data quality was insufficient to draw a firm 
conclusion.  Thus, many years later their experiment was repeated with an 8\,meter telescope
and 21$^{\rm st}$ century instrumentation.  No spectral features were detected (Figure \ref{fig8}), 
but the data were of sufficient quality to rule out a variety of hydrocarbon emission features 
associated with comets, the interstellar medium, and a variety of nebulae \citep{far08b}.

\section{THE INITIAL IMPACT OF {\em SPITZER}}

Prior to the launch of {\em Spitzer}, there was only a single previously published, mid-infrared 
study of white dwarfs; an {\em ISO} search for dust emission at 11 nearby white dwarfs, six of 
which have metal-enriched photospheres \citep{cha99}.  The {\em ISO} imaging exposures 
were executed in bandpasses centered near 7 and 15\,$\mu$m, and included observations 
of G29-38, G238-44, and vMa\,2.  Most of the white dwarfs were detected at 7\,$\mu$m but 
only a few were sufficiently bright to be seen at 15\,$\mu$m, yet all measured fluxes were 
consistent with photospheric emission with the sole exception of G29-38.  These data 
eventually formed part of the basis for the flat dust ring model hypothesized by \cite{jur03}.

\subsection{Infrared Capabilities of the {\em Spitzer Space Telescope}}

{\em Spitzer} opened up a previously-obscured space to white dwarf researchers, and a few
groups were primed to take advantage of its promised, unprecedented sensitivity to substellar 
companions and circumstellar dust at mid-infrared wavelengths.  The entire observatory was
cryogenically cooled to 5.5\,K which prevented the facility from being a significant source of 
thermal background radiation for its instruments.  {\em Spitzer} was launched in late 2003 with 
three instruments that had the capability to detect sources at the $\mu$Jy level \citep{wer04}.  
The Infrared Array Camera (IRAC; \citealt{faz04}) is a dual-channel, near- and mid-infrared 
imager with filters centered at 3.6, 4.5, 5.8, and 7.9\,$\mu$m.  The Infrared Spectrograph (IRS; 
\citealt{hou04} provided low and moderate-resolution spectroscopy between 5 and 40\,$\mu$m, 
plus limited field of view imaging at 16 and 22\,$\mu$m.  The Multi-Band Imaging Photometer for 
Spitzer (MIPS; \citealt{rie04}) produced imaging and photometry in three wide bandpasses at
24, 70, and 160\,$\mu$m.  At the time of writing, the cryogenic lifetime of the observatory has 
finished, and only near-infrared observations are possible during its warm mission.

\subsection{First Results}

Results from the first two years of {\em Spitzer} white dwarf observations did not emerge
chronologically as they were proposed or obtained, and were heavily influenced by a rapidly 
evolving field.  At the time observing proposals from the general community were solicited in 
late 2003 (and before the deadline arrived in early 2004), only the DAZ study of \citet{zuc03} 
had been published, and highly metal-enriched atmosphere of GD\,362 had not yet been 
discovered.  The ground-based disk searches and further metal-rich white dwarf discoveries 
published in 2005 both had an understandable impact on evolving programs and proposals 
to use the observatory.

\subsubsection{G29-38}

Readers will not be surprised to learn that the most highly sought {\em Spitzer} white 
dwarf target was G29-38.  Utilizing all three observatory instruments \citet{rea05} imaged 
the prototype dusty white dwarf at 4.5, 7.9, 16, and 24\,$\mu$m, and obtained a low-resolution 
spectrum between 5 and 15\,$\mu$m.  These data represented three major advances relative
to previous infrared observations of G29-38.  First, the 2\% to 5\% photometry at $3-8\,\mu$m
were by far the most accurate data obtained for the star.  Second, the mid-infrared photometry
included novel, longer wavelength coverage and the first observations at $\lambda\geq10\,\mu
$m with total errors better than 20\%.  Third and most significant scientifically, the spectroscopy 
over a wide range of mid-infrared wavelengths at S/N $>20$ was unprecedented.  

\begin{figure}
\figurenum{9}
\epsscale{0.9}
\plotone{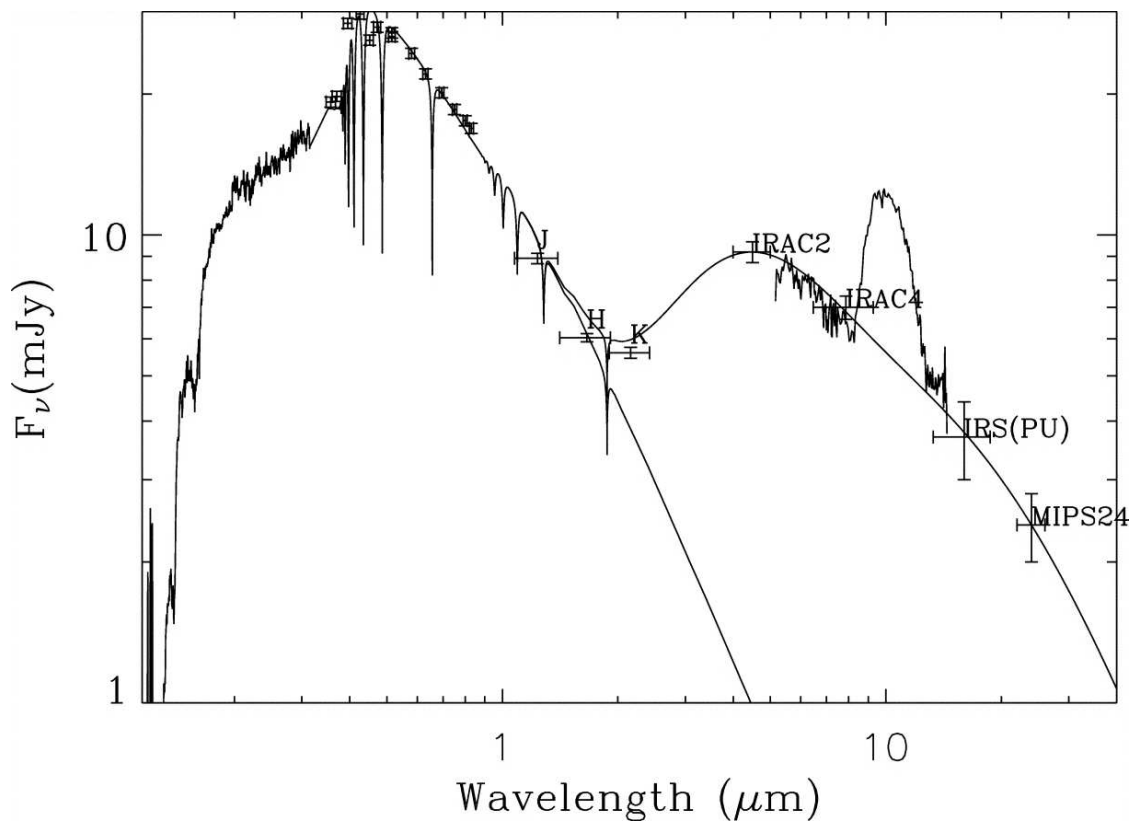}
\epsscale{1.0}
\caption{The ultraviolet through mid-infrared spectral energy distribution (SED) of G29-38,
including {\em Spitzer} photometric and spectroscopic observations \citep{rea05}.  Photometric 
data are shown as error bars, while solid lines plot an {\em International Ultraviolet Explorer (IUE)}
spectrum, a stellar atmosphere model from the ultraviolet through infrared, a model for the thermal
continuum, and the measured mid-infrared spectrum.  The {\em Spitzer} data reveal $T\approx900
$\,K dust emission and a strong silicate emission feature at $9-11\,\mu$m consistent with 
micron-sized olivines \citep{rea05}.
\label{fig9}}
\end{figure}

\begin{figure}
\figurenum{10}
\epsscale{0.8}
\plotone{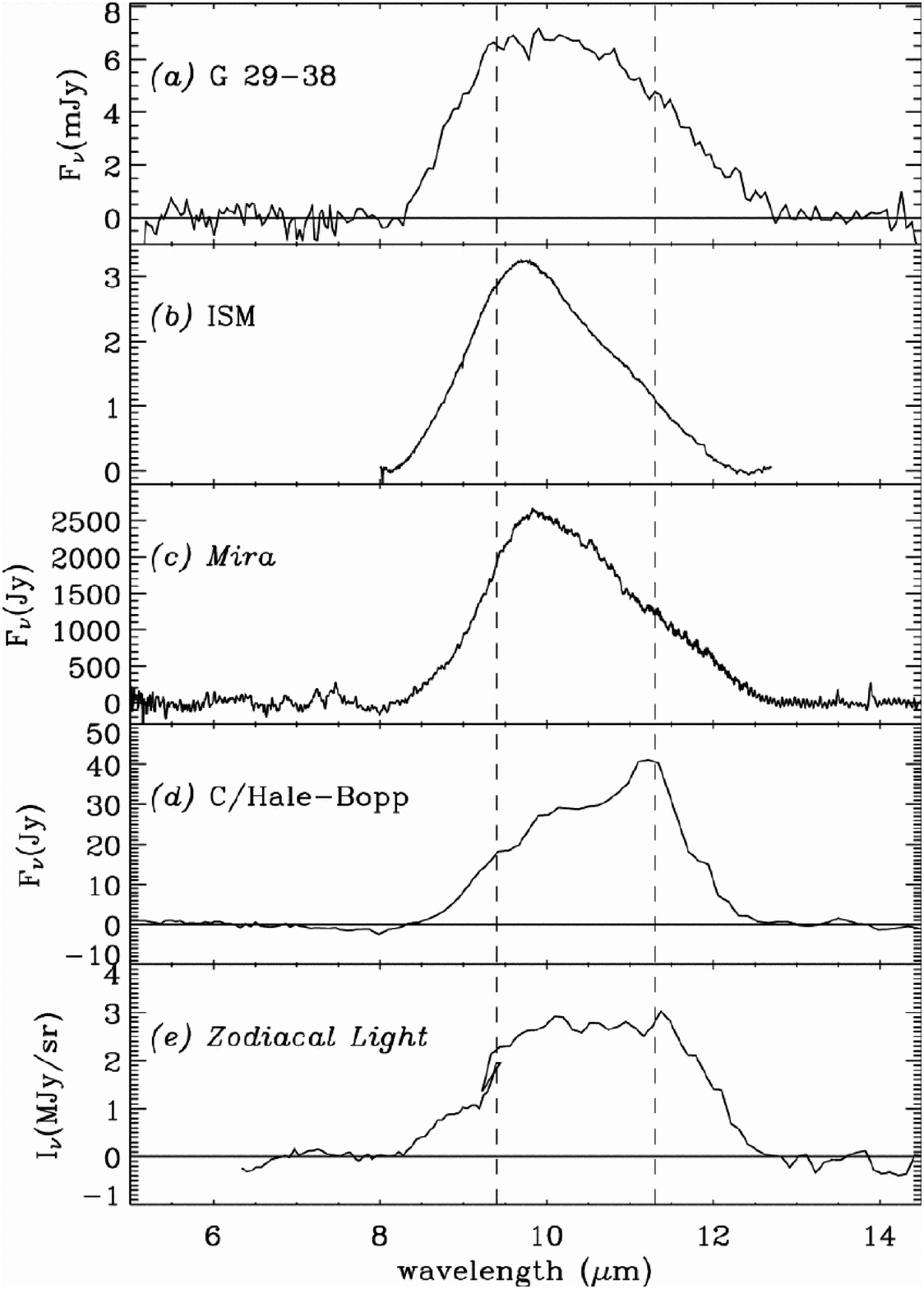}
\epsscale{1.0}
\caption{Comparison of the high S/N, continuum-subtracted, silicate emission feature observed
at G29-38 compared to various astronomical silicates.  The feature at G29-38 has a red wing 
distinct from interstellar silicates and most similar to the dust in the zodiacal cloud of the Solar 
System \citep{rea05}.
\label{fig10}}
\end{figure}

Figure \ref{fig9} displays these {\em Spitzer} data for G29-38, revealing both a strong thermal
continuum of $T\approx900$\,K, and a remarkable $9-11\,\mu$m silicate dust emission feature.
A comparison of the shape of its silicate emission to that observed in the interstellar medium, 
the envelope of the mass-losing giant star Mira, comet Hale-Bopp, and the zodiacal cloud was 
the first concrete evidence that the dust at G29-38 had a circumstellar, and hence planetary, 
origin.  Notably, its spectrum does not exhibit signatures of polycyclic aromatic hydrocarbon 
(PAH) molecules, which often dominate the mid-infrared spectra of the interstellar medium
\citep{dra03,all89}.  The appearance of the $9-11\,\mu$m feature (Figure \ref{fig10}) at 
G29-38 differs from interstellar silicates and also from silicates forming in the ejecta of Mira 
(particles that will eventually become part of the interstellar medium).  Of the four distinct 
sources compared to G29-38, its emission most resembles the emission from the rocky 
particles of the zodiacal cloud.  

\citet{rea05} modeled both the warm thermal continuum and the silicate emission with 
an optically thin, circumstellar cloud (i.e.\ a spherical shell or flattened disk) of dust with an 
exponentially decreasing radial profile.  The model invoked micron-sized olivine (plus some 
forsterite) dust to account for the strong emission feature and similarly small carbon grains 
to account for the thermal continuum.  Silicates are common in the debris of short-period 
comets and asteroids in the Solar System, hence the dust at G29-38 is consistent with 
planetary materials \citep{lis06}.  The authors did note that the lack of PAH features in 
the spectrum was potentially inconsistent with the presence of carbon grains at a ratio of 
3:1 to the silicate grains, as in their model.  It is likely that carbon for its featureless infrared 
spectrum, rather than a likely constituent of planetary debris, as externally polluted white 
dwarfs only rarely show signatures of carbon \citep{jur06}.  

The fundamental difference between this model and the disk model of \citet{jur03} is the 
assumed optical depth of the disk material at various wavelengths.  In the geometrically flat, 
vertically optically thick model of \citet{jur03}, the bulk of material is unseen and effectively 
shielded from stellar radiation.  Such a disk is heated by absorption of ultraviolet radiation 
-- often a major or the dominant source of radiant energy in white dwarfs -- and the warmest 
dust grains are located within a few tenths of a solar radius.  In order to absorb and re-emit 
up to 3\% of the stellar luminosity as does G29-38, a flat disk must be seen in a near face-on 
configuration.  For optically thick disk material, it is only possible to infer a minimum mass as 
most of the material is hidden by definition.  In contrast, the optically thin model proposed by 
\citet{rea05} yields a total disk mass from the infrared emission; for G29-38 this model 
predicts of order 10$^{18}$\,g of small dust grains, and potentially more mass contained 
in larger, inefficiently emitting, particles.  

Critically, optically thin dust grains are warmed by the full starlight of the host star and 
located at a few to several solar radii in the case of G29-38.  Poynting-Robertson (PR) or 
radiation drag will cause exposed dust particles to spiral in towards G29-38 on timescales 
of a few years \citep{rea05}.  Without a source of replenishment for this closely orbiting dust,
it is difficult to reconcile optically thin emission at G29-38 over an 18 year period between its 
discovery and the first {\em Spitzer} observations.

\subsubsection{GD\,362}

\begin{figure}
\figurenum{11}
\epsscale{0.8}
\plotone{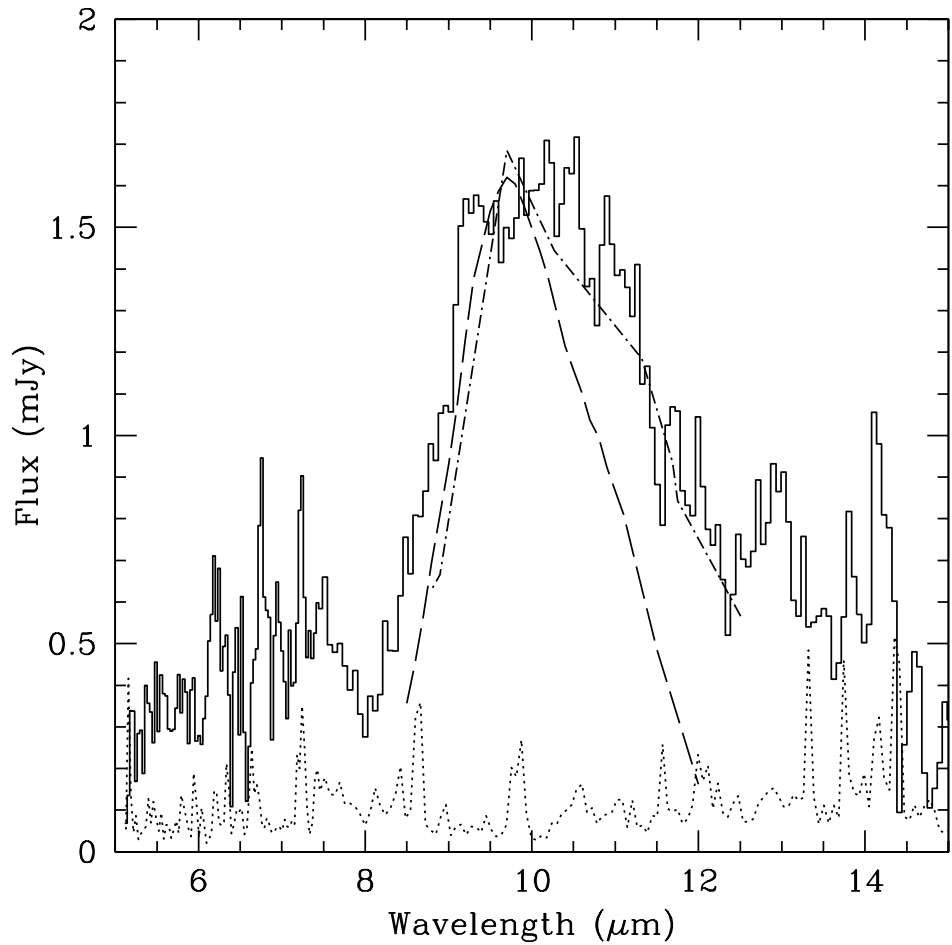}
\epsscale{1.0}
\caption{{\em Spitzer} IRS low-resolution spectrum of GD\,362 revealing a very strong silicate 
feature with a red wing extending to nearly 12\,$\mu$m \citep{jur07b}.  The solid line is the data 
while the dotted line represents the errors.  Overplotted with dashed and dashed-dotted lines are 
the emission profiles of interstellar silicates and from planetesimal dust at the main-sequence star
BD\,$+$20\,307 \citep{son05}, respectively.  The exceptionally strong emission feature at GD\,362 
reprocesses 1\% of the stellar luminosity.
\label{fig11}}
\end{figure}

Naturally, the discovery of infrared excess on top of spectacular metal-enrichment at GD\,362
made it an obvious {\em Spitzer} target.  \citet{jur07b} conducted observations of GD\,362 with 
IRAC $3-8\,\mu$m photometry, IRS $5-15\,\mu$m low-resolution spectroscopy, and MIPS $24\,
\mu$m photometry.  These observations detected a striking silicate feature sufficient in strength
to influence the 7.9\,$\mu$m IRAC photometry and by itself re-emit 1\% of the stellar luminosity 
(Figure \ref{fig11}).  While the $9-11\,\mu$m feature measured at G29-38 is quite strong, the 
emission detected at GD\,362 is simply {\em towering}; among mature stellar systems, only the 
very dusty main-sequence star BD\,$+20$\,307 \citep{son05}, has a comparably strong silicate
emission feature.  As in the case of G29-38, the 24\,$\mu$m photometry indicates a lack of
cool dust, and an outer...

\citet{jur07b} modeled the entire infrared emission with a more sophisticated version of the 
geometrically thin, vertically optically thick disk model used for G29-38.  The model consisted 
of three radially concentric and distinct regions within a flat disk geometry:  two inner, opaque 
regions and an optically thin outer region.  The innermost region was required to be vertically 
isothermal, and the middle region was modeled to have a temperature gradient between its 
top and middle layers.  The outermost region was then modeled to be the source of the silicate 
emission, and generally warmer than the middle region due to the change in optical depth 
\citet{jur07b}.  Importantly, the modeled disk at GD\,362 had a {\em finite} radial extent, and 
was contained entirely within 1\,$R_{\odot}$ of the white dwarf where rocky bodies such as 
large asteroids should be tidally-destroyed \citep{dav99}.  Remarkably, a strictly flat disk
model does not have sufficient surface area (i.e.\ cannot intercept enough starlight) to 
account for the prodigious, overall infrared emission from GD\,362; an warp or slight 
flaring in the (outer) disk was necessary to reproduce the data.

Again, the main difference between the models of \citet{jur07b} and \citet{rea05} is the 
relative transparency or opaqueness of the disks.  For a star like GD\,362 which contains 
more than 10$^{22}$\,g of metals in its convection zone \citep{koe09}, it is all but certain the 
disk is more massive than the minimum 10$^{18}$\,g of optically thin material required to
account for its silicate emission.

\subsection{The First {\em Spitzer} Surveys of White Dwarfs}

There were four programs approved in the first cycle of {\em Spitzer} that aimed to detect 
infrared excess at white dwarfs.  Most of these programs actively sought an excess from 
substellar or planetary companions in addition to searching for dust similar to that seen at 
G29-38.  On the one hand, the frequency of brown dwarfs and planets at the intermediate-
mass, main-sequence progenitors of white dwarfs constrains theories of star and planet 
formation.  The direct or indirect detection of such low-mass objects at A- and early F-type 
stars ranges from very difficult to impossible compared to their detection at the white dwarf 
descendants \citep{far08b}.  On the other hand, one possible origin for the photospheric 
metals in cool white dwarfs is wind capture from an unseen, low-mass stellar or substellar
companion \citep{zuc98,hol97}.  \citet{zuc03} reported a 60\% fraction of metal-enrichment 
among DA white dwarfs with close or very close (i.e.\ spatially unresolved from the ground
or known radial velocity variables), low-mass, main-sequence companions.  Therefore,
{\em Spitzer} was primed to detect various astrophysical sources that might account for
metal contamination observed in cool white dwarfs.

\subsubsection{An Unbiased Survey}

The largest {\em Spitzer} survey of white dwarfs to date was carried out in its first cycle and 
observed 124 stars selected for brightness at near-infrared wavelengths from their data in 
the 2MASS catalog \citep{mul07}.  No preference was given to metal-rich white dwarfs, but 
the brightness-selected sample included G29-38 and 11 other cool stars with photospheric 
metals.  For reasons of efficiency, each of the targets was observed with IRAC at 4.5 and 
7.9\,$\mu$m only, as IRAC is capable of simultaneous exposures at one near- and one 
mid-infrared bandpass.  Of the 12 externally polluted white dwarfs, only G29-38 (the IRAC 
photometry published by \citealt{rea05} was a party of this survey) and LTT\,8452 were 
found to have infrared excesses consistent with circumstellar dust \citep{von07}.  Perhaps
even more profound is the result that of 112 white dwarfs not considered to be externally
polluted, none had an infrared excess attributable to a disk.  One must keep in mind that
roughly one quarter of the stars in their search had effective temperatures above 25,000\,K,
and that the bulk of white dwarfs in the sample had not been observed with high-resolution 
spectroscopy necessary to detect modest metal abundances.  Therefore some caution is
warranted when interpreting this result, but tentatively speaking, no disks are detected at 
white dwarfs that are not metal-polluted.

The metal-enriched stars in their sample were rather diverse in effective temperature, 
calcium abundances, and basic atmospheric composition (i.e.\ both DAZ and DBZ stars), 
and the authors did not try to draw any statistical conclusions on the basis of these two 
disk detections.  Nevertheless, they demonstrated that the opaque, flat disk model fit the 
available photometric data on all known white dwarfs with disks at the time of publication 
(G29-38, GD\,362, GD\,56, and LTT\,8452; Figure \ref{fig13}).  Importantly, the authors 
concluded that the analogy with planetary rings suggests viscous spreading lifetimes on 
par with the Gyr cooling ages of white dwarfs, and therefore such a disk does not require 
replenishment as in the optically thin dust model.

\subsubsection{Surveys of DAZ White Dwarfs}

Together, both \citet{deb07} and \citet{far08b} targeted 18 cool DAZ stars from \citet{zuc03} 
with IRAC imaging observations at all four wavelengths, including the prototype DAZ white
dwarf G74-7.  Given the proliferation of disk discoveries occurring at that time, it was fairly
disappointing and somewhat surprising that neither of the two surveys identified any stars 
with infrared excess similar to G29-38 and other dusty white dwarfs.  \citet{deb07} suggested
that rapid dust depletion due to PR drag within the tidal disruption radius of the white dwarf
could be responsible for the absence of disks.  The four DAZ stars observed by \citet{deb07} 
have temperatures below 9000\,K and hence metal diffusion timescales longer than 100\,yr 
\citep{koe06}; a `recent' absence of dust is conceivable for these stars, if the accreted disk
material was originally optically thin.

\citet{far08b} surveyed several DAZ stars warmer than 9,000\,K and a few warmer than
10,000\,K where the observed metal abundances essentially require ongoing accretion and 
for which an absence of orbiting material would be difficult to reconcile.  They hypothesized
that collisions between grains in an developing disk could rapidly destroy dust particles while
preserving the circumstellar material -- in gaseous form, primarily -- necessary for the inferred, 
ongoing accretion.  A simple calculation showed that optically thin material within the Roche
limit of a white dwarf will collide on timescales 10 to 30 times faster than their PR timescales,
and thus a ring of gaseous debris might develop instead of circumstellar dust \citep{far08b}.

\subsubsection{G166-58}

However, an apparent infrared excess was identified at the metal-rich white dwarf G166-58, 
yet only at the two longer IRAC wavelengths of 5.8 and 7.9\,$\mu$m (\citealt{far08b}; Figure 
\ref{fig14}).  At the time of discovery, the available IRAC mosaicking software was limited to 
creating images with $1\farcs2$ pixels, and the image of G166-58 overlapped somewhat 
with a background galaxy $5\arcsec$ distant.  The background source complicated the flux 
measurements of the star, especially at 7.9\,$\mu$m where the galaxy appears brighter 
than the white dwarf.  Both point spread function (PSF) fitting photometry and radial profile 
analyses supported the conclusion measured excess originated in the point-like image of 
the star.\footnote{In later software released by the {\em Spitzer} Science Center, the ability to 
construct images with $0\farcs6$ pixels firmly corroborated the point-like infrared excess at 
G166-58 \citep{far10c}.}

Due to the diffraction limit of the telescope and the fact that most galaxies have steeply 
rising (power law) SEDs that peak at far-infrared wavelengths, {\em Spitzer} images of 
G166-58 at 16 or 24\,$\mu$m would almost certainly be confused with an even brighter 
galaxy.  Yet despite the absence of warmer dust at this star, the flux decrease towards 
7.9\,$\mu$m is consistent with a disk contained within the tidal disruption radius.  But it 
is the inner, dust-poor region that made this star unique upon discovery and begs for an 
explanation, which its discoverers lacked.

\subsubsection{Other Spitzer Disk Searches}

Also during the first cycle of {\em Spitzer}, two teams independently targeted the most 
massive white dwarfs to search for infrared excess \citep{far08a,han06}.  Two competing
hypotheses exist for the origin of white dwarfs with masses larger than roughly 1.0\,$\mu$m; 
remnants of single, high intermediate-mass stars, or mergers of two white dwarfs \citep{lie05,
fer05}.  In support of the first hypothesis, there is at least one, and possibly up to three massive 
white dwarfs that descended from single $M\ga5$\,$M_{\odot}$ main-sequence stars in the 
125\,Myr old Pleiades open cluster \citep{dob06}.  Such young white dwarfs are excellent 
targets for planets still warm from formation \citep{far08a,bar03}.  Evidence for mergers is 
rather tenuous, but an example is the comoving visual binary LB\,9802.  The system consists 
of two white dwarfs where the considerably hotter star is also substantially more massive, in 
stark contrast with expectations from the coeval evolution of two single stars  \citep{bar95}.  
Models of white dwarf mergers suggest massive disks (and possibly second-generation 
planets) should form to as a repository of shedded angular momentum \citet{han06}.

The {\em Spitzer} searches at massive white dwarfs conducted by \citet{han06} and 
\citet{far08a} produced no infrared excess candidates, leaving the question of white 
dwarf mergers still somewhat open.  If mergers occur with the simultaneous formation of 
a massive disk, it is quite possible they would dissipate rapidly as they are expected to be 
primarily gaseous, at least initially.  Although their composition is expected to be unusual, 
and composed largely of carbon (and oxygen), they may evolve similarly to the gas-rich, 
circumstellar disks observed at young stars and dissipate on similar, Myr timescales. The 
cooling age of a 50,000\,K, 1.2\,$M_{\odot}$ white dwarf is around 30\,Myr (tiny radii make 
them inefficient radiators) suggesting any disk could have vanished due to a combination 
of accretion, early phase radiation pressure, and dust/planetesimal formation.  It has been
suggested that the luminous, post-asymptotic giant R Coronae Borealis stars could be the
product of white dwarf mergers \citep{cla07}.  These stars are enshrouded in carbon-rich
dust \citep{lam01} and have energetic winds, implying their circumstellar material will not 
persist on Myr timescales.

\section{THE NEXT WAVE OF DISK DISCOVERIES}

A wealth of information about disks at white dwarfs emerged from the second, third, and 
fourth {\em Spitzer} cycles.  Roughly chronologically, the {\em Spitzer} observations of 
GD\,362 represented the beginning of something akin to a second generation of white 
dwarf disk discoveries, initially occurring in parallel with the first discoveries.  By 2007, all 
the disk discoveries and searches had been concentrated at DAZ stars, and this was about 
to change.  In 2006, a weak but definite helium absorption line was detected in a deep and 
high-resolution optical spectrum of GD\,362, demonstrating the star had an atmosphere 
dominated by helium \citep{zuc07}.  This discovery was announced to the white dwarf 
community at the biannual European white dwarf workshop held in 2006 at the University 
of Leicester, and foreshadowed in the paper reporting the {\em Spitzer} observations of its 
circumstellar dust. 

\subsection{The Second Class of Polluted White Dwarfs}

The attention paid to DAZ white dwarfs is understandable.  Hydrogen-rich atmosphere white 
dwarfs account for roughly 80\% of all white dwarfs at effective temperatures above 12,000\,K 
\citep{eis06}.  Typical timescales for heavy elements to diffuse below the outer, observable 
layers of a DA star are a few days for stars between 12,000 and 25,000\,K \citep{koe09}. In 
this temperature range, the convection zone or mixing layer of the star is incredibly thin, on the 
order of 10$^{-15}$ of its total mass.  As a DA star cools below 12,000\,K, its convection zone 
increases in depth rapidly and substantially, growing by five orders of magnitude as it reaches
10,000\,K, and another three orders of magnitude by 6500\,K.  The metal sinking timescales
grow commensurately, increasing to 100\,yr at 10,000\,K and 10$^{4}$\, year by 6500\,K 
\citep{koe09}.  Hence the existence of all but the coolest DAZ stars implies the recent or
ongoing accretion of heavy elements, and circumstellar disks are an obvious suspect.

In contrast, DBZ white dwarfs above 12,000\,K have helium atmospheres with only trace 
hydrogen abundances typically at the lower end of the range 10$^{-4}-10^{-6}$ \citep{vos07}.
It is thought these stars are the product of very efficient thermal pulses (helium flashes) that 
expel most of the superficial and primordial hydrogen in the final phases of asymptotic giant 
mass loss.  A white dwarf with a helium-dominated atmosphere is relatively transparent 
compared to its hydrogen-rich counterparts, facilitating the detection of trace amounts of 
heavy elements.  DB white dwarfs also have significantly larger convection zones than 
DA stars, roughly 4 to 5 orders of magnitude deeper at all but the coolest temperatures.  
The size of the convection zone determines the timescales for heavy elements to diffuse 
downward, and hence metals in DBZ stars can persist for up to 10$^{6}$\,yr beginning at 
temperatures of 12,000\,K \citep{koe09}.  Therefore, disk searches initially avoided the 
DBZ class because their photospheric metals could be traces of long past events.

The potential advantages of searching DBZ white dwarfs for circumstellar dust was highlighted
by \citet{jur06}.  He noted that their atmospheric transparency and significantly deep convection
zones yielded compelling compositional and mass limits on the polluting material.  Based on 
{\em IUE} spectroscopic observations of several helium- and metal-rich white dwarfs \citep{wol02}, 
\citet{jur06} identified three stars with measured or upper limit carbon-to-iron ratios that indicated 
the accretion of refractory-rich and volatile-poor (i.e.\ rocky) material:  GD\,40, Ross\,640, and 
HS\,2253+8023.  Furthermore, the mass of iron alone in the outer, mixing layers of these three 
stars ranges between 10$^{21}$ and 10$^{24}$\,g; masses comparable to large Solar System 
asteroids and Ceres.

The SPY and Hamburg Quasar surveys together uncovered more than one dozen DBZ stars 
including GD\,16, a white dwarf with a distinctive DAZ-type optical spectrum remarkably similar 
to GD\,362 \citep{koe05a,koe05b,fri00,fri99}.  Together with previously known white dwarfs in 
this class \citep{duf07,wol02,dup93b} and armed with the knowledge that their atmospheric 
compositions and total heavy element masses suggested the accretion of rocky material, the
DBZ stars made their way into {\em Spitzer} and ground-based searches for dust, alongside 
the DAZ stars.

\subsection{A Highly Successful {\em Spitzer} Search}

Recognizing the helium-rich nature and circumstellar disk of GD\,362, \citet{jur07a} 
disregarded basic atmospheric composition and selected a sample of 11 metal-polluted 
white dwarfs with potential excess flux in their 2MASS $K_s$-band photometry.  On this basis,
GD\,56 emerged as the strongest candidate for circumstellar dust, having a large apparent 
$K$-band excess\footnote{The selection of GD\,56, GD\,133, and PG\,1015$+$161 as Cycle 
2 {\em Spitzer} targets was made in early 2005, prior to the publication of their near-infrared 
spectra \citep{kil06}.}  All white dwarfs were observed using both IRAC $3-8\,\mu$m and 
MIPS 24\,$\mu$m imaging, and this was the first use of longer wavelength photometry in a 
survey of white dwarfs.  Four strong infrared excesses were identified in the program, at
GD\,40, GD\,56, GD\,133, and PG\,1015$+$161 (Figure \ref{fig15}).

The detection of circumstellar dust at GD\,40 was excellent confirmation that DBZ white dwarfs 
held important clues to the nature of their metal-contamination \citep{jur06}, in a different yet 
complimentary way to the DAZ stars.  [At the time of writing, GD\,362 and GD\,40 have revealed 
more about the nature of the circumstellar debris at metal-contaminated white dwarfs than any
other set of stars combined.]  Helium atmospheres are not only relatively transparent and thus
amenable to the detection of trace abundances of heavy elements, their sizable mixing layers 
provide a strict lower limit to the total mass of any accreted elements.  For stars with circumstellar 
dust such as GD\,40, the minimum, total mass of accreted metals places a lower limit on the mass 
of the asteroid whose debris now orbits the star.  Also, the DBZ stars contain only traces or upper 
limit hydrogen abundances; this is a sensitive diagnostic for a variety of accretion models.

The three DAZ white dwarfs found to have infrared excess were something of a cautionary 
tale.  For the first time perhaps, it became clear that ground-based observations up to 2.5\,$
\mu$m were sometimes insufficient to confidently identify circumstellar dust.  GD\,56 displays 
an unambiguous excess in $K$-band spectroscopy and photometry, similar to yet stronger 
than both G29-38 and GD\,362, and the IRAC photometry reveals particularly strong emission.  
While the $K$-band spectra of GD\,133 and PG\,1015$+$161 were inconclusive \citep{kil06}, 
their IRAC photometry reveals clear excess emission in each case.  It is worth remarking that
the reason these two white dwarfs lack notable $K$-band excesses is almost certainly {\em not} 
because they lack warm dust, as in the case of G166-58.  Rather, the strength of emission from 
a flat disk depends also on its solid angle with respect to the Sun \citep{jur03}.

In fact, the flat disk model is able to reproduce the infrared data at GD\,40, GD\,133 and
PG\,1015$+$161 very well using inner dust temperatures $T=1000-1200$\,K and near
to where grains should rapidly sublimate \citep{jur07a}.  However, the strong near-infrared 
emission at GD\,56 cannot be duplicated with a flat disk model, even with dust temperatures 
above 1200\,K.  The sharp rise requires more emitting surface than is available in the flat
disk model and can be reproduced with a 1000\,K blackbody \citep{jur07a}, implying some
portion of the disk is warped or flared.

The importance of the MIPS observations lies in the fact that the longer wavelength fluxes
constrain the detected disk material to lie well within the Roche limit of the white dwarf, and
therefore consistent with a parent body that was tidally destroyed.  The publication of these
results in 2007 brought the number of white dwarf disks observed at 24\,$\mu$m to six, with
none showing evidence for cool dust.  On the contrary the detections and upper limits at this 
wavelength implied the coolest grains had temperatures of several hundred K and orbited 
within 1.2\,$R_{\odot}$ (roughly 100 white dwarf radii).  If a circumstellar disk is formed by 
the gravitational capture of interstellar material at the classical Bondi-Hoyle radius, then
one might expect to detect cool dust as it approaches from this initial distance of several
AU \citep{koe06}.  The MIPS results of \citet{jur07a} demonstrated that emission from dust
captured at such distances was strictly ruled out, as the predicted emission would be tens 
to hundreds of times greater than that observed.

\subsection{The Detection of Gaseous Debris in a Disk}

At roughly the same time period, and to the amazement of the white dwarf community,
\citet{gan06} reported the discovery of a single, warm DAZ white dwarf with remarkably 
strong emission features from both calcium and iron.  Like many astronomical discoveries, 
the identification of metallic emission lines at SDSS\,J122859.93$+$104032.9 (hereafter 
SDSS\,1228) was accidental; its spectrum was flagged in a search for weak spectroscopic 
features due to very low mass (stellar or substellar) companions to apparently single white 
dwarfs (B.\ G\"ansicke 2010, private communication).  

The strongest emission at SDSS\,1228 is seen in the calcium triplet centered near 8560\AA 
\ (Figure \ref{fig16}) and these lines are rotationally broadened in a manner expected from 
Keplerian disk rotation.  While the detected features are directly analogous to hydrogen and 
helium emission from accretion disks in cataclysmic variables \citep{hor86}, the spectrum of 
SDSS\,1228 has strong hydrogen Balmer lines seen only in absorption, implying the emitting 
disk is essentially free of light gases.  The full optical spectrum of SDSS\,1228 is otherwise 
fairly typical of a warm DAZ star with a high metal abundance, showing strong magnesium
absorption, and similar to the spectrum of G238-44.

\citet{gan06} showed that the three calcium features were well-modeled by optically thick 
emission from a highly inclined disk, which together account for the shape of the central 
depressions within the emission features.  The peak-to-peak velocity broadening of $\pm\,630
$\,km\,s$^{-1}$ together with the steep outer walls of the feature limit the gas disk to a maximum 
radius of 1.2\,$R_{\odot}$.  This fact shows that the orbiting material is within the Roche limit of 
the star, and hence consistent with the tidal destruction of a large asteroid.  While disk models 
had been largely successful in reproducing the observed infrared emission at the white dwarfs
with circumstellar dust, and predicted disk radii generally within 1\,$R_{\odot}$, the gaseous 
metal emission line profiles were the first {\em empirical} evidence that circumstellar material 
at metal-enriched white dwarfs orbits within the Roche limit.

At the time of discovery, SDSS\,1228 was the first circumstellar disk identified at a metal-lined
white dwarf with $T_{\rm eff}>15,000$\,K.  The combined studies of \citet{von07} and \citet{kil06}
had targeted 11 DAZ white dwarfs warmer than this and speculated that their lack of infrared 
excess might be due to dust sublimation within the Roche limit of these higher temperature stars.  
At that time, this hypothesis was consistent with sublimated debris orbiting the 22,000\,K white 
dwarf SDSS\,1228 \citep{gan06}, but the pattern would soon be broken.  The infrared excess 
discovered at PG\,1015$+$161 was the first to buck the trend, and more examples would follow, 
including substantial dust at SDSS\,1228 itself.  Based on this and additional reasons, it is 
probable that the gaseous debris at SDSS\,1228 is the result of collisions rather than 
sublimation \citep{mel10}.

\subsection{Dust-Deficiency at DAZ Stars; Collisions?}

Based on the work of \citet{von07} and \citet{jur07a}, it became apparent that the DAZ white
dwarfs with the highest inferred accretion rates were most likely to harbor circumstellar dust.
Previous work had highlighted the correlation of dust disk frequency at DAZ stars with higher 
calcium-to-hydrogen ratios \citep{kil07,kil06}, but while metal abundance is correlated with 
its accretion rate, the size of the convection layer also plays a critical role \citep{koe06}.  It 
makes physical sense that DAZ stars accreting at the highest rates require the most massive 
reservoirs, and the more massive the supply of heavy elements, the better the chance of its 
detection in the infrared.  At the same time, several DAZ stars requiring relatively high metal 
accretion rates failed to show an infrared excess between 3 and 8\,$\mu$m, the wavelength 
range accessible with IRAC photometry.  An excellent example of this is G238-44, with one 
of the highest calcium-to-hydrogen abundances ($\log\,$(Ca/H)$=-6.7$) known and a diffusion 
timescale less than a day \citep{koe06}.

The idea that collisions within an evolving disk can grind dust grains into gaseous debris at
white dwarfs was first suggested by \citet{jur07a}, and can in principle account for a number 
of polluted white dwarfs where no infrared excess is detected by {\em Spitzer} out to 8 (and
even 24)\,$\mu$m.  The basic idea is that Keplerian velocities for particles orbiting a few to 
several tenths of a solar radius from a white dwarf are roughly between 400 and 800\,km\,s$
^{-1}$, and that small (i.e.\ a few percent) deviations from this can lead to collisions at speeds 
sufficient to `vaporize' dust grains.  Warm particles as small as 0.01\,$\mu$m are inefficient 
emitters and absorbers of infrared radiation ($2\pi a / \lambda \ll1$), while smaller particles 
are essentially the size of gas molecules.  

\citet{far08b} further showed that mutual collisions should dominate the initial temporal 
evolution of optically thin dust particles produced in a tidal disruption event.  For dust orbiting 
cool white dwarfs, collisional timescales for disk particles are at least ten times shorter than 
their PR timescales, implying gaseous debris produce via collisions is plausible. Furthermore, 
in order to {\em avoid} the self-erosion of dust in this manner likely requires a high disk surface 
density so that the particle spacing is comparable to the size of the grains \citep{far08b}, and 
collisions are efficiently damped.  Such a scenario is consistent the highest accretion rate stars 
exhibiting the infrared signature of dust most often, and with the (massive) optically thick disk 
models \citep{jur07a}.  

Finally, \citet{jur08} extended the idea of disk particle collisions to include multiple, smaller 
tidal disruption events, under the assumption that the mass distribution of a surviving asteroid 
belt scales similarly to the Solar System and is thus dominated by bodies a few to several km 
in size.  Such a process should more efficiently annihilate solid particles as an infalling, small 
asteroid impacts a pre-existing, low mass disk at a nonzero inclination.  In systems where 
multiple, smaller asteroids are destroyed on timescales much shorter than the often inferred
10$^5$\,yr disk lifetimes , the resulting debris should be primarily gaseous.  While the focus 
of this chapter is not theoretical, suffice to say that at this point in time, it is thought that some 
type of collisional scenario (i.e.\ vaporized debris) is responsible for the lack of observed 
mid-infrared excess at a number of metal-polluted white dwarfs, especially those where 
relatively high metal accretion rates are inferred.

\subsection{Expanding Searches to the DBZ Stars}

About this time, astronomers widened their gaze, turning to the DBZ stars in earnest and 
including longer wavelength observations of both classes of polluted white dwarfs.  \citep{kil08} 
targeted 20 DBZ white dwarfs with the near-infrared spectrograph that had successfully detected 
excess emission at several DAZ stars, but yielded no candidates within the helium-rich sample.
Surprisingly, GD\,40 failed to reveal an infrared excess in these observations, despite a potential 
$K_s$-band photometric excess in the 2MASS catalog; this was the reason it was selected as a 
{\em Spitzer} target by \citet{jur07a}.  The authors tentatively identified the relatively long diffusion 
timescales in these stars as the reason for the lack of dust, yet persistence of photospheric metals,
and estimated that dust disks had typical lifetimes around an order of magnitude shorter than the 
10$^6$\,yr required for metals to begin sinking in the bulk of DBZ stars \citep{koe09}.  At the same 
time, it became clear that ground-based observations had failed to detect 50\% of the known disks
by mid 2008 (G166-58, GD\,40 GD\,133, PG\,1015$+$161).  While LTT\,8452 has never been
observed with $K$-band spectroscopy, its disk signature is not revealed by the $H=14.00\pm
0.06$\,mag, $K_s=14.02\pm0.06$\,mag photometry available in 2MASS.

\citet{far09b} undertook a {\em Spitzer} Cycle 3 study comprising new observations as well as 
an analysis of all available archival data on metal-polluted white dwarfs as of the end of 2008.
Targets included 20 white dwarfs composed of roughly equal numbers of DAZ and DBZ types, 
including MIPS 24\,$\mu$m photometry of several stars previously observed only at shorter 
wavelengths.  New disks were detected around GD\,16 and PG\,1457$-$086, while a better 
characterization of the infrared emsission from LTT\,8452 was enabled by new photometry at 
both shorter and longer infrared wavelengths.  GD\,16 is strikingly similar in optical spectral 
appearance to GD\,362, exhibiting a DAZ-type spectrum while actually having a helium-rich 
atmosphere as evidenced by a very weak absorption features detected at high resolution with
VLT / UVES  \citep{koe05a}.  Together with GD\,362 and GD\,40, the discovery of dust at GD\,16
increased the number of disks at DBZ stars by 50\%.

The infrared fluxes at GD\,16 and LTT\,8452 are fairly typical, strong and well modeled by flat 
disks with inner and outer radii of roughly one and two dozen stellar radii, repsectively (Figure 
\ref{fig17}; \citealt{far09b}).  In contrast, the infrared excess of PG\,1457$-$086 is rather mild
in comparison to previously detected disks, with a fractional luminosity, $\tau=L_{\rm IR}/L=
0.0007$.  For comparison, G29-38 and GD\,362 have $\tau=0.03$, roughly 50 times brighter 
than the excess at PG\,1457$-$086 \citep{far09b}.  The observations and model for this star, 
along with the scientific implications of its existence are discussed in \S6.3.

Previous to this point in time, the DBZ stars had been understandably treated rather differently 
than the DAZ stars.  Because the metal sinking timescales are relatively rapid in the DAZ white 
dwarfs, a steady-state balance between accretion and diffusion is a reasonably safe assumption 
to make \citep{koe06}.  From this balance and the observed calcium abundance in each star, a 
total metal accretion rate can be calculated under the assumption that calcium accretes together
with other elements in solar proportions but without any hydrogen or helium.  This is equivalent 
to assuming a gas-to-dust ratio of 100:1 in the interstellar medium but that only dust is accreted 
\citep{jur07a}.  The analysis of \citet{far09b} attempted to put all metal-contaminated stars on 
equal footing by calculating time-averaged accretion rates for the DBZ stars in a similar manner.  
While physically unmotivated due to the long diffusion timescales in DBZ stars (i.e.\ a steady 
state cannot be assumed), a time-averaged accretion rate is still a useful diagnostic.

\subsection{Additional Disks with Gaseous (and Solid) Debris}

SDSS\,1228 was the first of three metal-enriched white dwarfs found to have circumstellar, 
gaseous debris via the detection of emission lines from the calcium triplet by the end of 2008,
the subsequent discoveries following in fairly rapid succession.  The second single DA white 
dwarf confirmed to manifest metallic emission was SDSS\,J104341.53$+$085558.2 (hereafter 
SDSS\,1043; \citealt{gan07}).  At 18,300\,K, SDSS\,1043 is a cooler star exhibiting the same 
phenomenon, and its spectrum also displays magnesium absorption indicating atmospheric 
pollution. 

With this second discovery, it was again suggested that the warm temperatures of SDSS\,1043 
and 1228 were consistent with solid debris becoming sublimated within their stellar Roche limits, 
and potentially dust-poor analogs of white dwarfs with infrared excess \citep{gan07}.  However, 
the dust disks at both PG\,1015$+$161 \citep{jur07a} and 1457$-$086 \citep{far09b} orbit stars 
of 19,300 and 20,400\,K, respectivey \citep{koe06}, and the optical spectrum of the former shows 
no evidence for calcium emission (B. G\"ansicke 2010, private communication).  \citet{gan07}
searched both GD\,362 and G238-44 for calcium emission lines, but the former exhibits only 
lines of absorption in the triplet, while the latter reveals only a stellar continuum.  

Importantly, an effort was made to identify additional DA white dwarfs with calcium emission 
from the vast SDSS DR4 catalog of \citet{eis06} and to place some limits on the frequency of 
these stars.  Using an automated routine to select stars with excess flux in the region of the 
calcium triplet, SDSS\,1228 and 1043 emerged as the only two candidates for emission among
over 400 DA white dwarfs with $g<17.5$\,mag, while another eight, relatively weak candidates
resulted from selecting among 7360 stars with $g<19.5$\,mag.  Clearly, the frequency of this
phenomenon is less than 1\% \citep{gan07}.

\citet{gan08} expanded their search for calcium emitting white dwarfs in the SDSS DR6 while
removing their restriction to DA stars.  Among over 15,000 likely white dwarfs and nearly 500
candidates for excess calcium flux, when visually inspected the bulk of spectra were found to 
suffer from poor night sky line subtraction resulting in an apparent excess in the triplet region.
The only stars to pass muster were SDSS\,1228, 1043, and SDSS\,J084539.17$+$225728.0;
a star previously cataloged as Ton\,345 and classified as an sdO star in the Palomar Green
survey \citep{gre86}.  Ton\,345 is a DBZ star with calcium triplet emission and photospheric 
absorption lines of calcium, magnesium, and silicon \citep{gan08}.  Remarkably, the emission
lines at Ton\,345 display a marked asymmetry, suggesting the disk of emitting material is not
circular, but has significant nonzero eccentricity in the range 0.2 to 0.4.  Additionally, temporal
monitoring of this star has revealed variability in the shape of the asymmetric line emission 
profiles (Figure \ref{fig18}), strongly suggesting disk evolution on roughly yr timescales that 
appears episodic rather than ongoing \citep{mel10,gan08}.

Overall, the white dwarfs with gaseous debris have provided significant insight into the 
composition, geometry, and evolution of circumstellar disks at white dwarfs.  Importantly, 
all three white dwarfs with gaseous debris also have infrared excess from dust \citep{mel10,
far10c,bri09} and atmospheres enriched with heavy elements.  The solid material is modeled 
to be spatially coincident with the gas, arguing against sublimation of dust interior to some 
critical radius.  In each case, the calcium triplet line profiles constrain the emitting material to 
lie within roughly 1\,$R_{\odot}$, while emission from hydrogen or helium is distinctly absent. 
Therefore, these stars belong to the same class of white dwarfs polluted by debris from tidally 
disrupted asteroids.  

While this is an evolving field, it appears unlikely that the temperature of the white dwarf plays 
a role in generating the gas, which is a natural result of collisions among solid particles as in 
the multiple asteroid scenario of \citet{jur08}.  However, stellar effective temperature dictates
how efficiently the circumstellar gas can be heated, and thus detected as it emits while cooling
\citep{mel10}.

\section{STUDIES AND STATISTICS}

Table \ref{tbl2} lists all 18 confirmed and suspected white dwarfs with circumstellar dust with
mid-infrared excess as of this writing (early 2010), ordered chronologically by year the infrared 
excess was published.  Also listed are the white dwarf effective temperature, estimated distance
from the Earth, apparent $K$-band magnitude, and the telescope which discovered the infrared
excess.

\subsection{Spectroscopic Confirmation of Rocky Circumstellar Debris}

Near the mid-point of {\em Spitzer} Cycle 3, there were ten white dwarfs known to have 
circumstellar dust, but infrared spectra had only been obtained for G29-38 and GD\,362.  
\citet{jur09a} used IRS to observe seven of the remaining dusty stars between 5 and 15\,$\mu
$m with the low resolution modules; G166-58 was deemed too problematic for spectroscopy 
due to its neighboring galaxy \citep{far08b}.  All but one of the IRS targets were detected in the 
observations; owing to a combination of intrinsic faintness and exposure time restrictions in a 
high background region of the sky, no signal was obtained at PG\,1015$+$161 \citep{jur09a}.

Table \ref{tbl1} lists all the white dwarfs with circumstellar dust targeted by IRS targets.  Also
listed are representative infrared fluxes and (unbinned) S/N estimations for the region covering
the silicate feature at $9-11\,\mu$m.  Despite the groundbreaking sensitiviy of {\em Spitzer}, the
bulk of stars have only modest  detections, and most were confidently detected only between 8 
and 12\,$\mu$m where the silicate emission peak is typically 20\% to 60\% stronger than the 
$6-8\,\mu$m thermal continuum (Figure 19).  Nonetheless, each of these stars exhibits strong
10\,$\mu$m emission with a red wing extending to at least 12\,$\mu$m.  The measured features 
are inconsistent with interstellar silicates (see Figure \ref{fig10}), but are instead typical of glassy 
(amorphous) silicate dust grains, specifically olivines typically found in the inner Solar System, 
and in evolved solids associated with planet formation \citep{lis08}.

IRS spectroscopy of all eight circumstellar dust disks reveals no evidence for emission from 
polycyclic aromatic hydrocarbons.  These (carbon-rich) molecular compounds are found in the 
infrared spectra of the interstellar medium \citep{dra03,all89}, some circumstellar environments 
\citep{jur06b,mal98}, and in comets \citep{boc95}, with strong features near 6, 8, and 11\,$\mu$m.
Hence, the dust at white dwarfs is intrinsically carbon-deficient \citep{jur09a}, and similar to the 
rocky material of the inner Solar System \citep{lod03}.  The findings from infrared spectroscopy
are consistent with, and almsot certainly mirrored by, the carbon-poor atmospheres established
at several metal-enriched white dwarfs such as GD\,40, GD\,61, and HS\,2253$+$8023 
\citep{jur08,des08,zuc07}.

The only white dwarf with circumstllar dust bright enough to be studied spectroscopically with 
{\em Spitzer} at wavelengths longer than 15\,$\mu$m is G29-38.  \citet{rea09} obtained a second 
low-resoltuion IRS spectrum of the prototype dust-polluted white dwarf, this time between 5 and 
35\,$\mu$m.  The repeat observations over the short wavelength range revealed no change in the 
shape of the continuum or silicate emission feature, but the longer wavelength data revealed an 
additional, weaker silicate feature (or combination of features) between 18 and 20\,$\mu$m (Figure
\ref{fig23}).  The 
entire infrared spectrum was reproduced using three physically-distinct models: 1) an optically thin 
shell, 2) a moderately optically thick and physically thick disk, and 3) an optically thick, physically 
thin disk with an optically thin layer or outer region.

While the first two models employed by \citet{rea09} are attractive as they permit the 
co-identification of various minerals and water ice with the observed infrared emission, they 
are in essence optically thin models that invoke no more than 10$^{19}$\,g of disk mass.  In 
contrast, the latter model is physically equivalent to the model of \citet{jur07b} for GD\,362,
implying the 10$^{19}$\,g of required optically thin material represents only a tiny fraction of 
the total disk mass.  As discussed earlier, the PR timescales for warm, optically thin dust at 
white dwarfs are very short; 15\,yr for 1\,$\mu$m silicate particles 1\,$R_{\odot}$ distant from 
G29-38.  Thus, for either of the optically thin models proposed by \citet{rea09}, a mechanism
is required to replenish 10$^{19}$\,g of material at G29-38 roughly every 15\,yr, yet its infrared
excess has been present for more than two decades \citep{zuc87}.  A disk whose total mass
is orders of magnitude greater can accomplish this readily, and the asteroid-sized masses of
heavy elements in DBZ stars such as GD\,40 argue strongly for commensurate disk masses.
\citet{rea09} find the flat disk model with an optically thin layer or outer region can reproduce
the entire infrared emission at G29-39, requiring only silicates.

The importance of these observations cannot be overstated.  More telling than the atmospheric
composition of the metal-contaminated white dwarfs themselves, this is the strongest evidence
that circumstellar dust at white dwarfs is derived from rocky planetary bodies \citep{jur09a}.

\subsection{First Statistics and The Emerging Picture}

By the end of 2008, a sufficient number of metal-polluted white dwarfs had been observed 
with {\em Spitzer} to merit statistical analysis; 52 stars with IRAC and 31 with MIPS.  Because 
no white dwarf has been detected at 24\,$\mu$m without a simultaneous and stronger detection
at IRAC wavelengths, the IRAC observational statistics better constrain the frequency of dust 
disks.  From these 52 IRAC observations comprising 11 dust disk detections, it was established 
that that dust disk frequency is correlated with 1) time-averaged accretion rate and 2) cooling 
age (Figure 20; \citealt{far09b}).  Based on the cumulative {\em Spitzer} IRAC observations of 
over 200 white dwarfs, infrared excess from circumstellar dust is only detected at those stars 
with atmospheric metal contamination \citep{far08a,mul07,han06}.

Applying the time-averaged metal accretion rate analysis to the IRAC dataset, \citet{far09b} 
found over 50\% of all cool white dwarfs with metal accretion rates $dM / dt \ga 3 \times 10^
8$\,g\,s$^{-1}$ have dust disks.  Furthermore, when these cool, metal-polluted stars are 
statistically accounted for as members of larger samples of white dwarfs from which they 
are drawn, it is found that between 1\% and 3\% of all white dwarfs with cooling ages less 
than around 0.5\,Gyr have both photospheric metals and circumstellar dust \citep{far09b}.
These results signify an underlying population of asteroids that have survived the post-main 
sequence evolution of their host star, and imply a commensurate fraction of main-sequence 
A- and F-type stars harbor asteroid belts, and probably build terrestrial planets.  Evidence is 
strong that white dwarfs can be used to study disrupted minor planets.

The MIPS dataset implies that dust is not observed outside the Roche limit of the metal-rich 
white dwarfs.  All white dwarf disks detected at 24\,$\mu$m have coexisting, strong $3-8\,\mu
$m IRAC excess fluxes, implying the dust is not drifting inward from the interstellar medium 
\citep{far09b}.  Dust grains captured near the Bondi-Hoyle radius of a typical metal-enriched 
white dwarf should have temperatures below 100\,K, warming as they approach the star under
PR drag.  Both the MIPS detections at white dwarfs with dust, and the nondetections for the 
remaining metal-rich stars, argue against the influx of interstellar material \citep{far09b,jur07a}.
Stars with infrared excess always display decreasing flux towards 24\,$\mu$m, indicating a 
compact arrangement of dust, consistent with disks created via the tidal disruption of rocky 
planetesimals.

Successful disk models are vertically optically thick at wavelengths up to 20\,$\mu$m, and 
geometrically thin \citep{jur03}.  There are two reasons such models are likely to be acccurate.  
First, particles in an opticlaly thin disk would not survive PR forces for more than a few days to 
years \citep{far08b}.  Second, the orbital periods of particles within 1\,$R_{\odot}$ of a white 
dwarf can be under 1 hr, implying the disk will relax into a flat configuration on short timescales.
Using this model, the warmest dust has been successfully modeled to lay within the radius 
at which blackbody grains in an optically thin cloud should sublimate rapidly.  Generally, the 
circumstellar disks have inner edges which approach the sublimation region for silicate dust 
in an optically thick disk; precisely the behavior expected for a dust disk which is feeding 
heavy elements to the photosphere of its white dwarf host. 

The majority of both DAZ and DBZ white dwarfs do not have infrared excesses when viewed
with {\em Spitzer} IRAC and MIPS \citep{far09b}.  Circumstellar gas disks are a distinct possibility 
at dust-poor yet polluted stars with metal accretion rates $dM / dt \ga 3 \times 10^8$\,g\,s$^{-1}$,
while fully accreted disks are a possibility for the DBZ stars with metal diffusion timescales near
10$^6$\,yr.  It is possible that a critical mass and density must be reached to prevent the dust 
disk from rapid, collisional self-annihilation, and when this milestone is not reached, a gas disk
results.

For $T_{\rm eff}\la 20,000$\,K, white dwarfs with younger cooling ages are more likely to be 
orbited by a dusty disk \citep{far09b}.  This observational fact is consistent with a picture where
a remnant planetary system gradually resettles post-main sequence \citep{deb02}.  Because
an asteroid must be perturbed into a highly eccentric orbit in order to pass within the stellar 
Roche limit, planets of conventional size are also expected to persist at white dwarfs.  The 
bulk of white dwarfs with dust have cooling ages less than 0.5\,Gyr, with only one older than 
1\,Gyr, suggesting the possibility that surviving minor planet belts tend to become stable or 
depleted on these timescales.

\subsection{Dust-Deficiency at DAZ Stars; Narrow Rings?}

During the final cryogenic {\em Spitzer} Cycle (5), two previously unusual infrared excess stars 
were found to have counterparts, potentially representing subclasses of circumstellar dust rings 
at white dwarfs.

\subsubsection{Disks with Subtle Infrared Emission}

The infrared excess at PG\,1457$-$086 is between 3 and $6\sigma$ above the predicted
photosopheric flux at the three shortest IRAC wavelengths where it is detected, and may not 
have been recognized without supporting near-infrared photometry that suggested a slight 
$K$-band excess \citep{far09a,far09b}.  Figure \ref{fig24} plots the SED of PG\,1457$-$086
together with a very mild infrared excess discovered at HE\,0106$-$3253 again at the $4-6
\sigma$ level in the IRAC bandpasses \citep{far10c}.  As a benchmark comparison, the IRAC
excess at G29-38 is between 15 and $20\sigma$.

\begin{figure}
\figurenum{12}
\epsscale{0.8}
\plotone{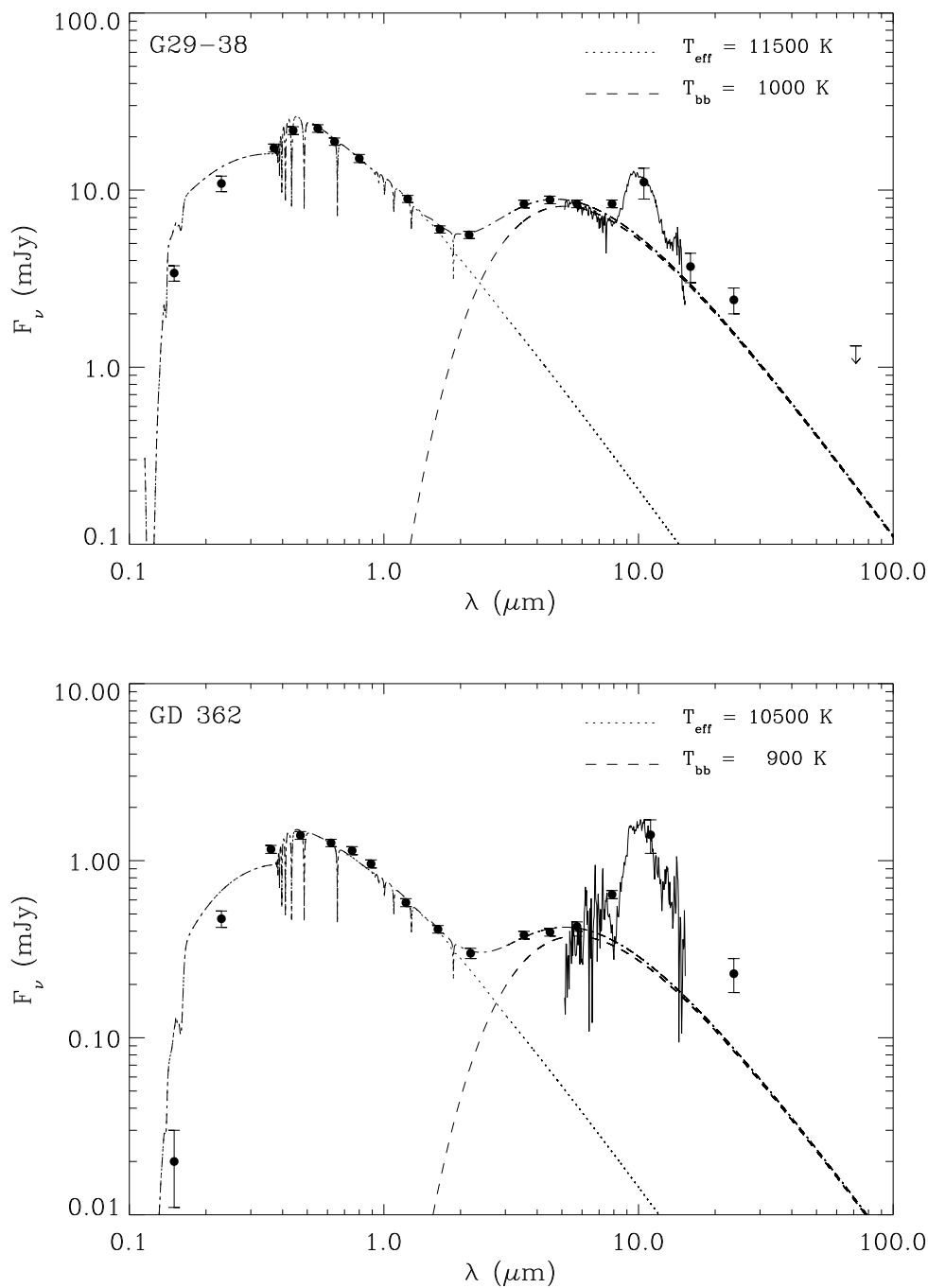}
\epsscale{1.0}
\caption{A side-by-side look at the full SEDs of G29-38 and GD\,362.  Available short wavelength 
data are shown together with {\em Spitzer} photometry and spectra (described in the text), plus 
10\,$\mu$m ($N$- or $N'$-band) photometry from the ground.  Also shown is an upper limit to 
the flux of G29-38 at 70\,$\mu$m from MIPS imaging.  Stellar atmosphere models are plotted as 
dotted lines, while simple blackbodies fitted to the near-infrared continua are shown as dashed 
lines.  The strength of the silicate emission at GD\,362 is plain in this figure, and its 24\,$\mu$m 
flux also appears strong relative to that at G29-38.
\label{fig12}}
\end{figure}

\begin{figure}
\figurenum{13}
\plotone{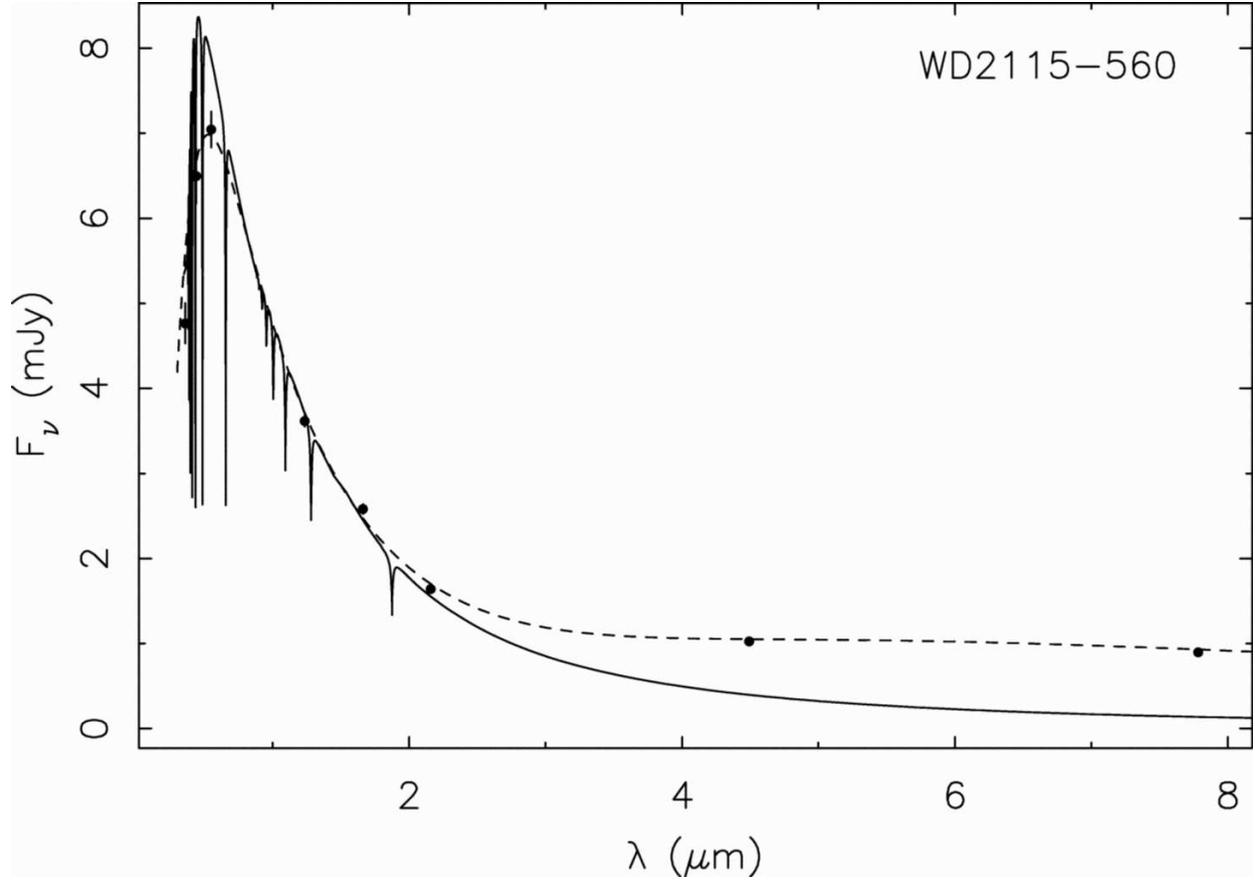}
\caption{Infrared excess at the DAZ white dwarf LTT\,8452 as measured by {\em Spitzer} IRAC.
Short wavelength photometry from the literature and IRAC fluxes are shown as dots with (small) 
error bars.  The solid line is a stellar atmosphere model, while the dashed line represents the 
addition of an optically thick, flat disk model for the circumstellar dust \citep{von07}.
\label{fig13}}
\end{figure}

\begin{figure}
\figurenum{14}
\plotone{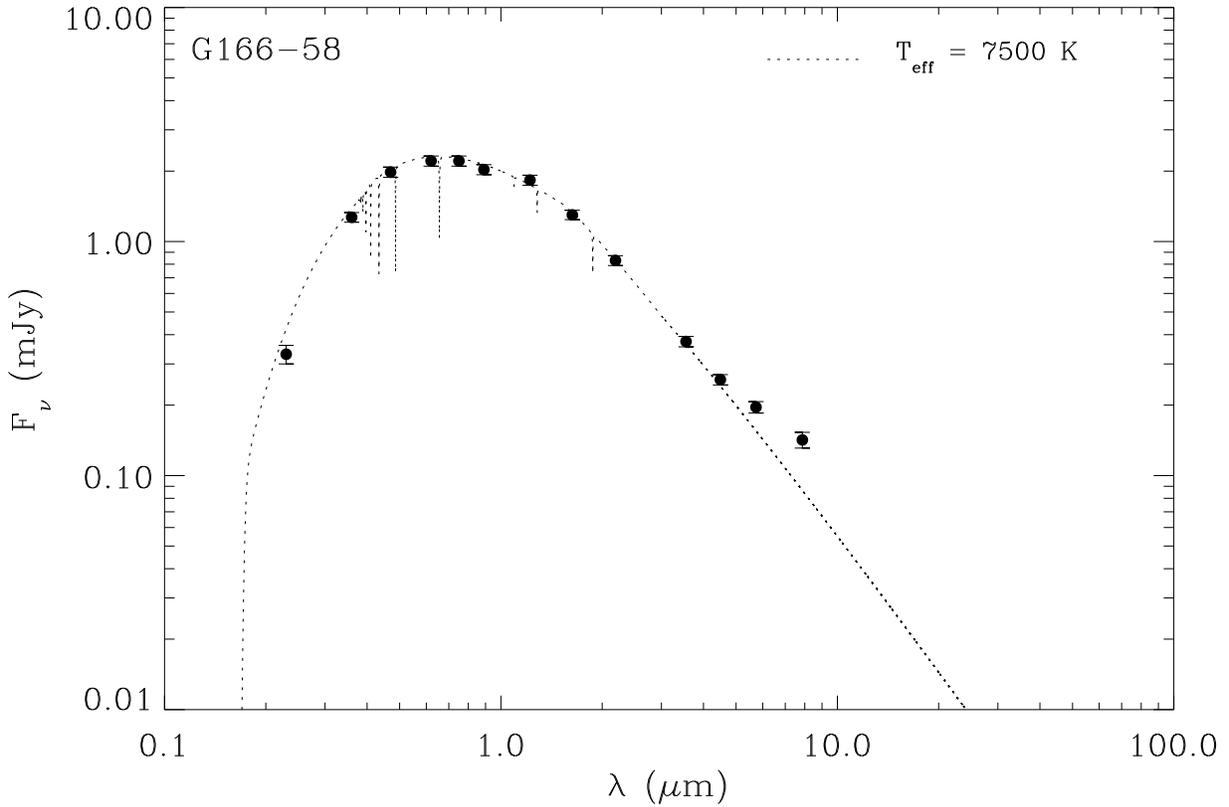}
\caption{G166-58 was the first white dwarf found to have an infrared excess at mid- but not
near-infrared wavelengths \citep{far08b}, and it remained anomalous among the first dozen
disks found at white dwarfs over a period of several years.  Available short wavelength data 
are shown together with IRAC photometry and a stellar atmosphere model.  The warmest 
dust at G166-58 is only $400-500$\,K compared to $1000-1200$\,K in other circumstellar 
disks at white dwarfs.
\label{fig14}}
\end{figure}

\begin{figure}
\figurenum{15}
\plotone{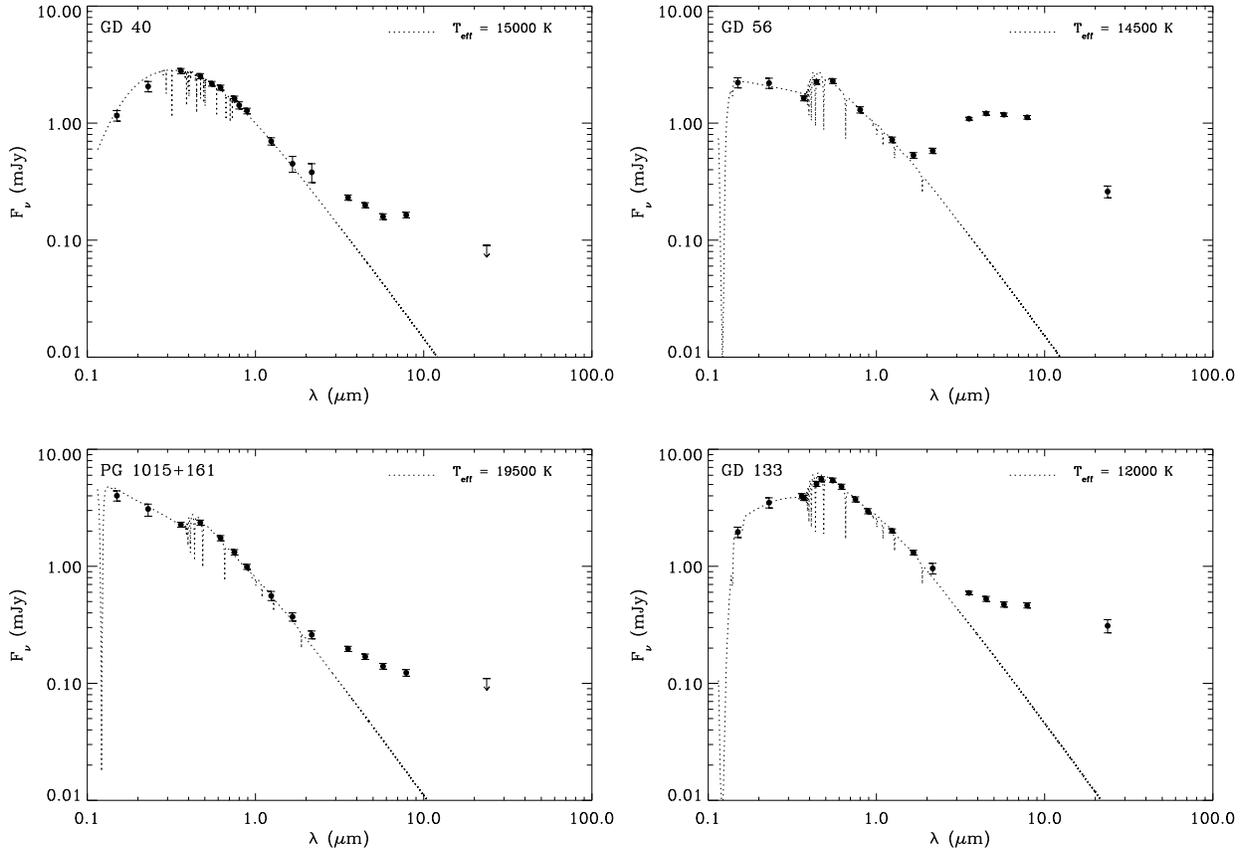}
\caption{Infrared excesses detected at GD\,40, GD\,56, GD\,133, and PG\,1015$+$161 with
{\em Spitzer} IRAC and MIPS.  Short wavelength photometric data from the literature are shown
together with appropriate stellar atmosphere models.  The downward arrows are $3\sigma$ 
upper limits for non-detections.  
\label{fig15}}
\end{figure}

\begin{figure}
\figurenum{16}
\plotone{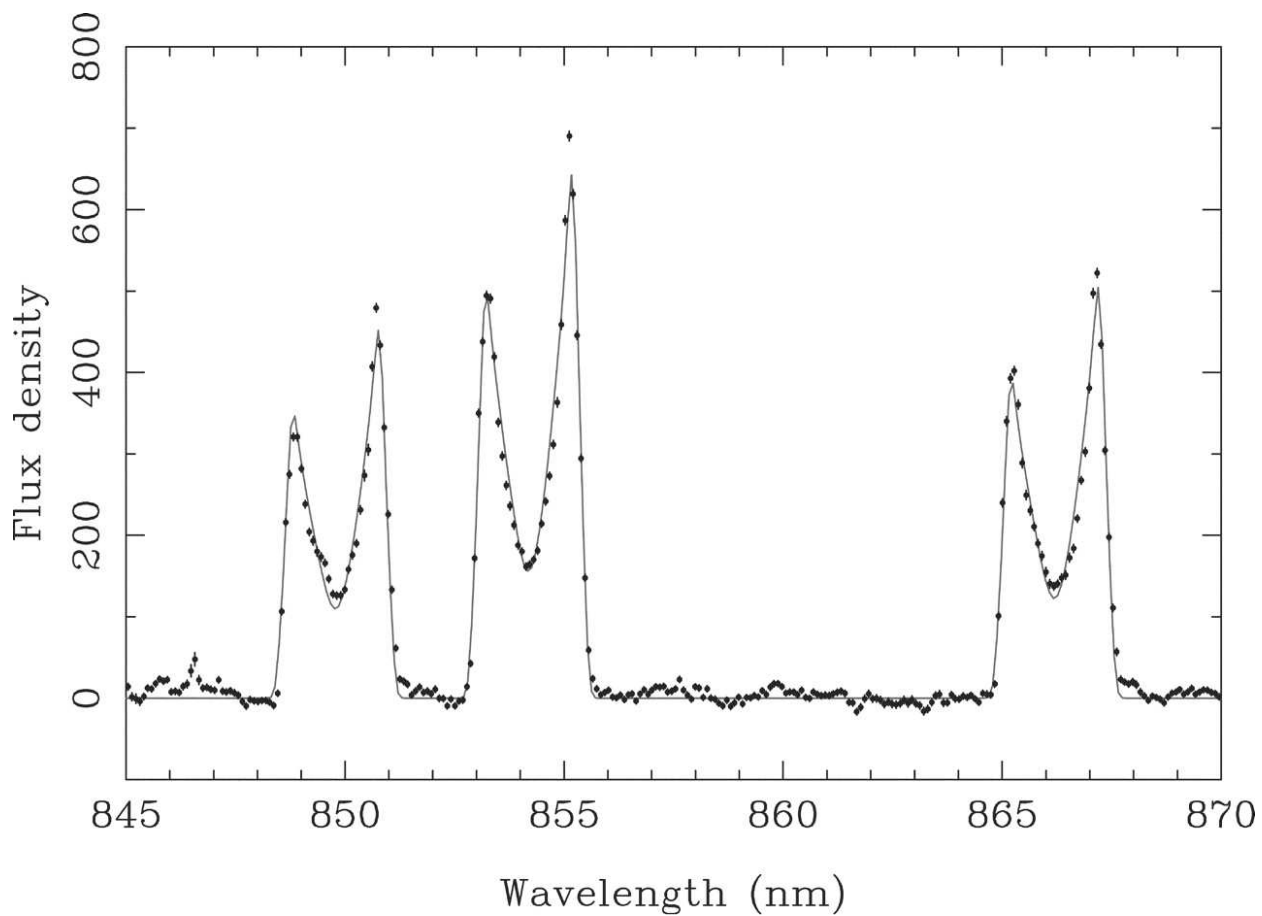}
\caption{Calcium emission in the optical spectrum of SDSS\,1228 measured taken at the William
Herschel Telescope with the ISIS spectrograph \citep{gan06}.  The data are shown as points with 
(small) error bars while the model for the emission is shown as a solid line.  The overall shape of 
the emission features is consistent with a high inclination (i.e.\ close to face on) and a gas disk 
radius no more than 1.2\,$R_{\odot}$ \citep{gan06}.  The asymmetry in the line profiles indicate 
a slightly elliptical orbit with $e=0.02$.
\label{fig16}}
\end{figure}

\begin{figure}
\figurenum{17}
\plotone{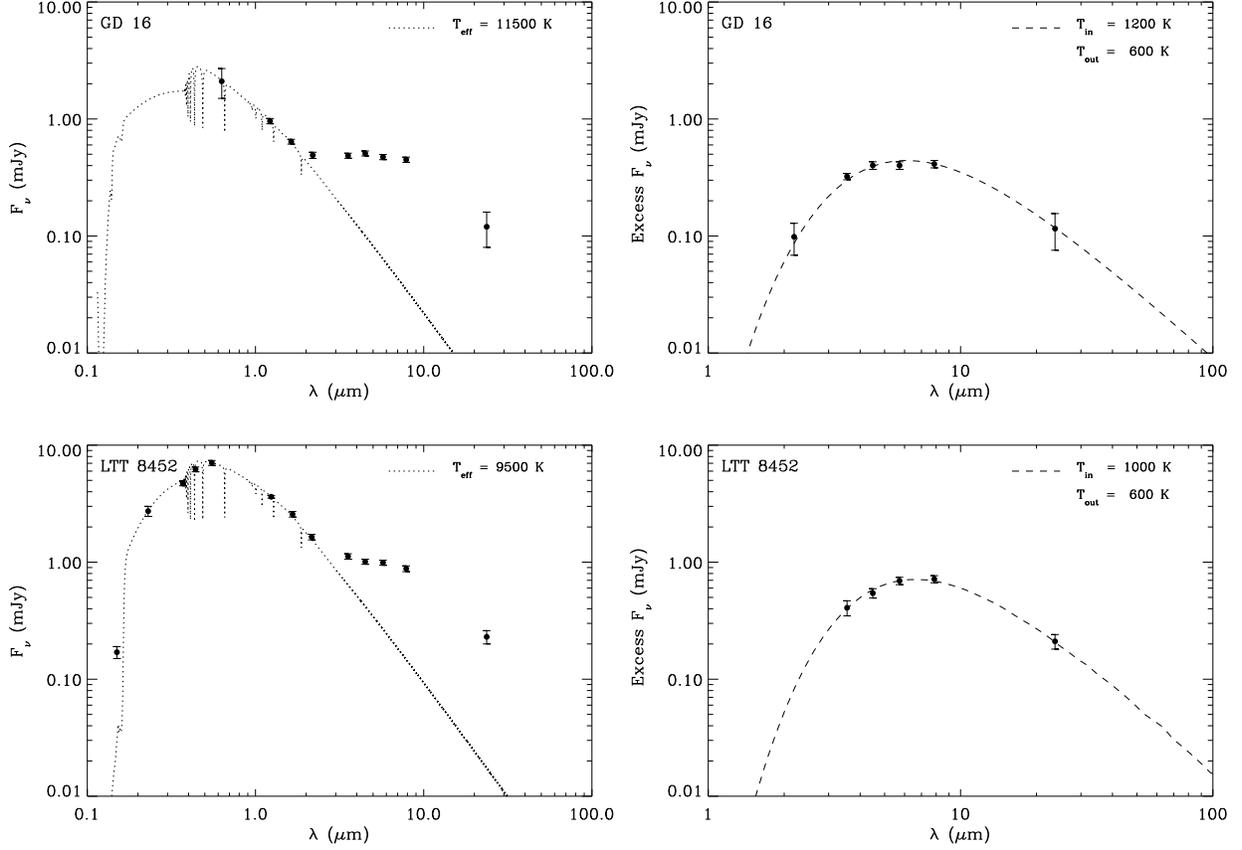}
\caption{{\em Spitzer} IRAC and MIPS $3-24\,\mu$m photometry for GD\,16 and LTT\,8452.  
The left hand panels show available short wavelength photometry togeher with the {\em Spitzer} 
data and reveal strong infrared excesses compared to photospheric models.  The right hand 
panels show the excess fluxes are well fitted by optically thick, flat disk models placing the 
warm dust within 0.5\,$R_{\odot}$ of the white dwarf \citep{far09b}.
\label{fig17}}
\end{figure}

\begin{figure}
\figurenum{18}
\includegraphics[scale=0.6,angle=-90]{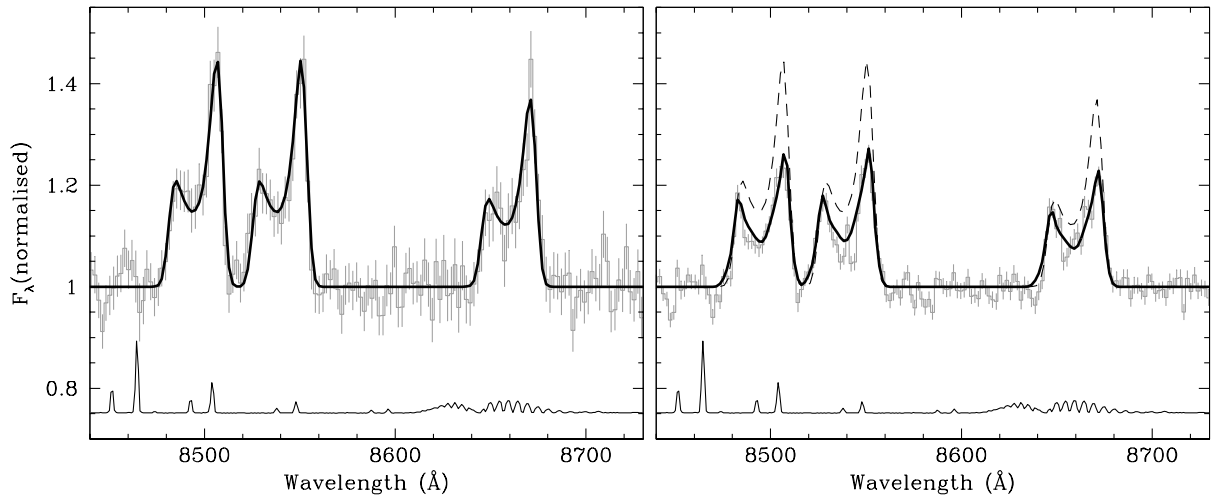}
\caption{Varying, asymmetric line profiles of the calcium emission from Ton\,345 \citep{gan08}.
The observed changes imply a shift in disk eccentricity from 0.4 (left panel, 2004 December) and 
0.2 (right panel, 2008 January).  Interestingly, no further changes to the line profile were observed 
in 2008 February and November \citep{mel10}.
\label{fig18}}
\end{figure}

Establishing just how much of the detected flux is excess and how much is photosphere can
be a major problem as many white dwarfs are not well-constrained by optical and near-infrared
photometry \citep{mcc99}.  Furthermore, it is probably the case that including IRAC 7.9\,$\mu$m
photometry in model fits, without the benefit of a MIPS 24\,$\mu$m constraint, will bias the outer 
disk radius towards larger values.  This is because the 7.9\,$\mu$m bandpass is sufficiently wide
to include flux from silicate emission, as clearly occurs for GD\,362 and to a lesser degree for 
G29-38 (see Figure \ref{fig12}).  \citet{far10c} fitted the $2-6\,\mu$m SEDs of all white dwarfs 
with infrared excess in a uniform manner to establish the best fractional infrared luminosity of 
their disks.  Figure \ref{fig25} plots these values and Table \ref{tbl2} lists these

There are at least three stars whose infrared excesses are sufficiently subtle that flat 
disk model fits to their IRAC data predict dust rings of radial extent $\Delta r<0.1\,R_{\odot}$:  
HE\,0106$-$3253, PG\,1457$-$086, and SDSS\,1043 all have $\tau<10^{-3}$ \citep{far10c,
far09b}.  Additional stars which may also have rings this narrow include HE\,0307$+$0746
and PG\,1015$+$161.  Since the inclination of these disks is unknown, the radial extent of 
the rings could be even smaller than predicted for nonzero inclination models.  For example,
if the disks at HE\,0106$-$3253 or PG\,1457$-$086 are near to face-on ($i=0\arcdeg$), their 
dust rings would have $\Delta r=0.01\,R_{\odot}$ or roughly an Earth radius and 10 times 
smaller than the rings of Saturn.

Narrow rings found at a few white dwarfs suggest that asteroid acccretion may be relevant to
additional and potentially many metal-contaminated stars without an obvious infrared excess
\citep{far10c}.  A dust ring of radial extent 0.01\,$R_{\odot}$ would be difficult to confirm via 
infrared photometry above an inclination of $i=50\arcdeg$ as it would produce an excess 
under $2\sigma$ for typical IRAC data.  At the same time, even such a narrow ring has the
potential to harbor over 10$^{22}$\,g of dust in an optically thick, flat disk configuration, and
supply metals at 10$^9$\,g\,s$^{-1}$ for nearly 10$^6$\,yr.  Circumstellar disks that produce 
subtle infrared excesses probably await detection, and may apply to the bulk of all metal-$
$polluted white dwarfs.

\subsubsection{Disks with Enlarged Inner Holes}

G166-58 displays an infrared excess that becomes obvious only at 5.7\,$\mu$m, while its
shorter wavelength IRAC data are consistent with the stellar continuum (Figure \ref{fig14} 
\citealt{far10c,far08b}).  As such, it is the only white dwarf with circumstellar dust where the
inner disk edge does not coincide with the region where silicate grains rapidly sublimate at 
temperatures near 1200\,K \citep{far09b}.  

However, in {\em Spitzer} Cycle 5 PG\,1225$-$079 was found to have a measured excess 
only at 7.9\,$\mu$m (Figure \ref{fig26}; \citealt{far10c}).  Without repeat observations or data 
at other wavelengths, it is difficult to assign any certainty to the measured excess at PG\,1225$
-$079.  If real (and confirmed in the future), then this coolmetal-rich white dwarf joins G166-58 
in having a dust disk with a relatively large inner region that is dust-poor.

The multiple asteroid model can account for disks with distinct regions dominated by either
dust or gas \citep{jur08}.  Hence it is plausible that the inner regions of dust disks at G166-58
and PG\,1225$-$079 were bombarded by small asteroids that vaporized solids there, but had
insufficient mass or orbital energy to destroy all the dust.  Additionally, the same model can
produce narrow or otherwise attenuated dust disks via impacts that annihilate rocky particles
in some, but not all originally dusty regions \citep{far10c}.

An alternative is that the inner circumstellar regions at these two stars are largely free of 
matter, having been near to fully accreted from the inside out.  This possibility suggests the 
extant photospheric metals in are residuals from prior infall of disk material.  G166-58 is by far 
the coolest white dwarf with an infrared excess, and the 2000\,yr timescale for metals to diffuse 
below the photosphere is relatively long for a DAZ white dwarf \citep{koe09}.  PG\,1225$-$079
is a DBZ white dwarf and therefore does not require accretion to be ongoing.  An accurate and
detailed determination of the lighter and heavier elements polluting these stars may be able to 
constrain their accretion history.

\subsection{The Composition and Masses of Asteroids at GD\,362 and GD\,40}

To date, the polluted stellar and circumstellar environments of two stars have been studied in 
great detail, facilitated by their relatively transparent, helium-rich atmospheres.  

\subsubsection{GD\,362}

Under a several hour exposure with Keck / HIRES, \citet{zuc07} detected 15 elements heavier 
than helium in the optical spectrum of GD\,362, positively shattering the record for number of 
metals detected in a cool white dwarf photosphere at any wavelength \citep{wol02,fri99,sio90b}.  
This remarkable star is spectacularly polluted and manifests detectable abundances of strontium 
and scandium, highly refractory elements that comprise about 1 part in 10$^9$ of the Sun 
\citep{lod03}.  

The array of ingredients polluting GD\,362 is rich in refractory and transitional elements, while 
relatively poor in volatiles \citep{zuc07}.  There are detectable abundances of six elements with 
condensation temperatures above 1400\,K present in the star; scandium, aluminum, titanium, 
calcium, strontium, and vanadium.  The transitional elements magnesium, silicon, and iron are
the most abundant elements in GD\,362 by a substantial margin; together with oxygen, these
three elements comprise 94\% of the bulk Earth \citep{all95}.  In fact, overall the pattern of heavy
elements in the white dwarf is most consistent with a combination of the bulk Earth and Moon 
(Figure \ref{fig21}; \citealt{zuc07}).  

\citet{jur09b} used X-ray observational upper limits to constrain the total mass accretion rate 
at GD\,362, noting that this helium-rich white dwarf has an anomalously high abundance of 
hydrogen ($\log\,$(H/He)$=-1.1$; \citealt{zuc07}).  All together, the mass of heavy elements 
in the convection zone of GD\,362 is $1.8\times10^{22}$\,g and comparable to the mass 
contained in a 240\,km diameter Solar System asteroid (e.g.\ Themis).  On the other hand, 
the mass of hydrogen in its outer layers is $7.0\times10^{24}$\,g \citep{koe09} and orders 
of magnitude larger than for all known helium-rich white dwarfs of comparable temperature 
except GD\,16 (which is also polluted with heavy elements from a circumstellar disk \citealt{jur09b,
far09b}).  

One possiblity for the atmospheric hydrogen in these two stars is delivery via water-rich 
planetary bodies.  Because surface ices will not survive the giant phases of stellar evolution, 
any extant water in asteroids at white dwarfs would have to be sufficiently buried.  The Solar 
System objects Ceres and Callisto are thought to have internal water that are roughly 25\% 
and 50\% of their total mass, respectively \citep{tho05,mcc05,can02}.  One possibility is that 
GD\,362 accreted its hydrogen in the form of water from a few to several hundred large 
asteroids (unlikely) or an even larger body with internal water.  If one supposes the current
accretion event is the result of the latter possibility, the all the observed properties of GD\,362
can be accounted for -- disk, atmospheric pollution, large hydrogen abundance, and X-ray
upper limit -- by the destruction and subsequent accretion of a parent body with a total mass 
between that of Callisto and Mars \citep{jur09b}.

\subsubsection{GD\,40}

In contrast, GD\,40 exhibits hydrogen-deficiency typical of DB stars in its temperature range, 
but is nonetheless interesting.  Already known to host magnesium, iron, silicon, and carbon 
from ultraviolet observations and calcium from optical spectroscopy \citep{fri99}, the discovery 
of its disk was in large part motivated by the recognition of its relative carbon-deficiency and 
the asteroid-sized mass of metals contained in the star \citep{jur06}.  Again using Keck / HIRES, 
\citet{kle10} detected an additional four elements and better constrained the abundnaces of all 
previously detected elements except carbon via multiple strong lines.  

From this optical dataset, seven of the eight heavy elements present in the atmosphere of 
GD\,40 form a subset of those detected in GD\,362, with the exception of oxygen.  Interestingly,
\citet{kle10} find that by assuming all of the atmospheric metals were delivered in their common
oxides and all the hydrogen was delivered in water, there is an excess of oxygen in the outer
layers of the star.  A solution to this conundrum is that the photospheric abundances differ from
the accreted abundances, and that GD\,40 has been accreting for a minimum of a few diffusion
timescales.  In this picture, the disk has a lifetime greater than 10$^6$\,yr, and the apparent
excess of oxygen is due to the fact that heavier elements such as iron sink more rapily than
oxygen \citep{koe09}.  

The mass of heavy elements currently in the convection zone of GD\,40 is $3.6\times10^{22}$\,g,
already an asteroid-sized mass.  However, with the caveat that the star must be in a steady state
and has been accreting for at least a few diffusion timescales, the minimum mass of metals in the 
destroyed parent body grows to $3\times10^{23}$\,g \citep{kle10} or about the size of Vesta, the 
second largest asteroid in the Solar System.  In a steady state accretion mode,  \citet{kle10} find 
that the water content of the now-destroyed minor planet can be no more than a few percent.
Lastly, GD\,40 exhibits a silicon-to-magnesium ratio significantly distinct from that found in stars 
and in the bulk Earth, hinting at the possibility that the parent body polluting GD\,40 was at least 
partially differentiated \citep{kle10}.

\subsection{A Last Look at the Interstellar Accretion Hypothesis}

The totality of published results and works in progress on disks and metal-pollution at white 
dwarfs circa 2009 strongly argues against interstellar accretion.  Yet even as the evidence for
circumstellar disks and rocky pollutants began accumulating, and subsequently disseminated 
at international conferences, the scientific community continued to cite the interstellar medium as 
a source for accreted metals in cool white dwarfs \citep{des08,duf07,koe06}.  Given the dearth of 
circumstellar disk detections at metal-rich white dwarfs cooler than 10,000\,K \citep{far10c,far09b}
and their 10$^4-10^6$\,yr metal diffusion timescales, it is perhaps understandable that some 
researchers remained skeptical.

The class of stars used most often to argue in favor of interstellar accretion are the coolest 
DBZ stars, spectrally classified DZ because helium transitions become difficult to excite below 
12,000\,K \citep{vos07,wes93}.  Because detectable metals can persist in DZ white dwarfs for 
a few to several Myr, it is conceivable that these stars are now located up to a few hundred pc 
from a region in which they became polluted by interstellar matter.  Classically, the glaring
problem with this picture is the lack of accreted hydrogen in these stars, as the interstellar
medium (cosmic or solar abundance) is 91\% hydrogen and 0.01\% elements heavier than
neon by number \citep{dap00,aan93}.  Calcium-to-hydrogen abundances in DZ stars are 
typically super-Solar, despite the fact that calcium continually sinks in their atmospheres 
\citep{far10a}.

\citet{far10a} used 146 DZ stars from the SDSS DR4 and with stellar parameters modeled by
\citet{duf07} to evaluate the interstellar accretion hypothesis by calculating several diagnostics.
No correlation is found between calcium abundance and tangential speed as expected if the 
stars were accreting at Bondi-Hoyle, fluid rates necessary to produce the observed pollution 
\citep{koe06,dup93a}.  More than half the sample white dwarfs are currently situated above the 
$\pm100$\,pc thick gas and dust layer of the Galaxy, with nearly one fifth of the stars located over 
200\,pc above the plane.  Furthermore, roughly half the sample are now moving back into the disk
of the Galaxy rather than away, implying they have been out of the interstellar medium for several 
to tens of Myr \citep{far10a}.

Comparing a commensurate number of cool, helium-rich SDSS white dwarfs with and without 
detected photopsheric metals also is a problem for instellar accretion.  The two classes of stars
appear to belong to the same population of disk stars in temperature, atmospheric composition,
Galactic positions and velocities.  Among all these helium-rich white dwarfs, there are pairs
within several pc of one another where: 1) only one star is metal-rich yet the pair share similar
space velocities, and 2) both stars are polluted yet their space velocities differ dramatically.

Perhaps the strongest evidence against interstellar accretion in these stars actually comes
from the same mechanism that permits the metals to reside for Myr timescales; their relatively
deep convection zones \citep{koe09,paq86}.  Because any atmospheric metals are thoroughly
mixed in the convective layers of the star, the measured abundances can be translated into
masses by knowing the mass of the convective envelope.  Figure \ref{fig22} plots the mass of
calcium detected in the 146 SDSS DZ stars as a function of effective temperature, revealing
these masses are typical of large Solar System asteroids.  Moreover, because metals in the
interstellar medium are locked up in dust grains, they will not be captured by a passing white 
dwarf at the fluid rate, but at a rate around 10 times the Eddington accretion rate \citep{alc80}.
The calcium masses plotted in Figure \ref{fig22} cannot be accounted for by the accretion of
interstellar dust grains, even assuming Myr passages within the densest of environments 
\citep{far10a}.

At present, there is no observational evidence in favor of interstellar accretion onto white
dwarfs, yet plentiful and compelling data supporting the accretion of rocky planetary material.
Given the existing observational data on metal-enriched white dwarfs, the interstellar accretion
hypothesis is no longer viable.

\subsection{Evidence for Water in Debris Orbiting White Dwarfs}

Starting from a default position that white dwarf atmospheric pollutants are delivered via 
planetary system remnants, some well-established observational properties then require
reexamination.  Perhaps most intriguing is the trace hydrogen seen in helium-rich white 
dwarfs. 

In short, if DB white dwarfs accreted interstellar hydogen at the fluid rate, they would quickly
become DA white dwarfs and therefore this does not occur \citep{wes79,koe76}.  Convective
mixing can dredge-up helium in DA stars with $T_{\rm eff}<12,000$\,K, resulting in a helium-$
$dominated atmosphere with traces of hydrogen \cite{tre08}, but hydrogen is found in DB and 
DBZ stars at warmer temperatures as well \citep{vos07}.  Lastly, trace hydrogen in helium-rich
white dwarfs can be either primordial or accreted from intersellar space at rates comparable 
to the Eddington rate (i.e.\ direct impact).  However, if helium atmosphere white dwarfs obtain 
trace metals and hydrogen from independent sources, then stars exhibiting both should be less 
frequent than stars exhibiting either.  This is not supported by the observations, suggesting that 
helium atmosphere white dwarfs with both metals and hydrogen are polluted by water-rich 
asteroids \citep{far10a}.

Of the four DBZ white dwarfs with circumstellar dust, GD\,16 and GD\,362 have more than $10^
{24}$\,g of atmospheric hydrogen, while GD,40 and Ton\,345 have abundances and upper limits 
four orders of magnitude lower \citep{jur09b}.  These stark differences are not due to sensitivity, 
as trace hydrogen is detectable in DBZ white dwarfs typically down to 1 part in 10$^5$ with 
moderate-resolution optical spectra \citep{vos07,duf07}.  The hydrogen-poor atmospheres of 
the latter two stars are fairly typical of DB white dwarfs, while the hydrogen contents of the 
former two stars are rather remarkable.

One possibility is these four stars represent extremes of water content in planetary bodies
post-main sequence.  Oxygen has yet to be detected in either GD\,16 or GD\,362, but as both 
stars have $T_{\rm eff}<12,000$\,K and only optical spectra to date, this is not really surprising.
It takes enormous oxygen abundances to produce detectable, optical absorption lines in cool 
white dwarfs \citep{gan10}.  Sensitive searches for photospheric oxygen in these two stars via
ultraviolet spectroscopy will be able to further constrain scenarios of pollution by parent bodies
with internal water.

Interestingly, there are currently two warm DBZ stars with marked oxygen detections made in 
the far-ultraviolet; GD\,61 and GD\,378 \citep{des08}.  \citet{jur10} recognized the significance 
of these data, noting that both atmospheres had O/C $>1$ and trace hydrogen in a temperature 
range where convection could not transform a DA into a DB.  Their carbon-deficiency rules out 
pollution by interstellar matter, and accounting for their multiple atmospheric metals as oxides 
leaves an excess of oxygen in both stars \citep{jur10}.  Because the heavy element diffusion
timescales in these DBZ white dwarfs are of the order 10$^5$\,yr, it remains uncertain what
fraction of the excess oxygen can be attributed to mineral oxides common to rocky planetary 
bodies, but whose metal components have sunk more rapidly than the oxygen.  Under some 
reasonable assumptions about their accretion history, \citet{jur10} finds that GD\,378 was likely 
polluted by a water-rich asteroid, while the case for water at GD\,61 is less compelling but still 
possible\footnote{mention {\em Nature} result here?}.

The near future will probably hold some confirmations of water in the circumstellar debris that 
orbits white dwarfs.

\section{RELATED OBJECTS}

The discussion in this chapter has been focussed on apparently single, cool and metal-enriched 
white dwarfs because only these stars are observed to have circumstellar dust.  Additionally, all 
stars in this class require external sources of pollution, for which circumstellar matter is currently 
the strongest candidate.  It has already been mentioned that {\em Spitzer} observations have not
identified infrared excess at over 150 white dwarfs without photospheric metals, but this needs
some qualification.  Below is a brief discussion of white dwarfs that are (potentially) related to the
stars of this chapter.

\subsection{White Dwarfs Polluted by Companions?}

\citet{zuc03} identified several DAZ stars among white dwarfs with low mass stellar companions, 
M dwarfs, in suspected or confirmed short period orbits.  The orbital separations of these systems 
are either established by radial velocity studies or constrained by direct imaging, and consistent
with detached, post-common envelope binaries \citep{hoa07,far05,sch96}.  Although these stars
are not interacting in the conventional sense of Roche lobe overflow, the relatively high frequency
of photospheric metal detections in their white dwarf components indicates that these DAZ stars
are capturing material from a stellar wind \citep{deb06,zuc03}.  Recent high-resolution imaging 
with the {\em Hubble Space Telescope (HST)} has strengthened this interpretation by finding 
all the DA$+$dM systems are visual pairs, while the DAZ$+$dM systems are typically spatially 
unresolved \citep{far10b}.

These observational results strengthened the hypothesis that apparently single, metal-enriched 
white dwarfs may be polluted by unseen companions \citep{dob05,hol97}.  However, any such
companions would have to be sufficiently low in both mass and temperature that they are not
detected in large near-infrared surveys such as 2MASS \citep{hoa07}.  Five years of {\em Spitzer}
IRAC studies have laid this hypothesis firmly to rest; the infrared SEDs of metal-rich white dwarfs
fail to reveal the expected signature of low mass companions down to 25\,$M_{\rm J}$, according
to substellar cooling models \citep{far09b,far08a,bar03,bur03}.  Therefore, metal-rich white dwarfs 
are not polluted by mass transfer or wind capture from unseen, substellar companions.

Interestingly, two of the five known brown dwarf companions to white dwarfs are in close 
orbits analogous to the DAZ$+$dM systems found by \citet{zuc03}; GD\,1400B and WD\,0137$
-$049B \citep{max06,far04}.  Neither of the two cool white dwarfs in these systems show signs
of atmospheric metal pollution from their close substellar secondaries, despite both stars having 
a large number of co-added, high-resolution spectra taken with VLT / UVES (R. Napiwotzki 2009,
private communication).  Thus, it  appears that the mid- to late-L dwarfs in these systems do not 
generate winds comparable to their higher mass, M dwarf counterparts.

\subsection{Dust in The Helix?}

\citet{su07} reported a strong infrared excess at the location of the central star of the Helix 
Nebula (NGC\,7293), detected at 24 and 70\,$\mu$m with {\em Spitzer} MIPS (Figure \ref{fig27}).
An analysis of the excess emission using photometry at all IRAC and MIPS wavelengths, together
with high-resolution IRS spectroscopy between 10 and 35\,$\mu$m, reveals a 100\,K continuum.
\citet{su07} attributee this emission to cold dust grains at several tens of AU from the star, which 
evolutionarily may be best described as a pre-white dwarf \citep{nap99}.

The 24\,$\mu$ flux source was reported to be partly spatially-resolved but potentially due to 
imperfect subtraction of the diffuse nebular emission, which is nonuniform and contains both
continuum and emission components \citep{su07}.  Assuming the detected excess is indeed 
point-like, the $6''$ size of the MIPS 24\,$\mu$m PSF only limits the physical size of the emitting 
region to 1300\,AU and herein lies one difficulty in the interpretation of the detected excess.
Additionally, the nebular emission is spatially complex and contains dust condensed within 
the expelled and cooling envelope of the giant star progenitor of the pre-white dwarf.  This
outflowing dust emits in the infrared and is a significant component of the overall emission 
in the Helix Nebula \citep{su07}. 

A distinct alternative to cold dust analogous to that detected at main-sequence stars is dust
captured by a companion star and heated by the pre-white dwarf.  \citet{bil09} reported the
results of a {\em Spitzer} survey for additional 24\,$\mu$m excesses at 72 hot and pre-white
dwarfs, about half of which are in planetary nebulae.  They find 12 new cases of strong 
excess, all of which are associated with planetary nebulae, and at least two of which have 
binary central stars.  Given that (asymmetric) planetary nebulae are likely to be intimately 
connected with binary star evolution \citep{dem09}, and in light of the findings of \citet{bil09},
it seems plausible the infrared excess in the Helix is not associated with a circumstellar dust 
disk at the pre-white dwarf.

\section{SUMMARY AND OUTLOOK}

\subsection{Remnant Planetary Systems}

The current picture painted by the observations of circumtellar dust at white dwarfs, and the
consequent atmospheric pollution, is of a surviving planetary system.  The asteroid analogs
currently favored by the data could be either primitive bodies and analogous to member of 
the Main Belt in the Solar System, or they might be fragments of a larger parent body such as 
a moon or major planet.  Regardless, the inferred masses and elemental abundance patterns 
of the observed debris is consistent with material condensed within the near environment of 
a main-sequence star and so far similar to the constitution of the terrestrial planets and their 
building blocks, the minor planets \citep{kle10,far10a,jur09b,zuc07,jur06}.

In order for an asteroid to pass within the Roche limit of a white dwarf, it must be in a highly
eccentric orbit requiring substantial gravitational perturbation.  Thus, major planets almost
certainly exist in metal-polluted white dwarf systems, especially in those with circumstellar 
dust disks.  Furthermore, it is energetically difficult to deposit the entire mass of an asteroid
in close orbit around a white dwarf.  This can potentially be achieved over many orbits as
collisions and viscous stirring, especially if significant amounts of gas are produced during 
a tidal disruption event.  Alternatively, a large but intact fragment of the shattered parent body 
might carry away most of the orbital energy and continue to interact with the evolving disk
with each orbital pass.

Evidence for such interactions may already exist.  The aforementioned change in the 
eccentricity of the gaseous debris at Ton\,345, as well as the eccentricity itself are most 
easily induced by gravitational forces \citep{mel10,gan08}.  Strong infrared continuum 
emission from stars like GD\,56 and GD\,362 cannot be accounted for by flat disks alone, 
implying some portion of their disks is warped or flared \citep{jur09a,jur07a}.  Again, these 
deviations from a fully relaxed disk require the injection of energy that can be provided by 
large bodies such as (Saturnian) ring-moon analogs near or within the disk, or major 
planets further out.

The inferred masses of circumstellar dust, supported by the observed heavy element 
masses within the polluted stars, should typically be only a fraction of the total mass of 
terrestrial bodies within these systems \citep{zuc10}.  If extrasolar planetesimal belts scale 
similarly to the Solar System, then smaller asteroids will be injected into the inner system 
relatively often but may result in a disk composed primarily of gaseous debris \citep{jur08}.  
More rarely, a large asteroid will approach a white dwarf and become tidally destroyed.  With 
sufficient mass it can overcome sputtering within any pre-existing, gas-dominated disks and 
self-annihilation via particulate collisions, resulting in an infrared excess \citep{jur08,far08b}.  
A combination of injection events may give rise to a diverse family of dust rings at white
dwarfs, including narrow or tenuous rings, and dusty disks large inner holes \citep{far10c}.

Given the correlation of dust disk frequency with cooling age, it seems reasonable that
dynamical resettling in planetary systems post-main sequence is driving the injection of
minor planets into the inner regions of white dwarf systems \citep{far09b}.  This picture is
strengthened by recent results indicating higher metal accretion for warmer white dwarfs
\citep{zuc10}, and predicts that rocky planetary bodies are destroyed within the Roche limit 
of hotter white dwarfs more often than those observed at cool white dwarfs.  High-resolution 
ulltraviolet spectra of hot DA white dwarfs with photospheric metals has revealed velocity-$
$shifted, circumstellar (and distinctly not interstellar), heavy element absorption features in 
several stars \citep{ban03}.  At $T_{\rm eff}\geq30,000$\,K, white dwarfs will sublimate even 
the most refractory materials within their Roche limits, suggesting the possibility that hotter 
white dwarfs may host disks composed only of gaseous heavy elements \citep{von07}.

Tentative evidence points to dust disks that are relatively long-lived and ultimately 
dissipated via accretion onto the white dwarf.  In this picture, the lifetime of a given dust 
disk may be largely determined by its initial mass.  Viscous spreading among solids and PR 
drag are orders of magnitude insufficient to generate the inferred metal accretion rates at DAZ 
stars, and it is likely that gas viscosity from sublimated particles at the inner dust disk edge 
drives these rates \citep{far10c,jur08,far08b,von07}.  Forces such as PR drag and radiation 
pressure are unable to significantly influence the lifetime of optically thick circumstellar dust 
at $T_{\rm eff}\la20,000$\,K white dwarfs.  Optically thin material necessary to account for 
the obseved silicate emission features at white dwarfs with circumstellar dust is naturally
supplied by an (optically thick) disk that is orders of magnitude more massive \citep{jur09a}.

The frequency of metal-polluted white dwarfs likely reflects the frequency of rocky planetary
systems at main-sequence, A- and F-type stars that are progenitors of the current population
of white dwarfs.  Very conservatively, the minimum frequency of remnant planetary systems 
at white dwarfs is around 3\% from those displaying both atmospheric metals and circumstellar 
dust \citep{far09b}.  However, evidence is growing that most if not all metal-contaminated white 
dwarfs are the result of the accretion of rocky planetesimals \citep{far10a}.  In this more liberal 
and likely more accurate view, the fraction of intermediate mass stars with terrestrial planetary 
systems that have survived partly intact into the white dwarf phase is $20-30$\% \citep{zuc10,
zuc03}.  It appears likely that some fraction of A- and F-type stars build planetary systems 
replete with water-rich asteroids, the building blocks of habitable planets and potentially 
analogous to the parent bodies that supplied the bulk of water in Earth's oceans 
\citep{mor00}.

\subsection{The Present and Near Future}

At the time of writing, {\em Spitzer} IRAC is still operating at 3.6 and 4.5\,$\mu$m with Cycle 
7 observations beginning in the latter part of this year (2010), with an eighth cycle is planned.
Observing at mid-infrared wavelengths similar to cryogenic {\em Spitzer}, the {\em Wide Field 
Infrared Survey Explorer (WISE)} has just completed its coverage of the entire sky.  While not
as sensitive as pointed observations with {\em Spitzer} IRAC, {\em WISE} has the capability to
detect bright (1\,mJy), warm dust disks at white dwarfs not yet known to be metal-polluted and 
hence not yet targeted by {\em Spitzer} or ground-based searches for infrared excess.

{\em Herschel} is currently observing in the far-infared and soliciting the first open round of 
proposals from the scientific community, but may not be sufficiently sensitive for dust studies
at white dwarfs.  The mid-infrared excesses observed by {\em Spitzer} are typically tens to 
hundreds of $\mu$Jy and falling towards longer wavelengths, while {\em Herschel} reports 
5 to 10\,mJy detection limits.  Unless white dwarfs harbor especially prominent excesses from
cold dust, around $1-10$ times their peak optical fluxes, {\em Herschel} will not detect dust in 
many white dwarf systems.  At somewhat longer wavelengths, the Atacama Large Millimeter 
Array (ALMA) is set to begin science operations (next year) in 2011.  ALMA should have both
$\mu$Jy sensitivity and sub-arcsecond spatial resolution, making it a promising tool for direct
imaging of spatially-resolved, cold debris at white dwarfs.

Further in the future, {\em JWST} will operate in the near- and mid-infrared with sensitivity
superior to {\em Spitzer}.  There are two leaps forward that can be envisioned with such
capacity.  First, better quality mid-infrared spectra of warm dust at white dwarfs will enable 
a more detailed analysis of its temperature, location, geometry, and composition.  Second,
infrared spectroscopy may reveal very subtle infrared excesses at metal-rich white dwarfs
where {\em Spitzer} IRAC photometry does not.  This technique was successfully employed
at main-sequence stars with {\em Spitzer} IRS, where high S/N spectroscopy revealed at
least two cases of spectacular yet subtle, warm dust emission \citep{lis07,bei06,bei05}.

On the other end of the electromagnetic spectrum, the recently-installed ultraviolet Cosmic 
Origins Spectrograph (COS) on the {\em HST} promises to be a useful instrument with which 
to study the elemental abundances of polluted white dwarfs.  Optimized for the far-ultraviolet,
COS will function best for metal-rich white dwarfs warmer than 16,000\,K, but near-ultraviolet
observations are possible with lower efficienty for all but the coolest white dwarfs with metals.
The science enabled by instruments like COS is really at the core of circumstellar dust studies
at white dwarfs.  The stellar atmosphere distills the accreted debris, providing the bulk chemical 
composition destroyed parent body.  In this manner, circumstellar dust studies at white dwarfs
have the potential to reveal more about the anatomy of extrasolar, terrestrial planetary bodies 
than any other technique.

\acknowledgments

I would like to thank my colleagues and collaborators for many years of helpful discussions 
and interesting projects, and the editor D. W. Hoard for the opportunity to contribute to this 
book.  The published and online\footnote{http://www.astronomy.villanova.edu/WDCatalog/index.html} 
versions of the white dwarf catalog assembled by G. McCook \& E. M. Sion have been an 
invaluable resource over the years and the white dwarf community owes the authors and 
caretakers an enormous debt of gratitude for their effort.  I also thank D. Koester for making
available the white dwarf atmosphere models used in this chapter.  Many of the plots and 
figures used here include data from the various sources: the {\em Galaxy Evolution Explorer 
(GALEX)}, the Sloan Digital Sky Survey (SDSS), the Two Micron All Sky Survey (2MASS), 
and the {\em Spitzer Space Telescope}.  {\em GALEX} is operated for NASA by the California 
Institute of Technology under NASA contract NAS5-98034.  The SDSS is managed by the 
Astrophysical Research Consortium for the Participating Institutions (http://www.sdss.org/).  
2MASS is a joint project of the University of Massachusetts and the Infrared Processing and 
Analysis Center / California Institute of Technology, funded by NASA and the National Science 
Foundation.  The {\em Spitzer Space Telescope}, is operated by the Jet Propulsion Laboratory, 
California Institute of Technology under a contract with NASA

\clearpage

\clearpage

\clearpage

\clearpage

\clearpage

\clearpage

\clearpage

\clearpage

\clearpage

\clearpage

\clearpage

\clearpage

\clearpage

\clearpage

\clearpage

\clearpage

\clearpage

\clearpage

\begin{figure}
\figurenum{19a}
\plotone{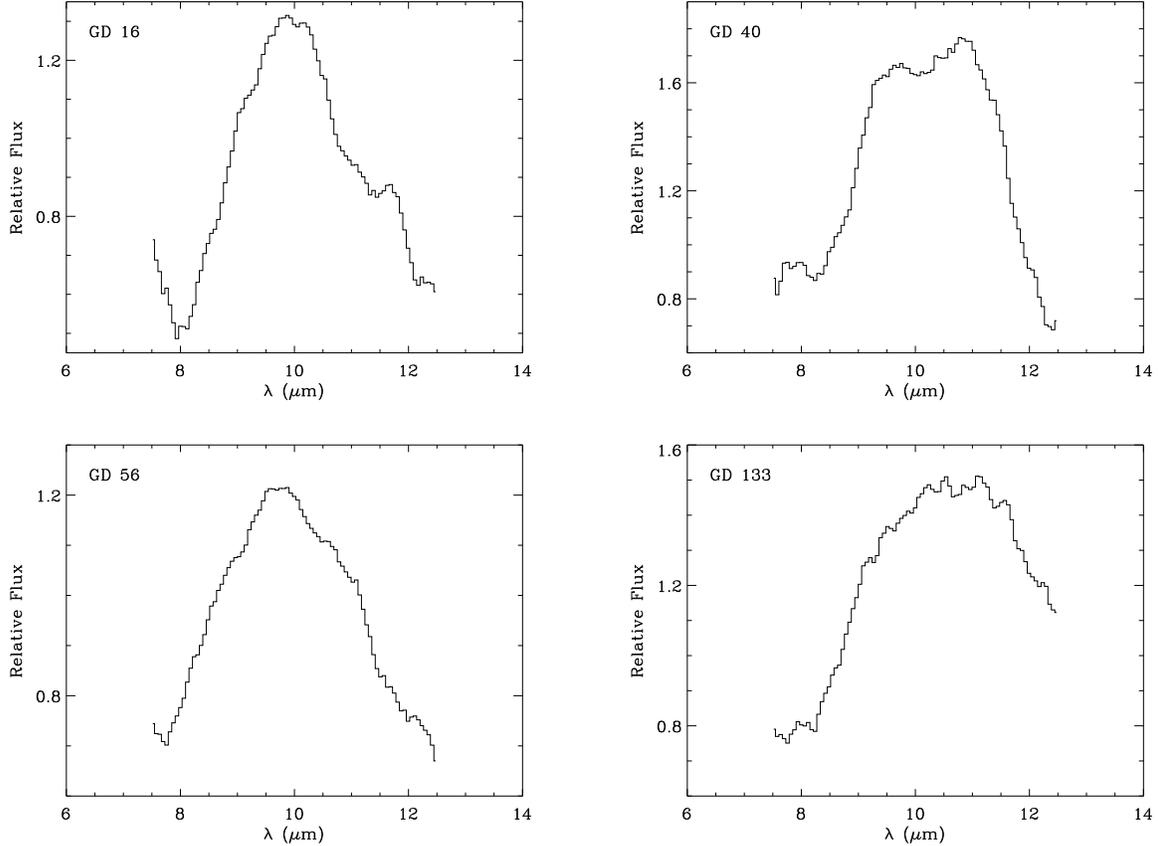}
\caption{Silicate emission features detected in all existing {\em Spitzer} IRS low-resolution 
observations of metal-rich white dwarfs with infrared excess.  Each spectrum has been 
normalized to the average of the 5.7 and 7.9\,$\mu$m IRAC fluxes, and smoothed by 15\,pixels 
(0.8\,$\mu$m).  The binned data points are highly correlated and structures within the broad 
silicate feature are probably not real. The detections are modest in most cases, and the data 
below 8\,$\mu$m are not shown as this region is typically noise-dominated.  Despite the low 
S/N in most cases, the binned data clearly show the features are broad, with red wings extending 
to 12\,$\mu$m and typical of minerals associated with planet formation \citep{jur09a}.  Note the 
strength of the feature at GD\,362 dwarfs all other detections.
\label{fig19a}}
\end{figure}

\clearpage

\begin{figure}
\figurenum{19b}
\plotone{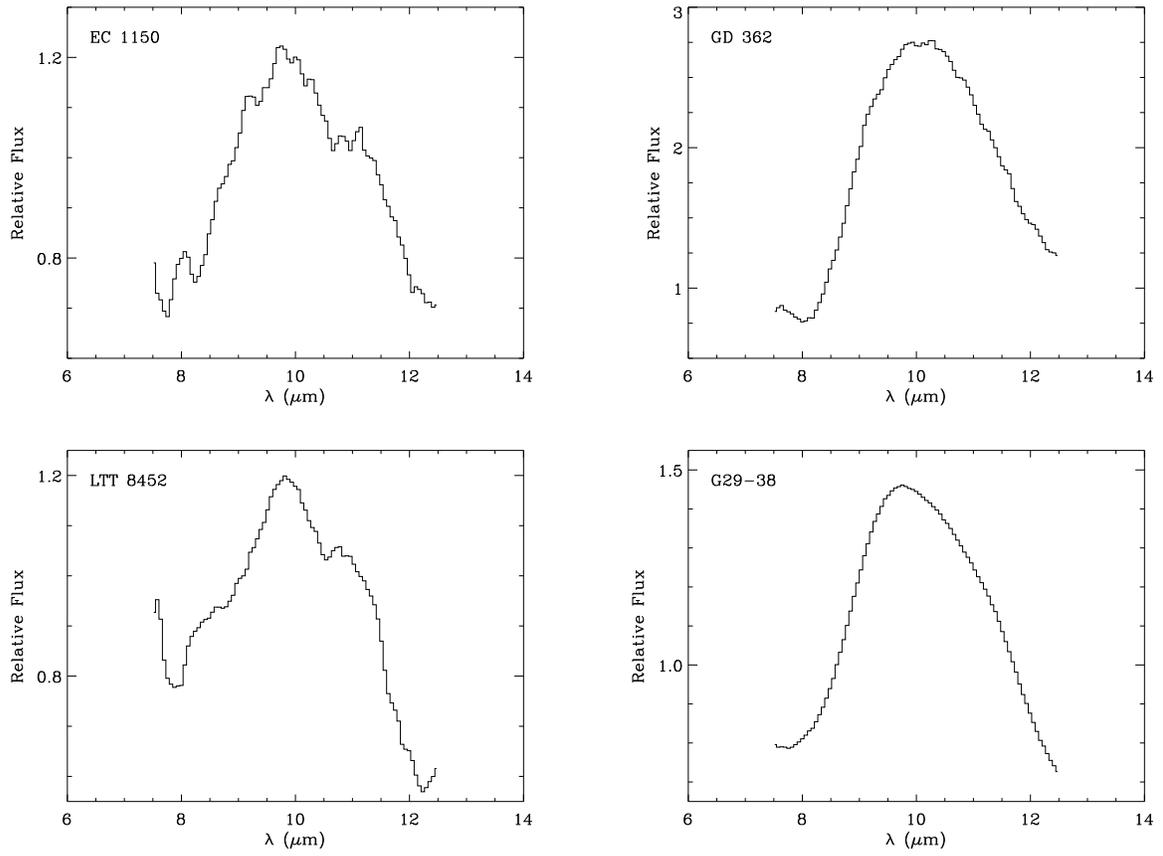}
\caption{{\em Continued}
\label{fig19b}}
\end{figure}

\clearpage

\begin{figure}
\figurenum{20}
\epsscale{0.8}
\plotone{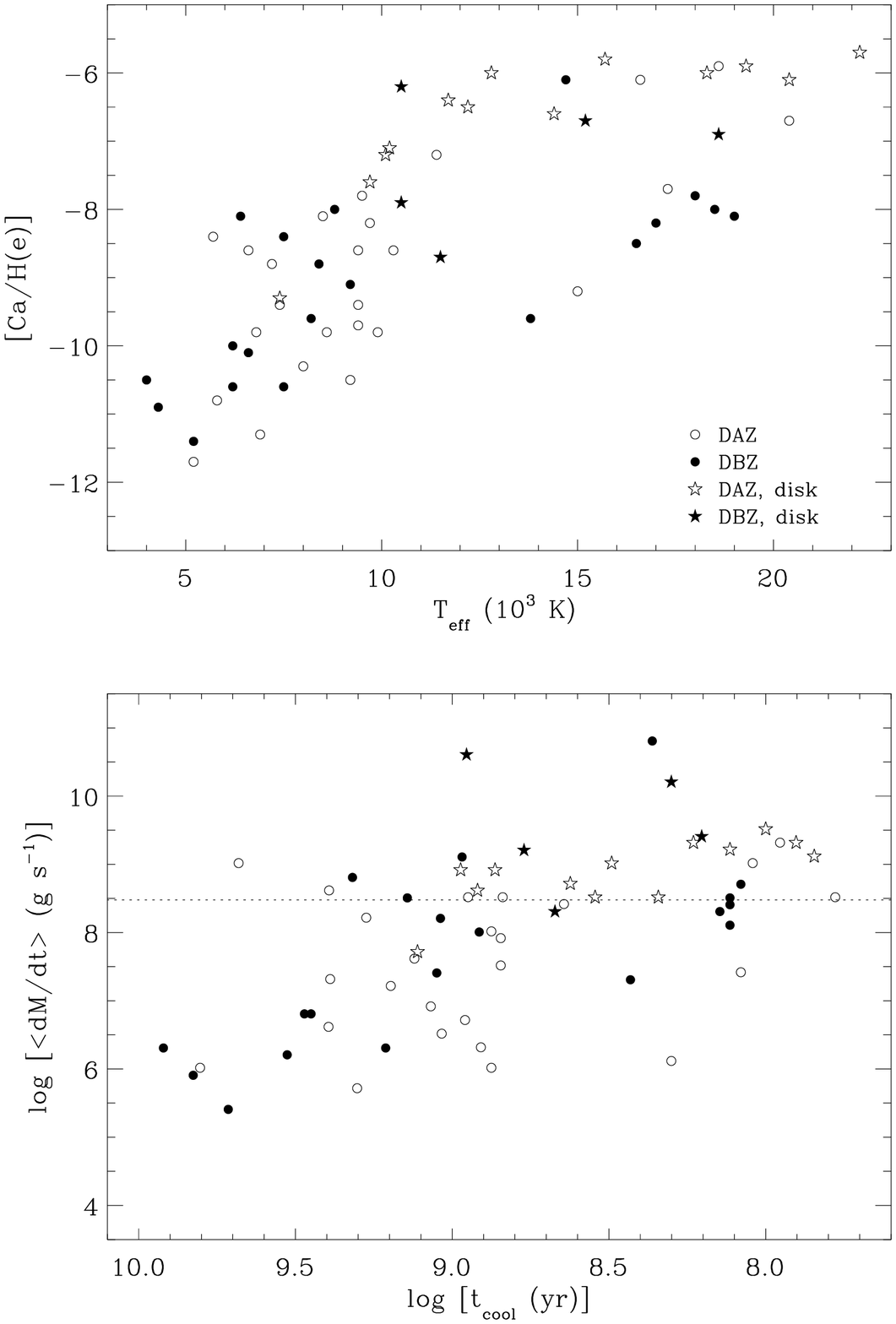}
\epsscale{1.0}
\caption{Dust disk frequency among all 61 metal-rich white dwarfs observed by {\em Spitzer} 
IRAC \citep{far10c}.  The upper panel uses a purely observational approach plotting both disk 
detections and non-detections versus calcium abundance and effective temperature.  The lower 
panel employs more physics by calculating the time-averaged metal accretion rate and cooling
age for each star.  The dotted line in the lower panel corresponds to $3\times10^8$\,g\,s$^{-1}$.
G166-58 is the only star with a disk that is located significantly below this accretion rate benchmark,
and with a cooling age beyond 1\,Gyr.
\label{fig20}}
\end{figure}

\clearpage

\begin{figure}
\figurenum{21}
\epsscale{0.8}
\plotone{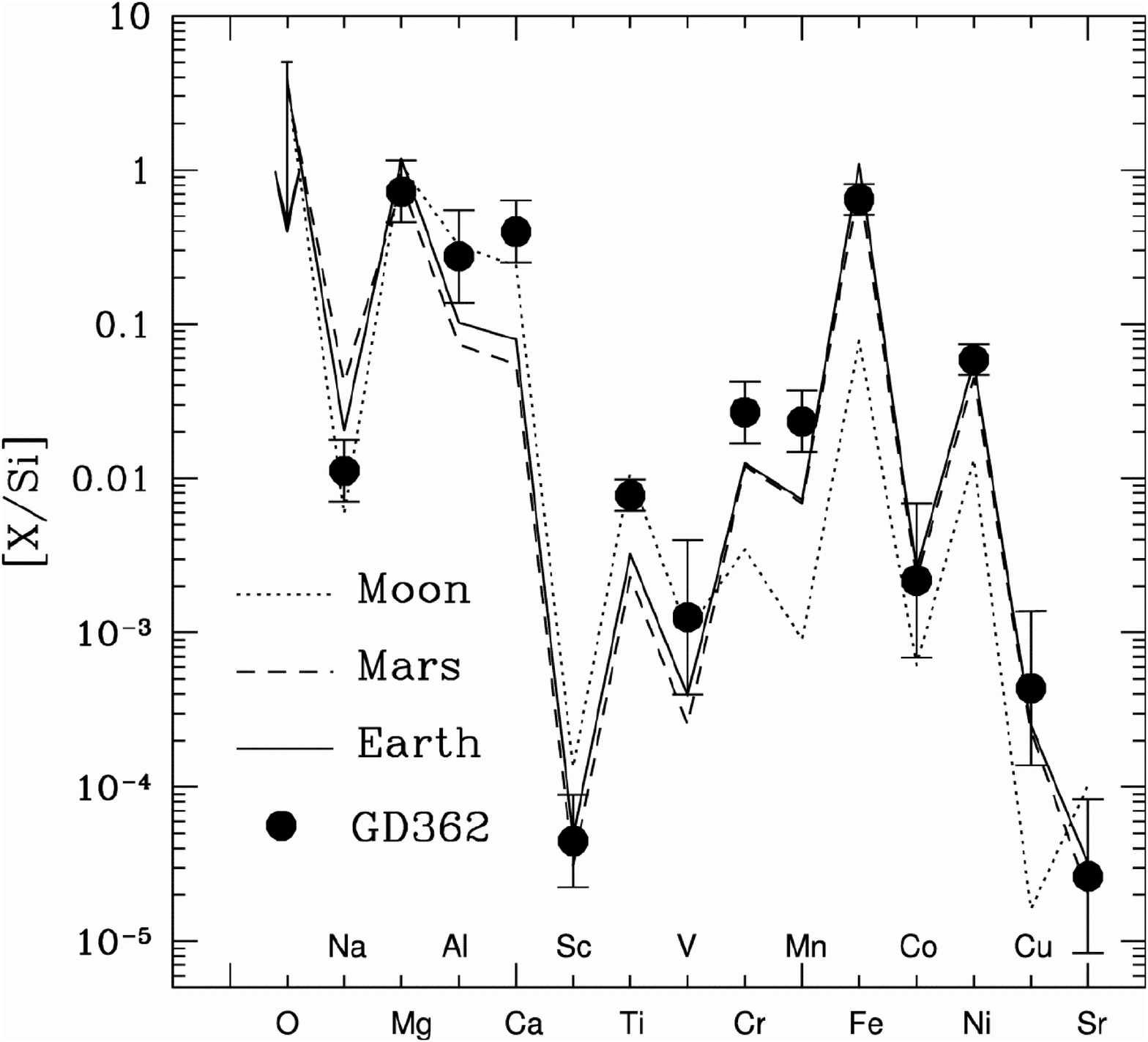}
\epsscale{1.0}
\caption{The remarkable heavy element abundances in the atmosphere of GD\,362 \citep{zuc07}.  
Plotted are the measured abundances for 14 detected metals (and an upper limit for oxygen) relative 
to silicon, together with the bulk composition of the Earth, Moon, and Mars .  The best overall fit is
achieved with a combination of the Earth and Moon.
\label{fig21}}
\end{figure}

\clearpage

\begin{figure}
\figurenum{22}
\plotone{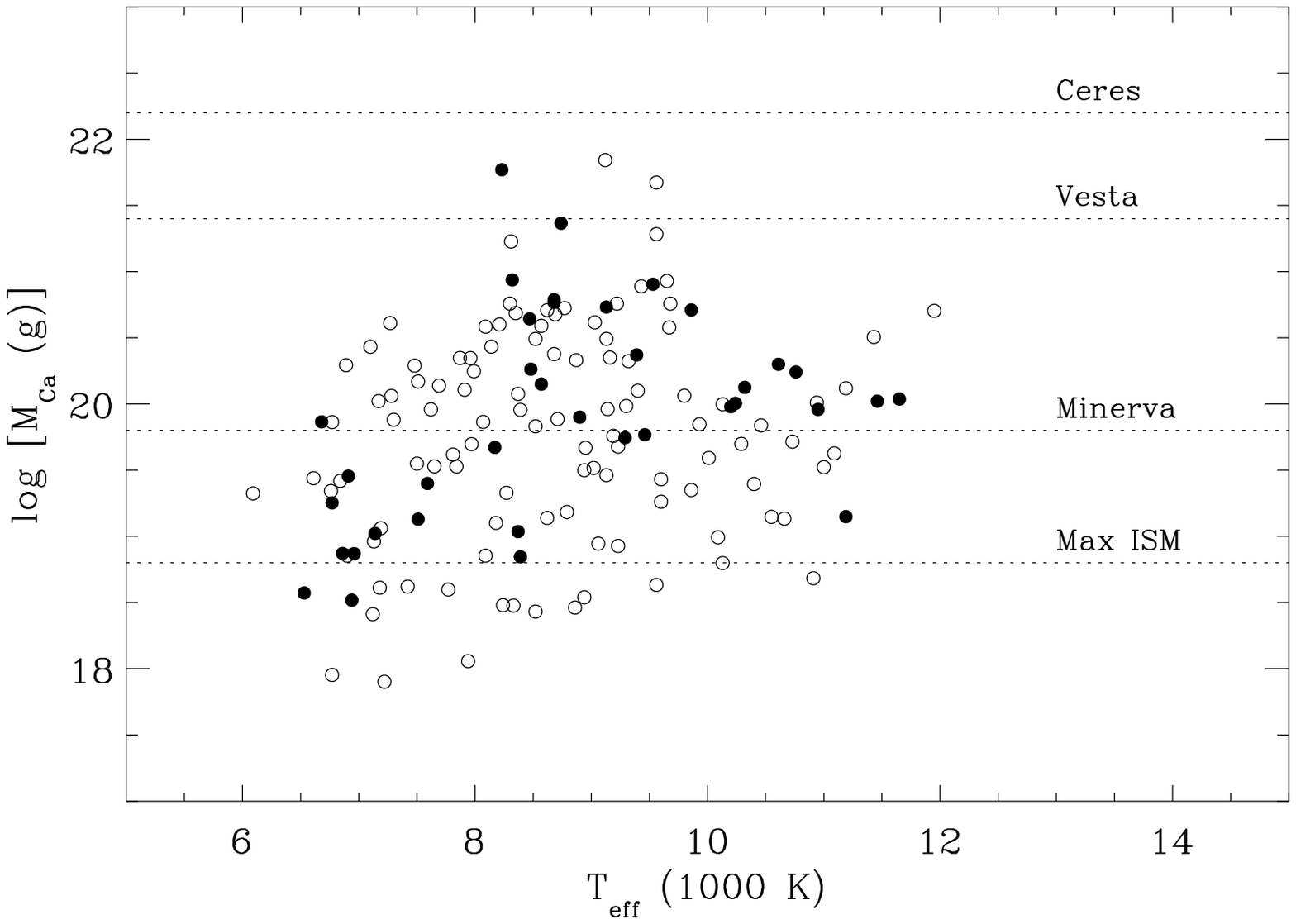}
\caption{Calcium masses in the convective envelopes of 146 cool and metal-polluted white 
dwarfs with helium atmospheres from the SDSS \citep{far10a}.  The open and filled circles 
represent stars with trace hydrogen abundance upper limits and detections, respectively.  The 
top three dotted lines represent the mass of calcium contained in the two largest Solar System 
asteroids Ceres and Vesta, and the 150\,km diameter asteroid Minerva, assuming calcium is 
1.6\% by mass as in the bulk Earth \citep{all95}.  The dotted line at the bottom is the maximum 
mass of calcium that can be accreted from interstellar dust by a 50\,km\,s$^{-1}$ cool white 
dwarf in a $\rho=1000$\,cm$^{-3}$ interstellar cloud over 10$^6$\,yr 
\citep{far10a}.
\label{fig22}}
\end{figure}

\clearpage

\begin{figure}
\figurenum{23}
\epsscale{0.9}
\plotone{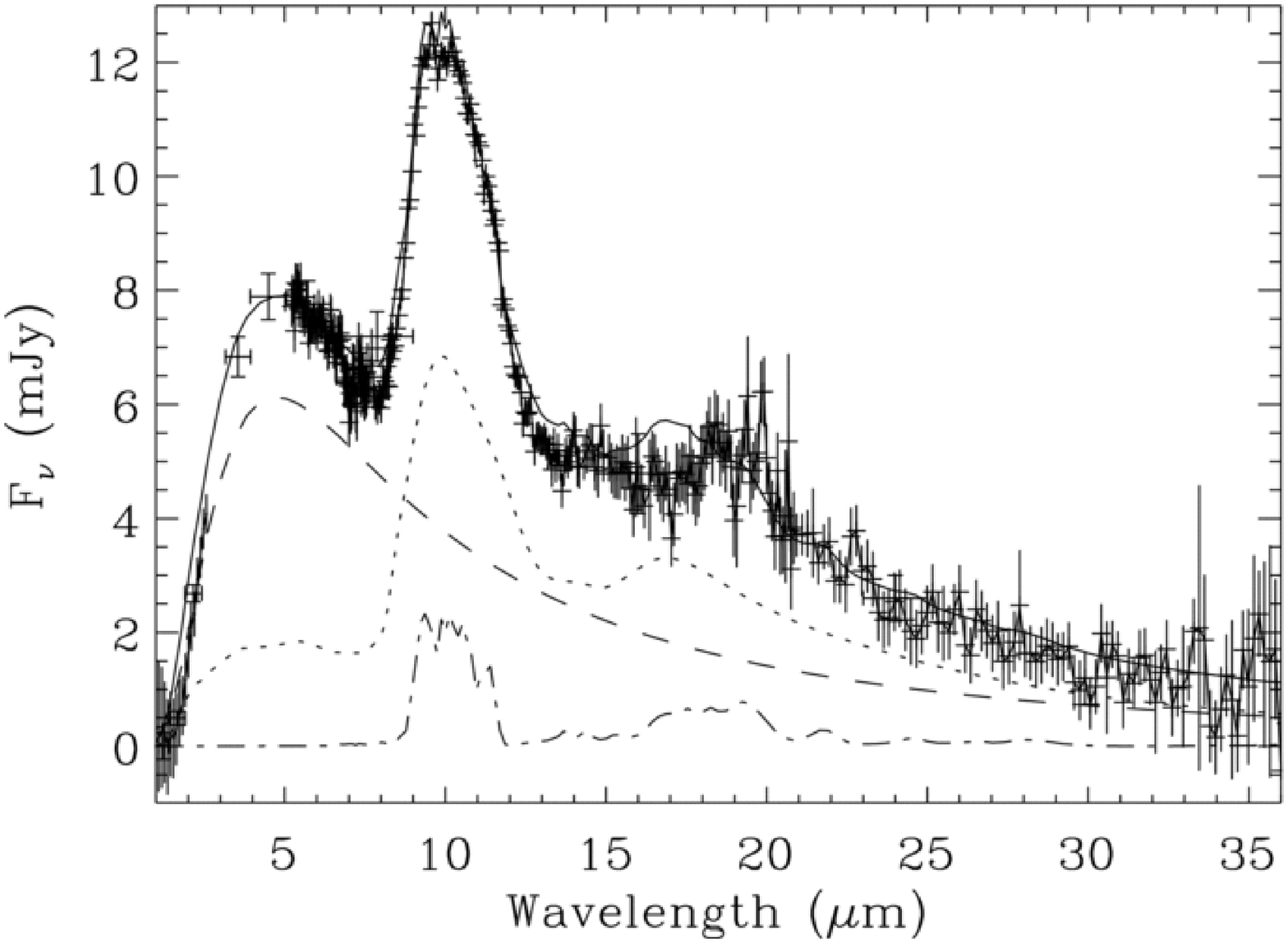}
\epsscale{1.0}
\caption{{\em Spitzer} IRS $5-35\,\mu$m spectrum of the circumstellar dust at G29-38 
\citep{rea09}.  Data are represented by crosses while the best fitting, flat disk plus optically 
thin layer, model is shown as a solid line.  The dashed line is the contribution of the optically 
thick, flat disk, while the emission from silicates is shown as dotted (olivine) and dash-dotted 
(pyroxene) lines.  Multiple model fits to the data produce equally good agreement \citep{rea09} 
but the model shown here is the most physically plausible.
\label{fig23}}
\end{figure}

\clearpage

\begin{figure}
\figurenum{24}
\epsscale{0.8}
\plotone{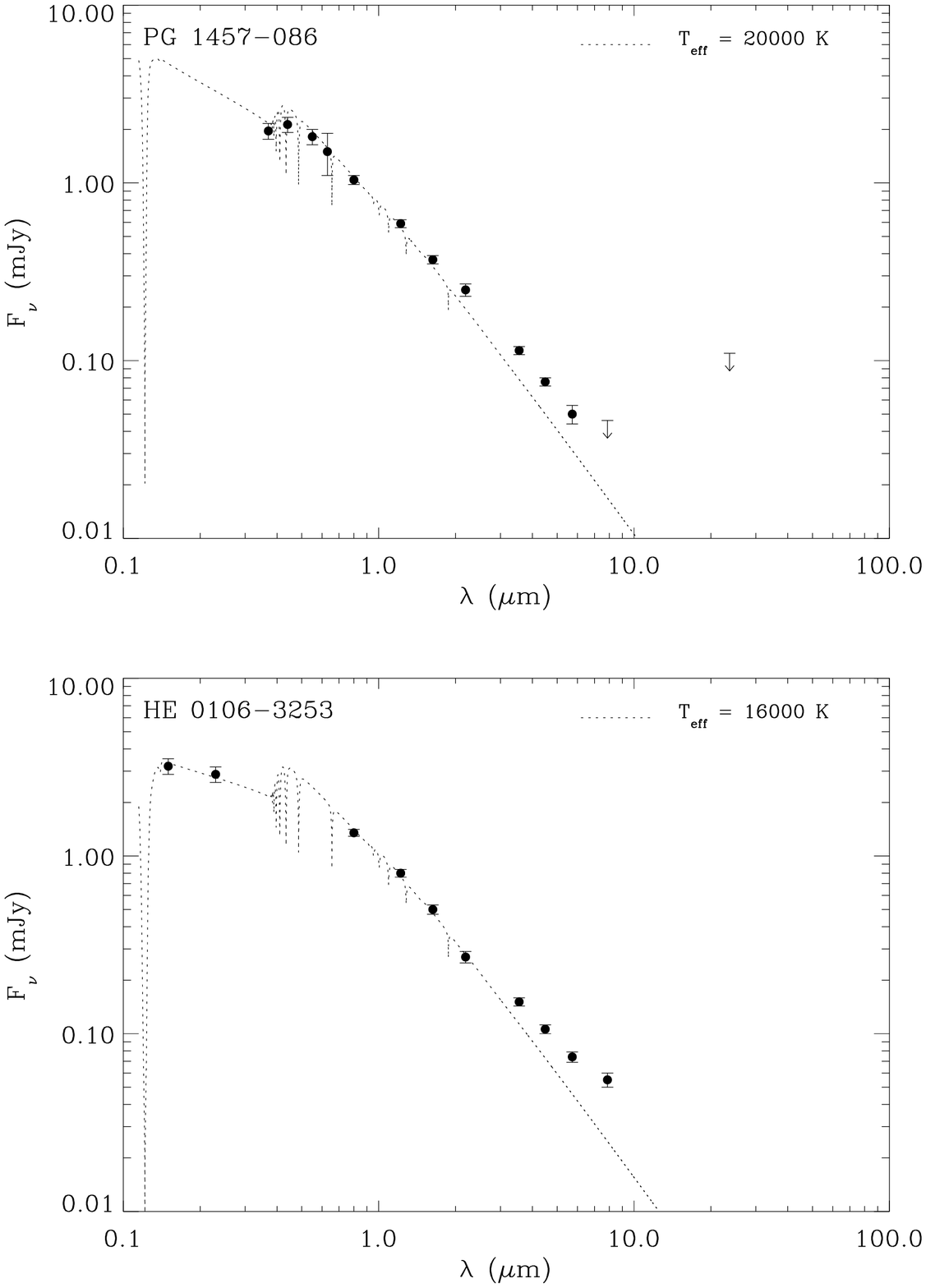}
\epsscale{1.0}
\caption{The two most subtle infrared excesses detected from narrow, circumstellar dust rings 
\citep{far10c,far09b}.  In each case, dedicated near-infrared $JHK$ photometry was instrumental 
in the recognition of the mid-infrared excess.  Flat disk models predict radial widths $\Delta r<0.1
\,R_{\odot}$ at modest inclinations, while for zero inclination (i.e.\ face-on) these rings shrink to 
only 0.01\,$R_{\odot}$ \citep{far10c}.
\label{fig24}}
\end{figure}

\clearpage

\begin{figure}
\figurenum{25}
\plotone{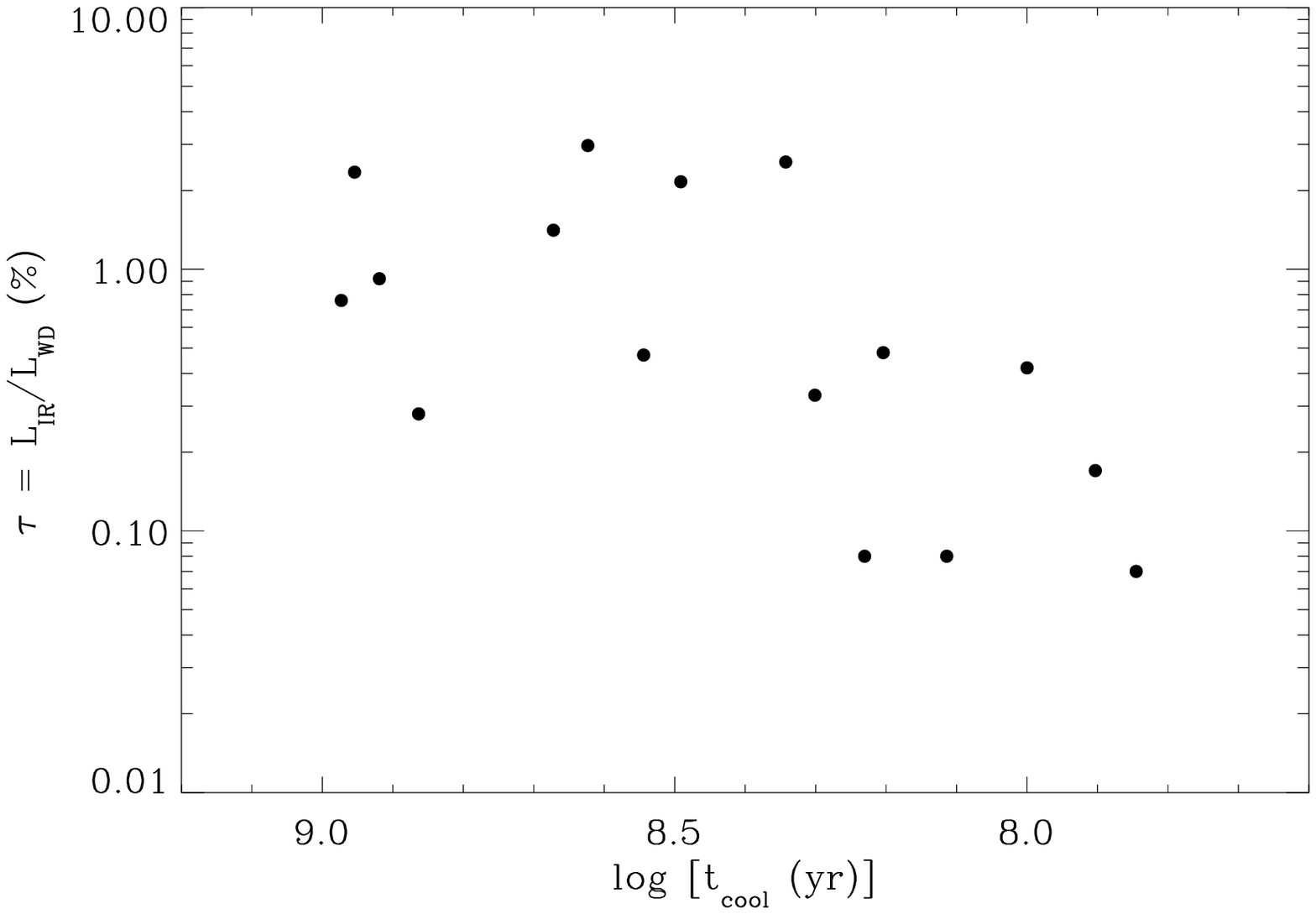}
\caption{Fractional infrared luminosity from the thermal continuum (see Table \ref{tbl3} of 
circumstellar dust disks at white dwarfs plotted versus cooling age.  There is insufficient data 
to confidently identify a trend, but it appears the narrowest rings are found at younger systems, 
while rings comparable to those at Saturn occur in more evolved systems.  If the trend is real, 
possible explanations include viscous spreading over long timescales \citep{far08b,von07} 
or a decrease over time in the frequency of additional asteroid impacts during disk evolution 
\citep{jur08}.
\label{fig25}}
\end{figure}

\clearpage

\begin{figure}
\figurenum{26}
\plotone{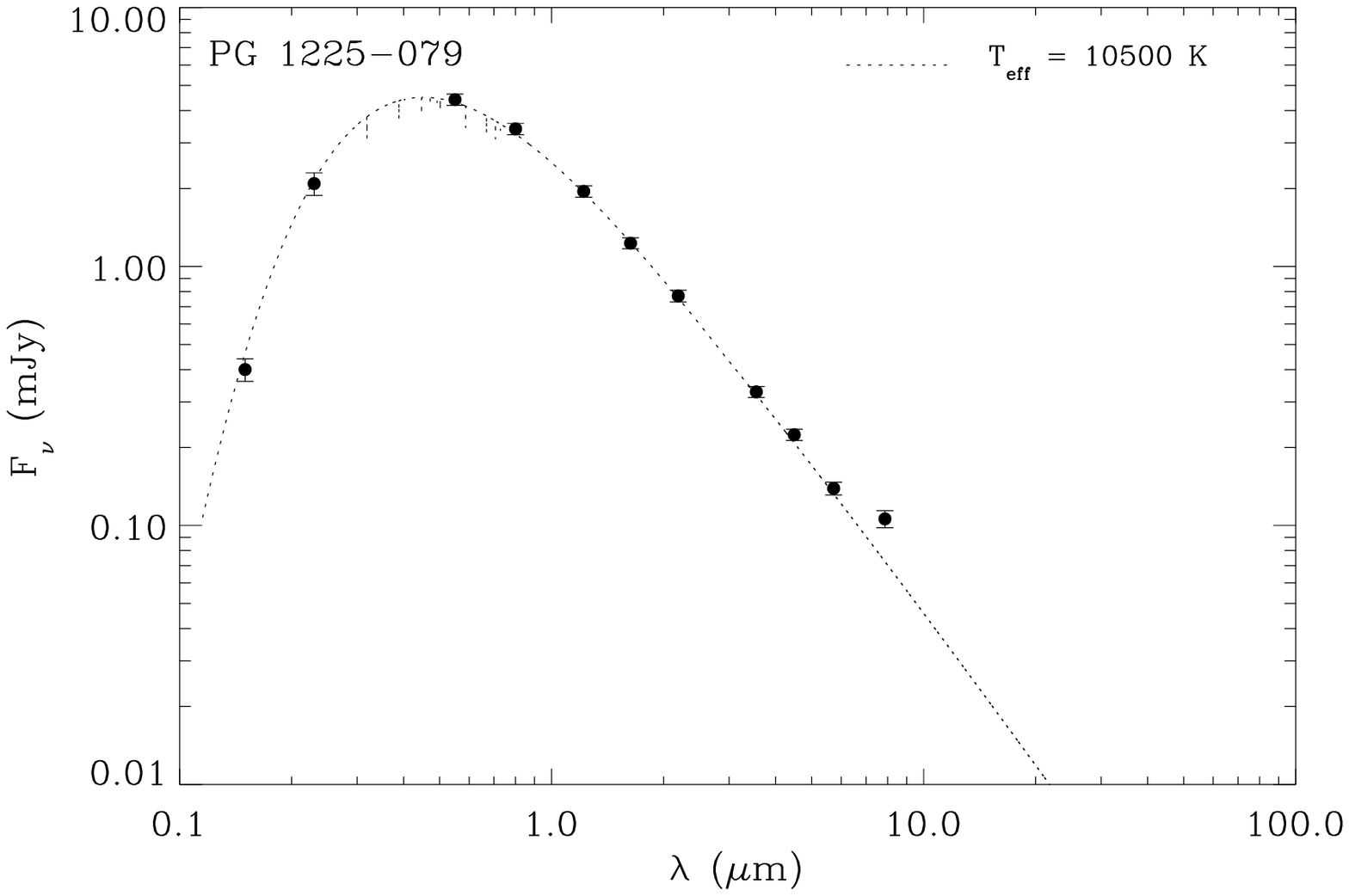}
\caption{SED of PG\,1225$-$079 with {\em Spitzer} IRAC flux measurements \citep{far10c}.  
All available photometric data are of sufficient quality to be confident of the measured excess at 
7.9\,$\mu$m, but without corroborating observations the single data point is somewhat uncertain.
\label{fig26}}
\end{figure}

\clearpage

\begin{figure}
\figurenum{27}
\plotone{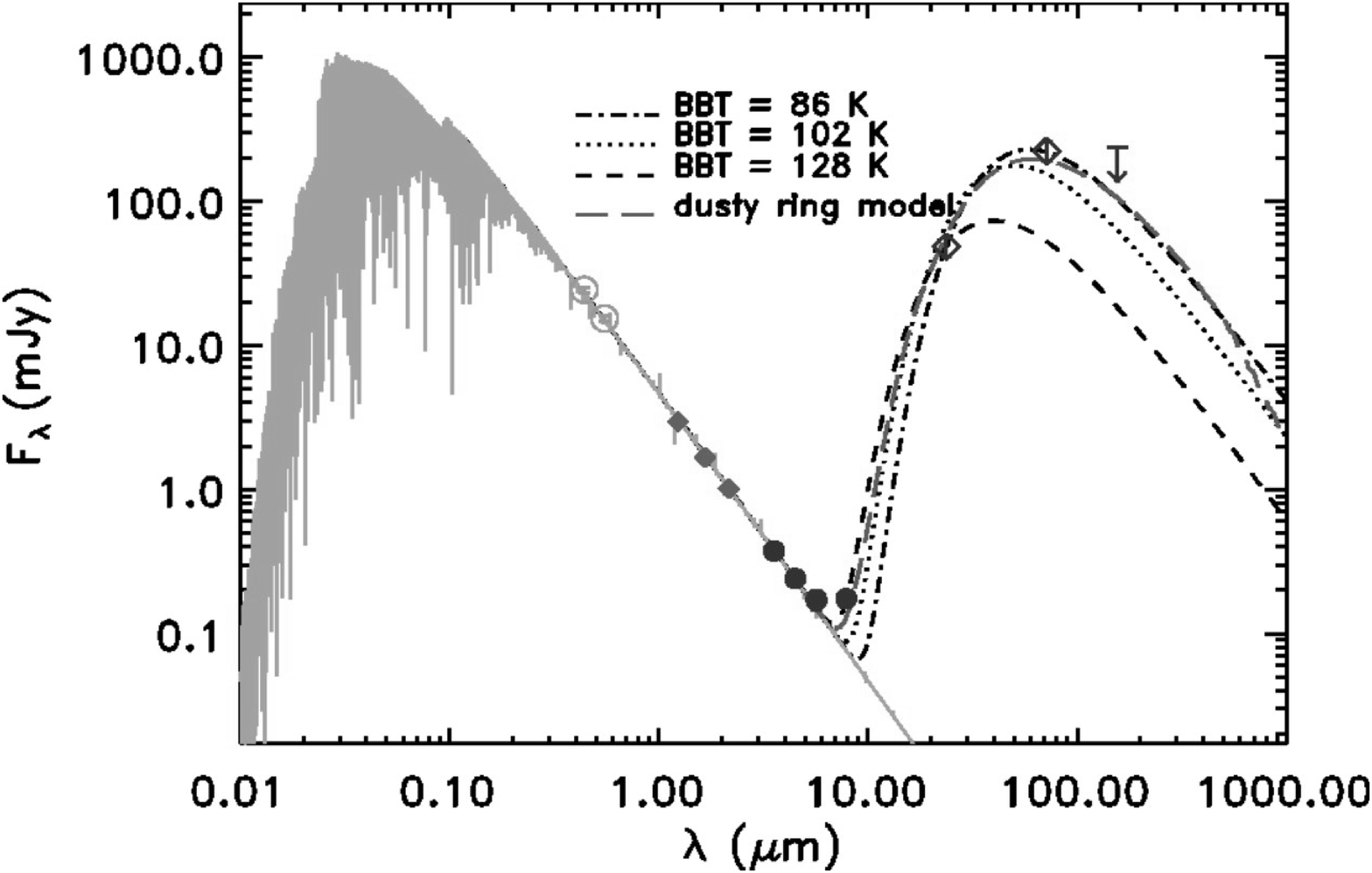}
\caption{The strong infrared excess detected by {\em Spitzer} at the location of the central 
star in the Helix Nebula \citep{su07}.  The combination of multiple, known emission sources
within the nebula and the low spatial resolution of MIPS beam makes the interpretation of the 
excess ambiguous.  A cold debris disk is one possibility but scenarios involving a companion 
star are perhaps more plausible \citep{bil09}.
\label{fig27}}
\end{figure}

\clearpage

\begin{deluxetable}{ccccccccc}
\tabletypesize{\footnotesize}
\tablecaption{Metal-Rich White Dwarfs Observed by {\em Spitzer} IRS at $5-15\,\mu$m\label{tbl1}}
\tablewidth{0pt}
\tablehead{
\colhead{WD}				&
\colhead{Name}			&
\colhead{$F_{2.2\mu{\rm m}}$}	&
\colhead{$F_{7.9\mu{\rm m}}$}	&
\colhead{Raw S/N}\\

\colhead{}					&
\colhead{}					&
\colhead{($\mu$Jy)}			&
\colhead{($\mu$Jy)}			&
\colhead{}}

\startdata

0146$+$187		&GD\,16		&0.49	&0.45	&3.4\\
0300$-$013		&GD\,40		&0.38	&0.16	&4.7\\
0408$-$041		&GD\,56		&0.58	&1.12	&8.0\\
1015$+$161		&PG			&0.26	&0.12	&\nodata\\
1116$+$026		&GD\,133		&0.96	&0.46	&6.0\\
1150$-$153		&EC			&0.37	&0.61	&3.7\\
1729$+$371		&GD\,362		&0.30	&0.64	&6.8\\
2115$-$560      	&LTT\,8452	&1.64	&0.88	&5.5\\
2326$+$049      	&G29-38		&5.59	&8.37	&22\\

\enddata

\end{deluxetable}

\clearpage

\begin{deluxetable}{ccccccccc}
\tabletypesize{\footnotesize}
\tablecaption{The First 18\tablenotemark{a} White Dwarfs with Circumstellar Dust Disks\label{tbl2}}
\tablewidth{0pt}
\tablehead{
\colhead{WD}			&
\colhead{Name}		&
\colhead{SpT}			&
\colhead{$T_{\rm eff}$}	&
\colhead{$d$}			&
\colhead{$K$}			&
\colhead{Publication}	&
\colhead{Discovery}		&
\colhead{Reference}\\

\colhead{}				&
\colhead{}				&
\colhead{}				&
\colhead{(K)}			&
\colhead{(pc)}			&
\colhead{(mag)}		&
\colhead{Year}			&
\colhead{Telescope}		&
\colhead{}}

\startdata

2326$+$049      	&G29-38			&DAZ		&11700	&14		&12.7	&1987	&IRTF			&1\\
1729$+$371		&GD\,362			&DBZ		&10500	&57		&15.9	&2005	&IRTF/Gemini		&2,3\\
0408$-$041		&GD\,56			&DAZ		&14400	&72		&15.1	&2006	&IRTF			&4\\
1150$-$153		&EC\,11507$-$1519	&DAZ		&12800	&76		&15.8	&2007	&IRTF			&5\\
2115$-$560      	&LTT\,8452		&DAZ		&9700 	&22		&14.0	&2007	&{\em Spitzer}		&6\\
0300$-$013		&GD\,40			&DBZ		&15200	&74		&15.8	&2007	&{\em Spitzer}		&7\\
1015$+$161		&PG				&DAZ		&19300	&91		&16.0	&2007	&{\em Spitzer}		&7\\
1116$+$026		&GD\,133			&DAZ		&12200	&38		&14.6	&2007	&{\em Spitzer}		&7\\
1455$+$298		&G166-58			&DAZ		&7400	&29		&14.7	&2008	&{\em Spitzer}		&8\\
0146$+$187		&GD\,16			&DBZ		&11500	&48		&15.3	&2009	&{\em Spitzer}		&9\\
1457$-$086      	&PG				&DAZ		&20400	&110	&16.0	&2009	&{\em Spitzer}		&9\\
1226$+$109		&SDSS\,1228 		&DAZ		&22200	&142	&16.4	&2009	&{\em Spitzer}		&10\\
0106$-$328		&HE\,0106$-$3253	&DAZ		&15700	&69		&15.9	&2010	&{\em Spitzer}		&11\\
0307$+$077		&HS\,0307$+$0746 	&DAZ		&10200	&77		&16.3	&2010	&{\em Spitzer}		&11\\
0842$+$231		&Ton\,345 		&DBZ		&18600	&120	&15.9	&2008	&{\em AKARI} 		&11\\
1225$-$079		&PG				&DBZ		&10500	&34		&14.8	&2010	&{\em Spitzer}		&11\\
2221$-$165		&HE\,2221$-$1630 	&DAZ		&10100	&70		&15.6	&2010	&{\em Spitzer}		&11\\
1041$+$091		&SDSS\,1043 		&DAZ		&18300	&224	&17.7	&2010	&CFHT/Gemini		&12\\

\enddata

\tablenotetext{a}{1225$-$079 is listed tentatively as the 18$^{\rm th}$ discovery}

\tablerefs{
(1) \citealt{zuc87}
(2) \citealt{bec05};
(3) \citealt{kil05};
(4) \citealt{kil06};
(6) \citealt{kil07};
(6) \citealt{von07};
(7) \citealt{jur07a};
(8) \citealt{far08b};
(9) \citealt{far09b};
(10) \citealt{bri09};
(11) \citealt{far10c};
(12) \citealt{mel10}}

\end{deluxetable}

\clearpage

\begin{deluxetable}{ccccc}
\tabletypesize{\small}
\tablecaption{Thermal Continuum Excess at White Dwarfs with Dust \label{tbl3}}
\tablewidth{0pt}
\tablehead{
\colhead{WD}						&
\colhead{Name}					&
\colhead{Stellar Model}				&
\colhead{$2-6\,\mu$m Blackbody$^a$}	&
\colhead{$\tau=L_{\rm IR}/L_{\rm WD}$}\\

\colhead{}							&
\colhead{}							&
\colhead{$T_{\rm WD}$ (K)}			&
\colhead{$T_{\rm IR}$ (K)}			&
\colhead{}}

\startdata

0106$-$328	&HE\,0106$-$3253		&16\,000		&1400		&0.0008\\
0146$+$187	&GD\,16				&11\,500		&1000		&0.0141\\
0300$-$013	&GD\,40				&15\,000		&1200		&0.0033\\
0408$-$041	&GD\,56				&14\,500		&1000		&0.0257\\
0307$+$077	&HS\,0307$+$0746		&10\,500		&1200		&0.0028\\
0842$+$231	&Ton\,345				&18\,500		&1300		&0.0048\\
1015$+$161	&PG					&19\,500		&1200		&0.0017\\
1041$+$091	&SDSS\,1043			&18\,500		&1500		&0.0008\\
1116$+$026	&GD\,133				&12\,000		&1000		&0.0047\\
1150$-$153	&EC\,11507$-$1519		&12\,500		&900		&0.0216\\
1226$+$110	&SDSS\,1228			&22\,000		&1000		&0.0042\\
1225$-$079:	&PG					&10\,500		&300:		&0.0005:\\
1455$+$298	&G166-58				&7500		&500		&0.0015\\
1457$-$086	&PG					&20\,000		&1800		&0.0007\\
1729$+$371	&GD\,362				&10\,500		&900		&0.0235\\
2115$-$560	&LTT\,8452			&9500		&900		&0.0092\\
2221$-$165	&HE\,2221$-$1630		&10\,100		&1000		&0.0076\\
2326$+$049	&G29-38				&11\,500		&1000		&0.0297\\

\enddata

\tablecomments{Measured infrared excess from thermal continuum emission between 2 and 
6\,$\mu$m, as most stars lack spectroscopic data on any potential silicate emission in the $8-12\,
\mu$m region.}

\tablenotetext{a}{This single temperature is a zeroth order approximation of the true disk SED.}

\end{deluxetable}

\end{document}